\documentclass
{jfm}

\usepackage{amsmath,amsfonts,latexsym,array,lscape,amssymb,
cases,empheq}
\usepackage{graphicx}
\usepackage{bm}
\usepackage{multirow}
\usepackage{array}
\usepackage{setspace}
\usepackage{xcolor}
\usepackage{natbib}
\usepackage{pxfonts}
\usepackage{caption}
\usepackage{graphicx,tikz,caption,subcaption}

\newcommand{\be}{\begin{equation}}
\newcommand{\ee}{\end{equation}}

\usepackage{url}  

\ifCUPmtlplainloaded \else
  \checkfont{eurm10}
  \iffontfound
    \IfFileExists{upmath.sty}
      {\typeout{^^JFound AMS Euler Roman fonts on the system,
                   using the 'upmath' package.^^J}%
       \usepackage{upmath}}
      {\typeout{^^JFound AMS Euler Roman fonts on the system, but you
                   dont seem to have the}%
       \typeout{'upmath' package installed. JFM.cls can take advantage
                 of these fonts,^^Jif you use 'upmath' package.^^J}%
      }
  \else
  \fi
\fi

\ifCUPmtlplainloaded \else
  \checkfont{msam10}
  \iffontfound
    \IfFileExists{amssymb.sty}
      {\typeout{^^JFound AMS Symbol fonts on the system, using the
                'amssymb' package.^^J}%
       \usepackage{amssymb}%
       \let\le=\leqslant  
       \let\ge=\geqslant  \let\geq=\geqslant
      }{}
  \fi
\fi

\ifCUPmtlplainloaded \else
  \IfFileExists{amsbsy.sty}
    {\typeout{^^JFound the 'amsbsy' package on the system, using it.^^J}%
     \usepackage{amsbsy}}
    {\providecommand\boldsymbol[1]{\mbox{\boldmath $##1$}}}
\fi

\title[Wavefronts and modal structure of long surface and interfacial ring waves]{Wavefronts and modal structure of \\ long surface and internal ring waves on\\ a parallel shear current}

\author[Curtis Hooper, Karima Khusnutdinova, Roger Grimshaw]%
{CURTIS \ns HOOPER $^{1,2}$,  %
KARIMA \ns KHUSNUTDINOVA $^1$, %
\and \\
ROGER \ns GRIMSHAW $^{1,3}$ \ns %
	\thanks{Email address for correspondence: K.Khusnutdinova@lboro.ac.uk}
}

\affiliation{
\small $^1$ Department of Mathematical Sciences, Loughborough University, UK\\
\small $^2$ Wolfson School of Mechanical, Electrical and Manufacturing  Engineering,\\ 
\small Loughborough University, UK \\
\small $^3$ Department of Mathematics, University College London, UK
}

\date{?; revised ?; accepted ?. - To be entered by editorial office}

\begin{document}

\maketitle

\begin{abstract}
 We study long surface and internal ring waves propagating  in a stratified fluid over a parallel shear flow.  The far-field modal and amplitude equations for the ring waves are presented in dimensional form. 
We re-derive them from the formulation for plane waves tangent to the ring wave, which opens a way to obtaining important characteristics of the ring waves (e.g. group speed) and to constructing more general `hybrid solutions' consisting of a part of a ring wave and two tangent plane waves.   The modal equations constitute a new spectral problem, and are analysed for a number of examples of surface ring waves in a homogeneous fluid and internal ring waves in a stratified fluid.  The detailed analysis is developed for the case of a two-layered fluid with a linear shear current where we 
study their wavefronts and two-dimensional modal structure. Comparisons are made between the modal functions of the surface waves in a homogeneous and two-layered fluids, as well as the interfacial waves described exactly and in the rigid-lid approximation. We also analyse the wavefronts of surface and interfacial waves for a family of power-law upper-layer currents, which can be used to model wind generated currents, river inflows and exchange flows in straits. A global and local measure of the deformation of wavefronts are introduced and evaluated.
\end{abstract}

\section{Introduction}
The Korteweg-de Vries (KdV) equation and its generalisations such as the Gardner, Ostrovsky and Kadomtsev-Petviashvili (KP) equations are well known as good weakly-nonlinear models describing long surface and  internal waves that are commonly observed in the oceans, see, for example,  
\cite{
	GOSS, HM, 
	GPTK, AB, GHJ}.
Solitary wave solutions of a more general extended KdV (eKdV) model, including embedded solitons and their interactions with the usual solitons, have been recently reviewed and studied by \cite{KST} (see also the references therein). These models apply to the waves with plane or nearly-plane fronts.  

Waves generated in straits, river-sea interaction areas and by tidal interaction with localised topographic features often look like a part of a ring, e.g. \cite{NM, VSS, VSPI}.  Asymptotic theory describing surface ring waves in a homogeneous fluid has been developed from the Boussinesq equations, and without a shear flow,  by \cite{M} and from the Euler equations, including the waves propagating over a parallel depth-dependent shear flow, by \cite{J80,J90}. The generalisation for surface and internal ring waves  in a stratified fluid has been developed, without a shear flow by \cite{L, WV} and with a shear flow by \cite{KZ16a}.  The respective models capture basic balance between nonlinearity and dispersion, describing waves with cylindrical divergence in the KdV regime. Alternative approaches to such problems, and important experimental work,  have been developed, in particular, for surface waves, by \cite{E14a, E14b, SK, AKZ, AE, LE, SEE}, and for internal waves, by \cite{ASK,Gr, BV15, BV20} (see also the references therein).  General approaches to the solution of initial-value problems with the help of cylindrical Korteweg-de Vries - type models have been discussed by \cite{WZ, RRS, MS, KZ16b, G19}.

The generalisation developed by \cite{KZ16a} was based on the existence of a suitable far-field linear modal decomposition, which had more complicated structure than the known modal decomposition for the plane waves. The developed linear formulation provided, in particular,  a description of the distortion of the shape of the wavefronts of surface and internal ring waves in a two-layer fluid by the piecewise-constant current. The wavefronts of surface and interfacial ring waves were described in terms of two branches of the envelope of the general solution of the derived nonlinear first-order differential equation, constituting further generalisation of the well-known Burns  \citep{B}  and generalised Burns \citep{J90} conditions. The two branches of this solution have been described in parametric form. An explicit analytical solution  was developed for the wavefront of the interfacial mode in the rigid-lid approximation for a sufficiently weak current, when a part of the ring wave can propagate in the upstream direction ({\it elliptic} regime), while solutions for stronger currents were developed in \cite{K} ({\it parabolic}  and {\it hyperbolic} regimes). 

The constructed solutions have revealed the qualitatively different behaviour of the wavefronts of surface and interfacial waves propagating over the same piecewise-constant current. Indeed,  while the wavefront of the surface ring wave was elongated in the direction of the flow, the wavefront of the interfacial wave was strongly squeezed in this direction. This phenomenon was linked to the presence of 
long-wave instability  of plane waves tangent to the ring wave and propagating in the downstream and upstream directions for a sufficiently strong current (see \cite{O, BC, BM, LM, KZ16a, K}). 

The aims of the present paper are twofold. Firstly, we briefly present the dimensional form of the modal and amplitude equations for the ring waves in a fluid with arbitrary stratification and depth-dependent parallel shear flow. The equations are derived from the Euler equations written in the cylindrical coordinate system (Section 2). We do that in order to facilitate their use in oceanographic and laboratory studies, similarly to the widely used formulation for the plane waves, and to provide the necessary equations for the derivation of the cylindrical Benjamin-Ono and intermediate-depth type models, which can be obtained using the same modal decomposition. 
Next, we re-derive the modal equations from the formulation for the plane waves tangent to the ring  wave (Section 3), working within the framework of the local wave vector and local wave frequency. Thus, we establish a useful link between the descriptions of obliquely propagating plane waves tangent to a ring  wave, and the ring wave, which allows us to obtain  useful characteristics of the ring waves and to outline a construction of more general {\it hybrid solutions}  formed by a part of a ring wave and two tangent plane waves.   Similarly looking hybrid solutions can be seen, for example, on satellite images of internal waves. Secondly, we aim to analyse the modal equations - a new spectral problem which is at the heart of the theory. We consider several configurations motivated by the modelling of geophysical fluid flows, and introduce new global and local quantitative tools for the description of the deformations of the wavefronts of ring waves propagating over various shear currents. The detailed analytical study is developed for the geometry of the wavefronts and vertical structure of the three-dimensional ring waves in a two-layered fluid with a linear shear current  (Section 4). We compare the exact solutions for surface and interfacial modes with the results obtained in the approximations of the homogeneous fluid for the surface mode, and in the rigid-lid approximation for the interfacial mode. We also discuss surface and interfacial modes for a family of power-law upper-layer currents, in which case solutions can be constructed in terms of the hypergeometric function (Section 5).  Significant squeezing similar to that described for a piecewise-constant current can take place for some currents in the family.
Such currents are close to river inflows and exchange flows in straits, while for wind-generated - type currents the wavefronts appear to be elongated in the direction of the current.
We conclude in Section 6.

\section{Dimensional modal and amplitude equations for ring waves}
The derivation of the modal and amplitude equations described in this section  briefly overviews that given in \cite{KZ16a} but it is developed in dimensional form.  
Also, we reformulate the boundary conditions assuming that the bottom is at $z = -h$ and the undisturbed surface is at $z=0$, which is customary in oceanographic applications. These modifications aim to make the theory directly applicable in oceanographic contexts.

We consider a ring wave propagating in an inviscid incompressible fluid, described by the full set of Euler equations with the free surface and rigid bottom boundary conditions. Assuming that the waves are long we neglect surface tension. 
We assume that $u,v,w$ are the velocity components in the $x,y,z$ directions respectively, $p$ is the pressure, $\rho$ is the density of the fluid, $g$ is the acceleration due to gravity, $z=\eta(x,y,t)$ is the height of the free surface (with $z=0$ at the unperturbed surface, and $z=-h$ at the flat bottom) and $p_a$ is the atmospheric pressure at the surface. 
The vertical particle displacement $\zeta$ is used as an additional independent variable, which is defined by the equation
\begin{equation}
	\zeta _t + u\zeta_x +v\zeta_y + w\zeta_z = w
	\label{zeta}
\end{equation}
and  the surface boundary condition
\begin{equation}
	\zeta = \eta \qquad at \hspace{0.5cm} z=\eta(x,y,t).
	\label{zeta1}
\end{equation}
The fluid is in the following basic state:
\be
u_0=u_0(z),\ \  v_0=w_0=0,\ \  p_{0z}=- \rho_0 g,\  \ \zeta = 0.
\ee
Here $u_0(z)$ is a horizontal shear flow in the $x$-direction and $\rho_0=\rho_0(z)$ is a stable background density stratification.

We introduce the cylindrical coordinate system moving at a constant speed $c$, and use the same notations for the projections of the velocity field on the new coordinate axes:
\begin{eqnarray}
	x\rightarrow ct+r\cos\theta, \hspace{1cm}y\rightarrow r\sin\theta, \hspace{1cm}z\rightarrow z, \hspace{1cm} t\rightarrow t, \\
	u\rightarrow u_0(z)+ u\cos\theta -v\sin\theta, \hspace{1cm} v\rightarrow u\sin\theta+v\cos\theta, \\
	w\rightarrow w, \hspace{1cm} p\rightarrow p, \hspace{1cm}\rho \rightarrow \rho_0+\rho.
\end{eqnarray}
Then, the equations take the form
\begin{eqnarray}
	(\rho_0 + \rho) \left [u_t + u u_r + \frac{v}{r} u_{\theta} + w u_z  - \frac{v^2}{r}  + ((u_0-c) u_r + u_{0z} w) \cos \theta  \right .\nonumber \\
	\hspace{6cm} \left .  - (u_0-c) (u_{\theta}-v) \frac{\sin \theta}{r}\right ] + p_r = 0, \label{1a}\\
	(\rho_0 + \rho) \left [v_t + u v_r + \frac{v}{r} v_{\theta} + w v_z + \frac{uv}{r}  + (u_0-c) v_r \cos \theta \right .  \nonumber \\
	\hspace{4.5cm} \left .  - \left ((u_0-c)\bigg(\frac{v_{\theta}}{r} + \frac{u}{r}\bigg) + u_{0z} w\right ) \sin \theta \right ] + \frac{p_{\theta}}{r} = 0, \label{2a}\\
	(\rho_0 + \rho) \left [ w_t + u w_r + \frac{v}{r} w_{\theta} + w w_z  + (u_0-c) \left ( w_r \cos \theta - w_{\theta} \frac{\sin \theta}{r} \right ) \right ] 
	\nonumber  \\
	\hspace{9cm} 
	+ p_z + g \rho = 0, \label{3a}\\
	\rho_t + u \rho_r + \frac{v}{r} \rho_{\theta} + w \rho_z  + (u_0-c) \left (\rho_r \cos \theta - \rho_{\theta} \frac{\sin \theta}{r} \right ) + \rho_{0z} w = 0, \label{4a} \\
	u_r + \frac{u}{r} + \frac{v_{\theta}}{r} + w_z = 0, \label{5a}\\
	w = \eta_t + u \eta_r + \frac{v}{r} \eta_{\theta}  + (u_0-c) \left (\eta_r \cos \theta - \eta_{\theta} \frac{\sin \theta}{r} \right ) \quad \mbox{at} \quad z =  \eta, \label{6a}\\
	p = \int_{-h}^{\eta} g \rho_0(s) ds \quad \mbox{at} \quad z = \eta, \label{7a} \\
	w = 0 \quad \mbox{at} \quad z=-h, \label{8a}
\end{eqnarray}
with the vertical particle displacement satisfying the following equation and boundary condition:
\begin{eqnarray}
	\zeta_t + u \zeta_r + \frac{v}{r} \zeta_{\theta} + w \zeta_z + (u_0-c) \left ( \zeta_r \cos \theta - \zeta_{\theta} \frac{\sin \theta}{r} \right ) = w, \label{9a}\\
	\zeta = \eta \quad \mbox{at} \quad z = \eta. \label{10a}
\end{eqnarray}

The derivation by \cite{KZ16a} was based on the observation that the linearised equations in the far field ($r\sim$ \cal{O}($\varepsilon^{-1}$)), where $\varepsilon$ is a small amplitude parameter) admit the modal decomposition (separation of variables) of the form
\begin{eqnarray}
	\displaystyle \zeta_1 =A(\xi, R, \theta) \phi(z,\theta), \label{zeta_1} \\
	\displaystyle u_1 = -A\phi u_{0z}\cos\theta - \frac{m \hat F}{m^2+m'^2}A\phi_z, \label{LOS1}\\
	\displaystyle v_1 = A\phi u_{0z}\sin\theta - \frac{m' \hat F}{m^2+m'^2}A\phi_z, \\
	w_1 = A_{\xi} \hat F\phi,\\
	\displaystyle p_1 =\frac{\rho_0}{m^2+m'^2}A \hat  F^2\phi_z, \\
	\rho_1 = -\rho_{0z}A\phi, \\
	\eta_1 =A\phi \hspace{0.3cm} \text{at} \hspace{0.3cm} z=0, \label{LOS6}
\end{eqnarray}
where  $\xi=m(\theta) r - st$, $R = \varepsilon r m(\theta)$
and $s$ was defined to be the wave speed in the absence of any shear flow (with $m=1$). 
The function $\phi = \phi(z;\theta)$ is non-dimensional, and it satisfies the following modal equations:
\begin{equation}
	\Bigg(\frac{\rho _0 \hat  F^2}{m^2 + m'^2} \phi_z \Bigg)_z + \rho_0 N^2 \phi = 0, 
	\label{eq:ME1}
\end{equation}
\begin{equation}
	\frac{\hat  F^2}{m^2+m'^2}\phi _z - g \phi = 0 \hspace{0.4cm} \text{at } \quad z=0, 
	\label{eq:free_surface_cond}
\end{equation}
\begin{equation}
	\phi = 0 \hspace{0.4cm} \text{at } \quad z=-h,
	\label{eq:MEC2}
\end{equation}
where
\begin{equation}
	\hat  F = \hat  F(z; \theta)  = -s+(u_0 -c)(m\cos \theta -m'\sin \theta), \quad N^2 = - \frac{g \rho_{0z}}{\rho_0},
	\label{eq:F}
\end{equation}
$m=m(\theta)$ and $m'=dm/d\theta$. We fixed the speed of the moving coordinate frame $c$ to be equal to the speed of the shear flow at the bottom of the fluid. 

We will refer to the non-dimensional function $m(\theta)$ as the {\it speed modifying function} (or simply as {\it modifying function}) for the speed of the ring wave in a particular direction compared to the speed $s$ in the absence of the shear flow, and we shall refer to the corresponding differential equation for this function as the {\it directional adjustment equation} (or simply as {\it adjustment equation}).  Indeed, the modified speed of a linear long wave propagating at an angle $\theta$ to the current is $\displaystyle \frac{s}{m(\theta)}$, and the adjustment equation can be regarded as the {\it 2D long-wave dispersion relation}.

The derivation of the nonlinear amplitude equation was then developed using an asymptotic multiple-scale expansion around this leading-order far-field solution. 
It was based on the existence of two small parameters; the amplitude parameter $\epsilon =a/h$ and the wavelength parameter $\delta = h/ \lambda$, where $a$ and $\lambda$ were the characteristic amplitude and wavelength.  The `maximal balance' condition used to derive the nonlinear amplitude equation is $\delta^2=\epsilon$. 
The dimensional form of the equation for the amplitude function $A(\xi, R, \eta)$  is given in Appendix A.

To leading order, the shape of a wavefront in the far-field at a distance $r$ from the origin at a fixed moment of time is given by the equation $m(\theta) r - st = const$ and we require that $m(\theta)\ > 0$ considering an outward propagating ring wave. Both $s$ and $m(\theta)$ are to be determined from the solution of the modal equations.
The vertical structure of the wave field is also defined by the modal equations. Moreover, the coefficients of the amplitude equation depend on the solution of the modal equations (see Appendix A).   Thus, the system of modal equations (\ref{eq:ME1}) - (\ref{eq:MEC2}) constitutes an important new spectral problem. Solutions for various configurations of the basic stratification and shear flow need to be found in order to make progress in the study of the three-dimensional ring waves and their generalisations (see Section 3). 
Therefore, our present paper is devoted to the analysis of the modal equations.

\begin{figure}
	\begin{center}
		\includegraphics[width=0.5 \textwidth]{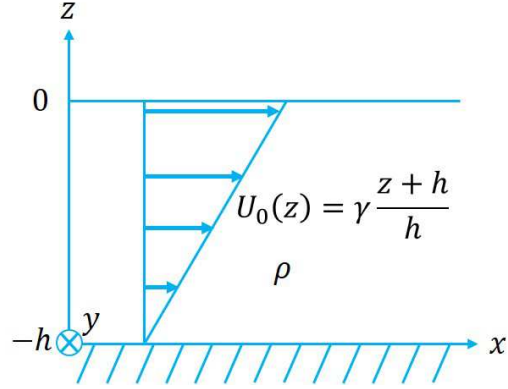}
		\caption{Homogeneous fluid with a linear shear current.}
		\label{fig:homo}
	\end{center}
\end{figure}

To illustrate, let us first re-consider an example of surface ring waves in a homogeneous fluid with a shear flow  \citep{J90, J_book} from the viewpoint of the generalised formulation (\ref{eq:ME1}) - (\ref{eq:MEC2}), and  in dimensional form.   In particular, let us choose the linear shear flow shown in Figure \ref{fig:homo}. We take the density of the fluid, $\rho _0$, to be a constant, whilst the shear flow is given by $\displaystyle u_0 (z) = \gamma \frac{z+h}{h} $ where $\gamma$ is also a positive constant. 

We will use this example in order to formulate a rather general  sufficient condition ensuring the absence of critical layers, and to  introduce quantitative measures for the description of the deformation of the wavefront of a ring wave on a shear current. We will also examine the two-dimensional vertical structure of the ring waves described by the modal functions. This analysis is new since the formalism developed in \citep{J90, J_book} was not based on the ideas of modal decomposition. 

\begin{figure}
	\begin{center}
		\includegraphics[width=0.49 \textwidth]{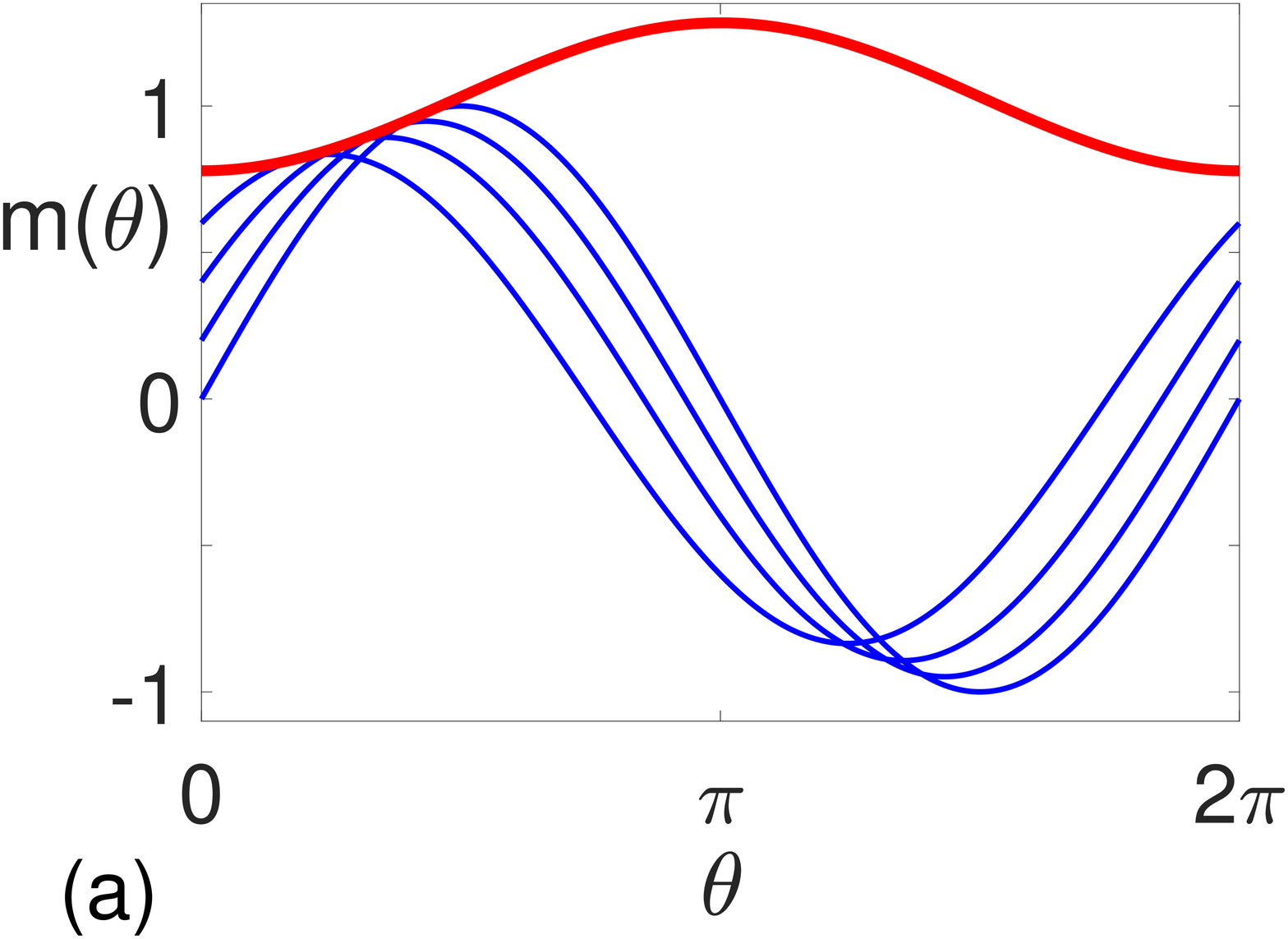}
		\includegraphics[width=0.48 \textwidth]{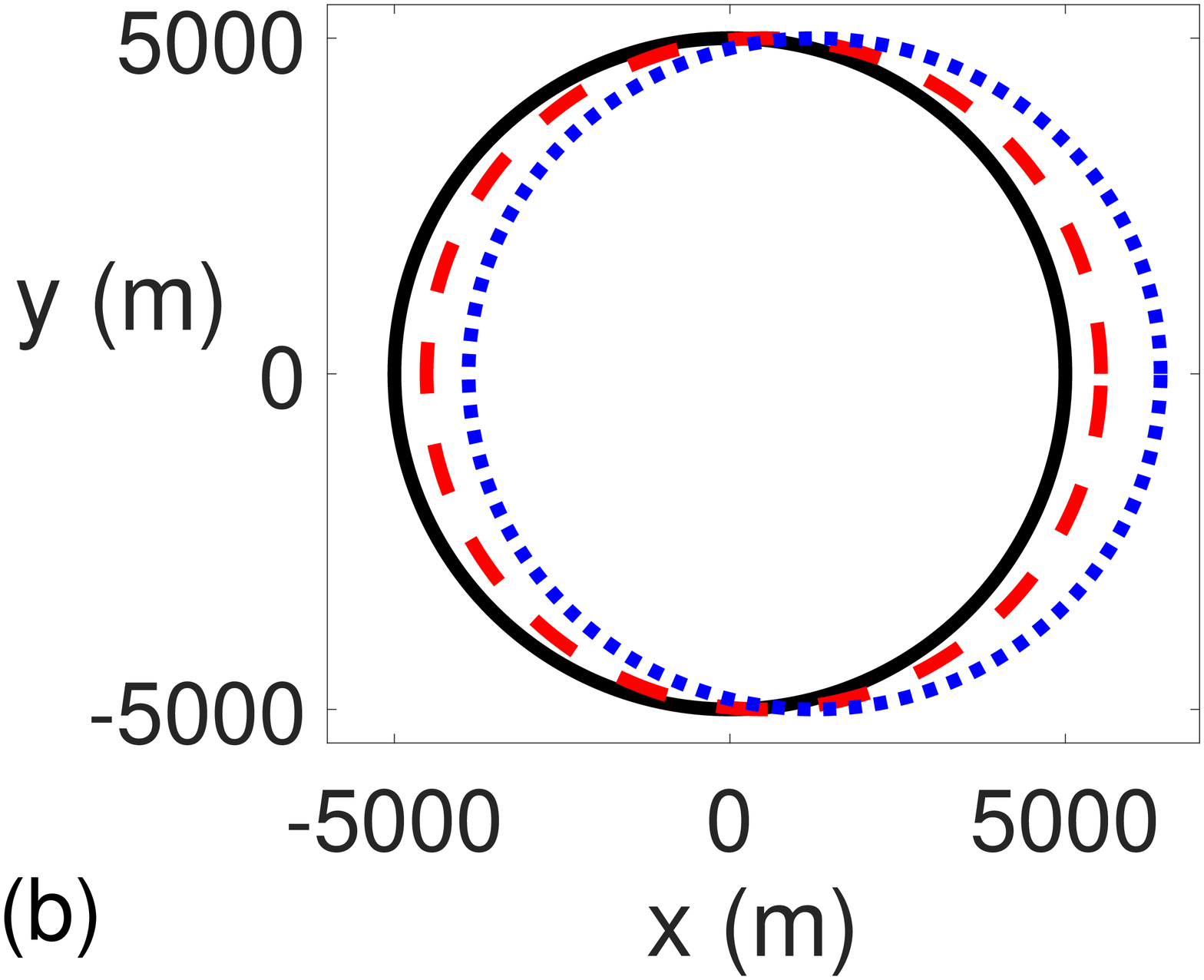}
		\caption{(a) The general solution \eqref{eq:GS} for $a =0,\ 0.2,\ 0.4,\ 0.6$ (blue, thin) with its envelope \eqref{eq:ss1} (red, thick) for $\gamma = 5\ m s^{-1}$. 
			(b) Wavefronts of surface ring waves. The black (solid) curve is for $\gamma = 0 \ m s^{-1}$, the red (dash) curve for $\gamma = 2\ m s^{-1}$ and the blue (dot) curve for $\gamma = 5\ m s^{-1}$. Here, $g = 9.8\ m s^{-2}$,  $h = 10\ m$ and $r m(\theta) = 5000\ m$.  
		}
		\label{fig:ss2}
	\end{center}
\end{figure}

On solving \eqref{eq:ME1} subject to \eqref{eq:MEC2} we find
\begin{align}
	\phi = \frac{\Lambda(m^2 + m'^2)}{\rho}\int_{-h}^z \frac{1}{\hat F^2} dz 
	= \dfrac{\Lambda (m^2 + m'^2)(z + h)}{\displaystyle  \rho s \left [s - \gamma \frac{z+h}{h} (m\cos\theta - m' \sin\theta)\right ]},
	\label{eq:phi_with_shear}
\end{align} 
where $\Lambda$ is a parameter which may depend on $\theta$. Then, to satisfy the condition \eqref{eq:free_surface_cond}, we find that the speed modifying function $m$ must satisfy the differential equation 
\begin{equation}
	m^2 + m'^2 = \frac{s}{gh} [s - \gamma  (m\cos\theta - m' \sin\theta)].
	\label{eq:m_ode}
\end{equation}
Assuming the absence of a shear flow by setting $\gamma = 0$ and $m = 1$, we have from \eqref{eq:m_ode} that
\begin{equation}
	s^2 = gh,
	\label{eq:s}
\end{equation}
thus \eqref{eq:m_ode} becomes
\begin{equation}
	m^2 + m'^2 =  1 - \frac{\gamma}{s} (m \cos\theta - m' \sin \theta).
	\label{eq:gammaza}
\end{equation} 
This coincides with the generalised Burns condition for this linear shear flow \citep{J90, J_book} but is given in dimensional form. It is a nonlinear first-order differential equation which has a general solution of the form
\begin{equation}
	m(\theta) = a \cos \theta + b(a) \sin \theta, \quad \mbox{where} \quad b(a) = \pm \sqrt{1 - \frac{\gamma}{s}  a - a^2}.
	\label{eq:GS}
\end{equation}
It was shown that the solution that describes a ring wave is in fact the envelope of the general solution (so-called {\it singular solution} of (\ref{eq:gammaza})) \citep{J90, J_book}. This solution is found by requiring that  $\displaystyle \frac{dm}{da} = 0$, which implies
$
\displaystyle b'(a) = -\frac{1}{\tan \theta}
$
and allows one to find the singular solution in the form
\begin{equation}
	m(\theta) = -\frac{\gamma}{2s}  \cos\theta \pm \sqrt{1+\frac{\gamma^2}{4s^2}}.
	\label{eq:ss1}
\end{equation}
The upper sign should be chosen for the outward propagating ring wave, so that $m > 0$ for all values of $\theta$. From this we can recover that in the absence of a  shear flow, with $\gamma = 0$, $m(\theta)=1$, a condition that we stated for concentric waves. The general solution (\ref{eq:GS}) and its singular solution (\ref{eq:ss1}) are shown in Figure \ref{fig:ss2}(a). The wavefronts for $\gamma = 0\ m s^{-1}$, $\gamma = 2\ m s^{-1}$ and $\gamma = 5\ m s^{-1}$ are shown in Figure \ref{fig:ss2}(b). Naturally for $\gamma=0\ m s^{-1}$, in the absence of a shear flow, the wavefront takes the form of a circle. Increasing the value of $\gamma$ elongates the wavefronts of surface waves in the direction of the current. 

A critical layer occurs when $\hat F(z, \theta) = 0$. Consider $\hat F_{\theta} = - (u_0 - c) (m + m'') \sin \theta$. Let us assume that $u_0 - c > 0$, i.e. there are no current reversals, which is  the case in the example above and in all subsequent examples. We shall consider a singular solution satisfying $m + m'' > 0$, i.e. an outward propagating wave. Indeed, such a wave in the absence of any current is described by $m = 1$, therefore $m+m'' > 0$ for a sufficiently weak current, by continuity. Then, $\hat F_\theta < 0$ if $\theta \in (0, \pi)$ and $\hat F_\theta > 0$ if $\theta \in (\pi, 2 \pi)$. Therefore, $\hat F$ will reach a maximum at $\theta = 0$. Thus, to avoid critical layers we require
\begin{equation}
	\hat F \le \hat F|_{\theta = 0} = - s + (u_0 - c) m(0) < 0,
\end{equation}
implying $(u_0-c) m(0) < s$.  Here, $\displaystyle \frac{s}{m(0)} \ge s$ is the downstream wave speed, implying
\begin{equation}
	u_0 - c < s \le \frac{s}{m(0)}.
\end{equation}
Thus, the inequality $u_0 - c < s$ is a simple, and rather general sufficient condition for the absence of critical layers, generalising the condition formulated by \cite{KZ16a} for a piecewise-constant current.  The inequality $\displaystyle u_0 - c < \frac{s}{m(0)}$ is less restrictive, and this is the necessary and sufficient condition. It requires the knowledge of $m(0)$. Both conditions are applicable to all examples in this paper, and there are no critical levels. In particular, for the current example, $\displaystyle u_0 = \gamma \frac{z+h}{h}$, and we obtain the sufficient condition
\be
\gamma < \sqrt{gh}.
\ee
The necessary and sufficient condition reads 
\be
\gamma < \frac{\sqrt{gh}}{\displaystyle - \frac{\gamma}{2 \sqrt{gh}} + \sqrt{1+ \frac{\gamma^2}{4 gh}}}.
\ee

\begin{figure}
	\begin{center}
		\includegraphics[width=0.49 \textwidth]{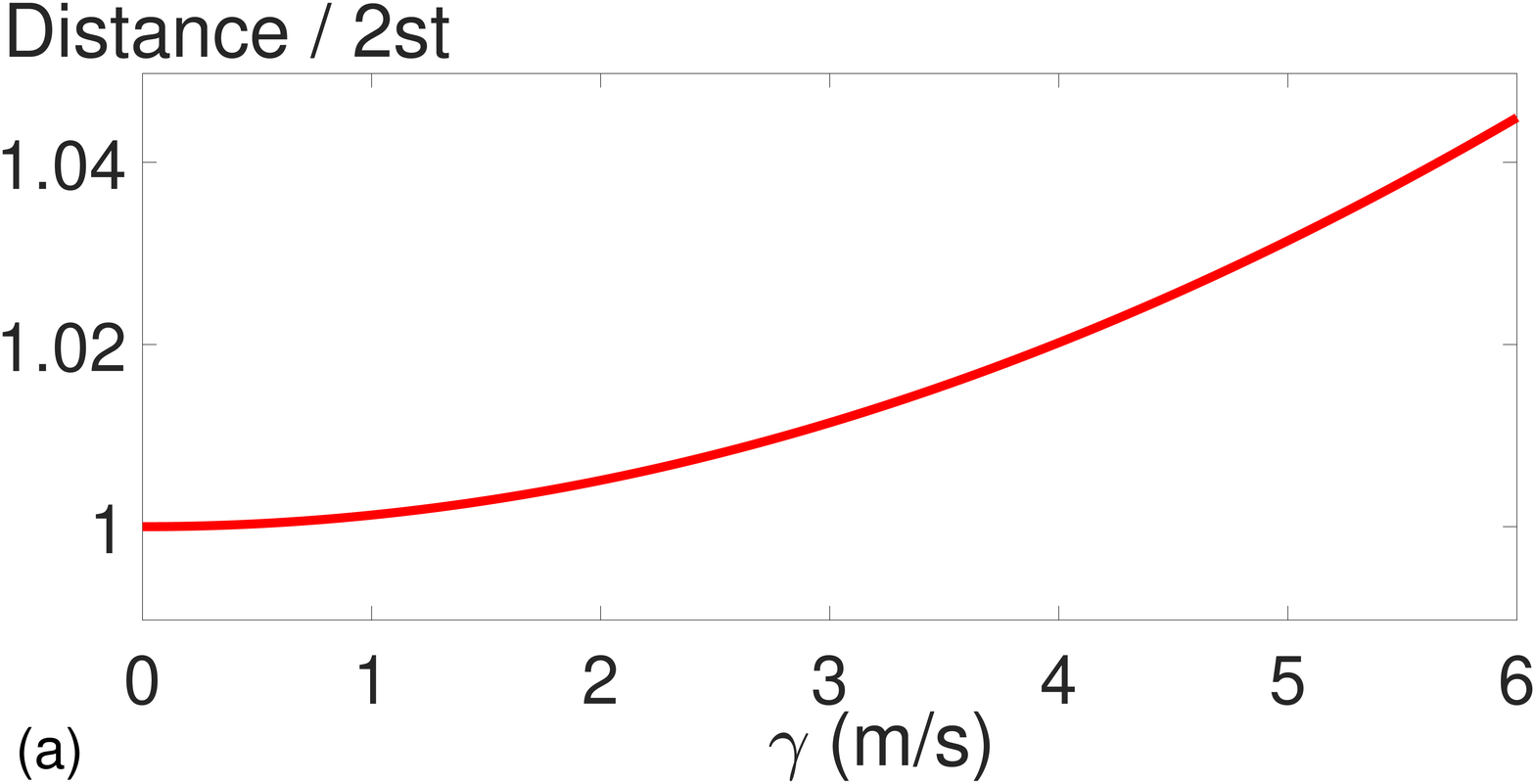}
		\includegraphics[width=0.49 \textwidth]{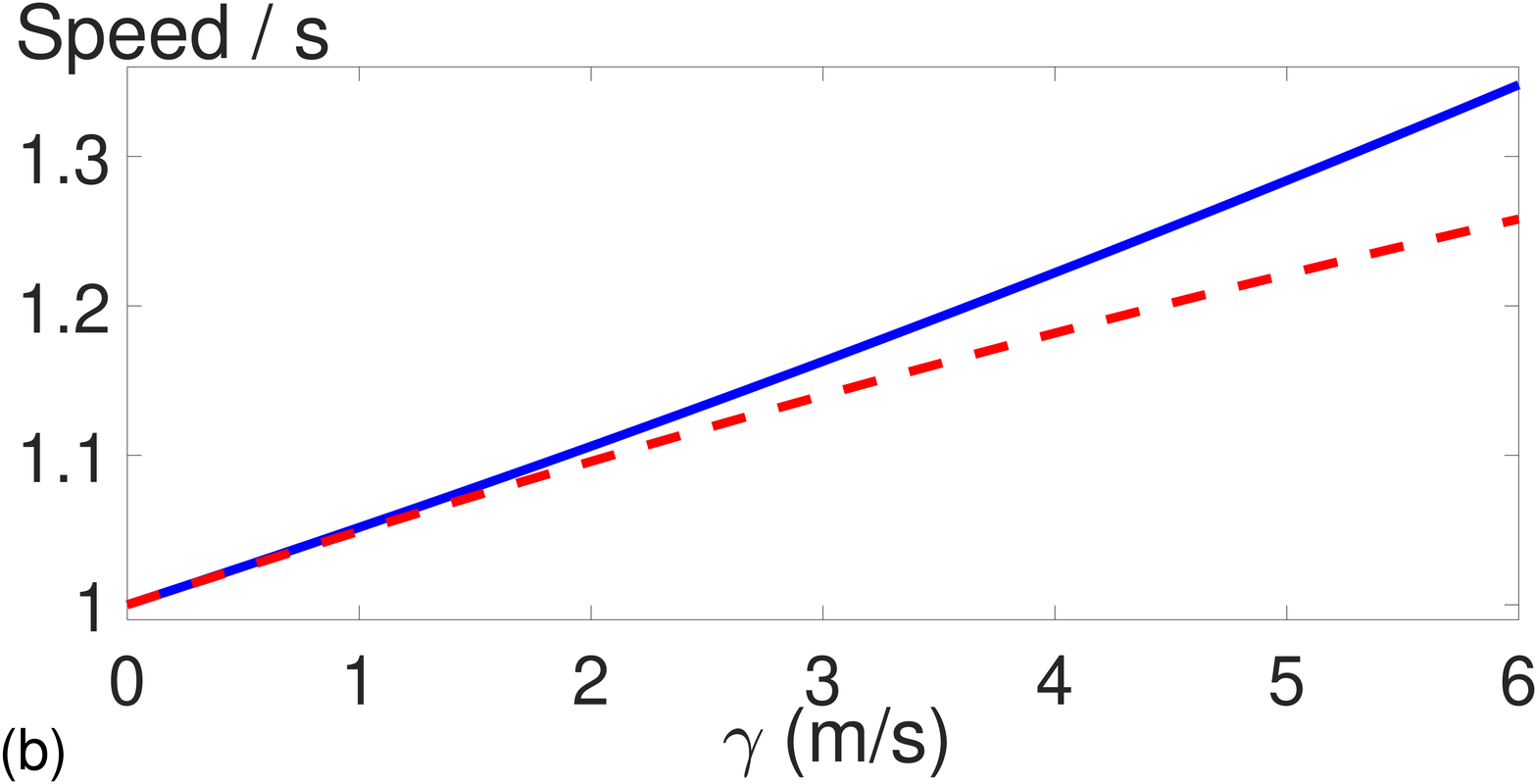}
		\caption{(a) Relative distance between  the points on the wavefronts of surface ring waves in downstream and upstream directions as a function of $\gamma$. 
			(b) Relative speed of the wavefronts of surface ring waves in downstream and upstream directions as a function of $\gamma$. The blue (solid) curve is for $\theta = 0$ (downstream) and the red (dash) curve for $\theta = \pi$ (upstream).  Here, $g = 9.8\ m s^{-2}$ and $h = 10\ m$. 
		}
		\label{fig:distance}
	\end{center}
\end{figure}

Next, we shall introduce a measure of the deformation of the wavefront. This can be done globally, for the whole ring wave, considering the distance between the points on the wavefront in the downstream and upstream directions, i.e.
\begin{equation}
	D_{\gamma} = \frac{st}{m(0)} + \frac{st}{m(\pi)},
	\label{D}
\end{equation}
and comparing this distance to a similar distance in the absence of the shear flow, $D_0 = 2st$. It is natural to consider the ratio
\begin{equation}
	\frac{D_{\gamma}}{D_0} = \frac 12 \left (\frac{1}{m(0)} + \frac{1}{m(\pi)} \right ).
	\label{D1}
\end{equation}
Here, $m(\theta)$ is given by (\ref{eq:ss1}).  The plot of this global measure as a function of the strength of the current is shown in Figure \ref{fig:distance}(a).
It might be also helpful to show the plots of the ratios of the speeds of the wavefront at $\theta = 0$ and  $\theta = \pi $ (downstream and upstream directions) to the speed $s $ in the absence of any current. This can be seen in Figure \ref{fig:distance}(b), where the upstream speed has been shifted upwards by 2 units, for a better comparison with the downstream speed. Both plots give a clear indication of a significant elongation of the wavefront in the direction of the current.


However, in many satellite images the oceanic internal waves propagate as a part of a ring and not the whole ring (see, for example, Figure 1 in \cite{KZ16a} or Figure 13 in \cite{A}). Therefore, it is desirable to introduce a local measure of the deformation of the wavefront. This can be done by introducing the geometric curvature of the wavefront, which in polar coordinates is given  by (e.g. \cite{DC})
\begin{equation}
	k_{\gamma} = \frac{|r^2 + 2 r'^2 - r r''|}{(r^2 + r'^2)^{3/2}}, \quad \mbox{where} \quad  r = r(\theta).
\end{equation}
Applying this formula to $\displaystyle r(\theta) = \frac{st}{m(\theta)}$ we obtain
\begin{equation}
	k_{\gamma} = \frac{|m + m''|}{st \left [1 + \left ({\displaystyle  \frac{m'}{m}} \right )^2 \right ]^{3/2}}.
	\label{k}
\end{equation}
In the absence of any current, $m(\theta )= 1$ and $\displaystyle k_0 (\theta) = (st)^{-1}$. 
Taking the ratio, 
\begin{eqnarray}
	\frac{k_{\gamma}}{k_0}  &=& \frac{|m + m''|}{\left [1 + \left ({\displaystyle  \frac{m'}{m}} \right )^2 \right ]^{3/2}},
	\label{Curv}
\end{eqnarray}
which in the present example gives
\begin{eqnarray}
	\frac{k_{\gamma} (\theta)}{k_0 (\theta)}  =  \frac{\sqrt{\displaystyle 1 + \frac{\gamma^2}{4 s^2}}}{ G^{3/2}}, \quad \mbox{where} \quad
	\label{k} 
	G = \displaystyle  1 + \frac{\gamma^2}{4 s^2} \frac{\sin^2 \theta}{\displaystyle  \left ( \sqrt{1 + \frac{\gamma^2}{4 s^2}} - \frac{\gamma  \cos \theta}{2 s} \right )^2 }.
\end{eqnarray}

\begin{figure}
	\begin{center}
		\includegraphics[width=0.55 \textwidth]{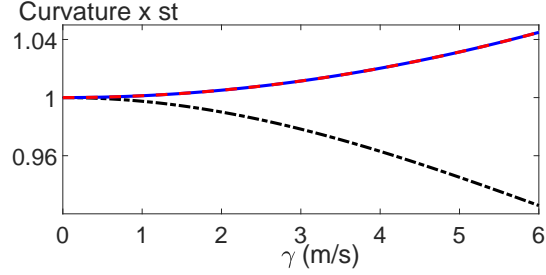}
		\caption{Relative curvature of the wavefronts of surface ring waves in different directions as a function $\gamma$. The blue (solid) curve is for $\theta = 0$, the black (dash-dot) curve for $\displaystyle \theta = \frac{\pi}{2}$ and the red (dash) curve for $\theta = \pi$. Here, $g = 9.8\ m s^{-2}$ and $h = 10\ m$.  
		}
		\label{fig:curvature}
	\end{center}
\end{figure}

The plot of this local measure as a function of the strength of the current is shown for $\displaystyle \theta = 0, \frac{\pi}{2}$ and $\pi$ in Figure \ref{fig:curvature}. We can see that the curvature is growing in the upstream and downstream directions, while it is decreasing in a transverse direction, giving a clear indication (and quantitative measure) of the elongation of the wavefront compared to the concentric wavefront in the absence of the current.

Note, that in accordance with the Gauss-Bonnet theorem for the closed convex curves (e.g. \cite{DC}), the curvature (a time-dependent function) satisfies the relation
\begin{equation}
	\oint  k_{\gamma}  ds = 2\pi,
	\label{GB}
\end{equation}
which yields the hidden {\it  total curvature conservation law} in the following general form:
\begin{equation}
	\oint  k_{\gamma}  ds = \int_0^{2 \pi} k_{\gamma} (\theta) \sqrt{r^2 + r'^2} d\theta  = \int_0^{2 \pi} \frac{m\  |m + m''|}{m^2 + m'^2} d\theta = 2\pi.
	\label{GB1}
\end{equation}

\begin{figure}
	\centering
	\begin{subfigure}[b]{0.48\textwidth}
		\includegraphics[width=\textwidth]{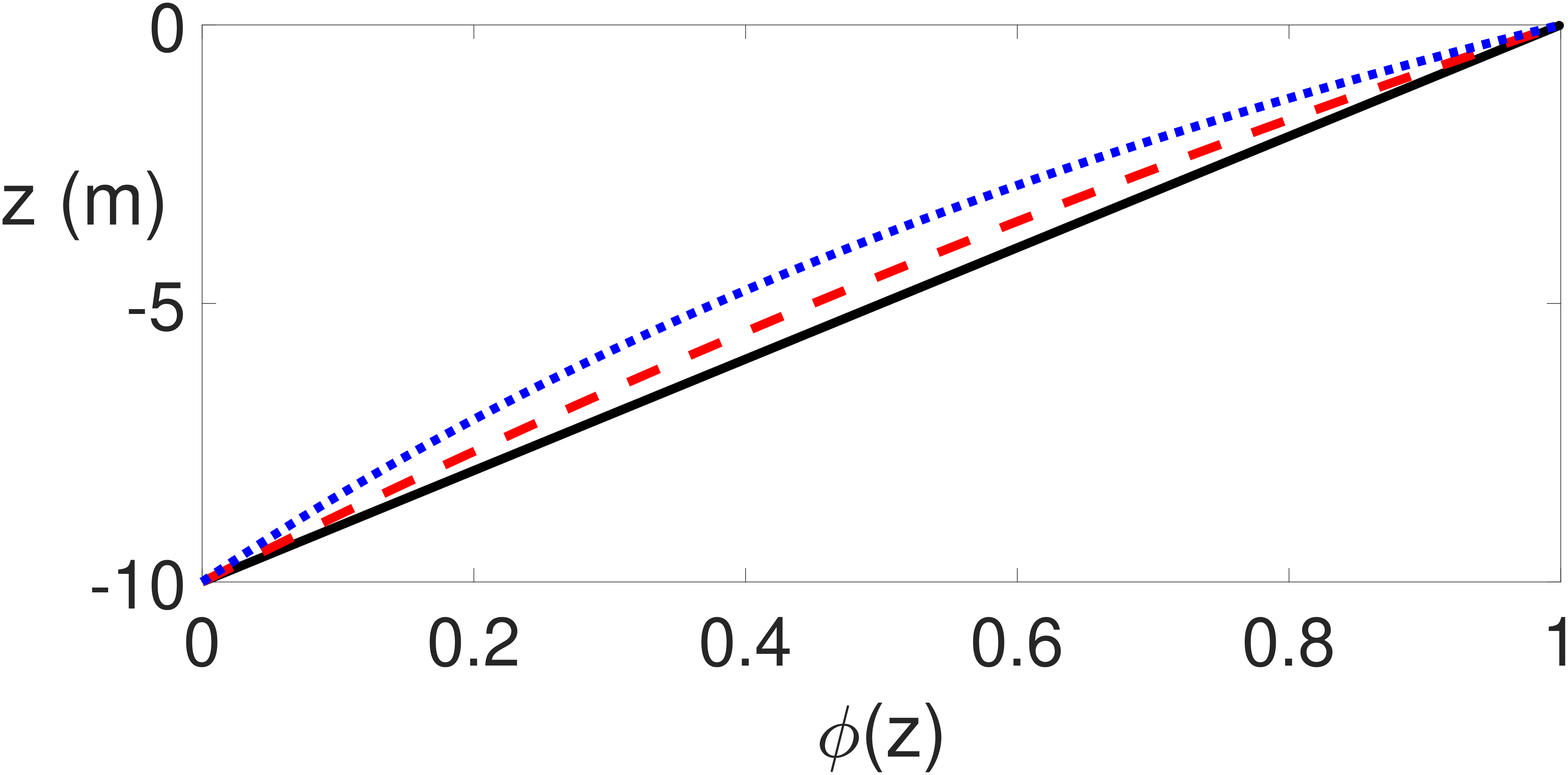}
		\caption{$\theta = 0$ (downstream)}
		\label{fig:theta_0}
	\end{subfigure}
	~ 
	\begin{subfigure}[b]{0.48\textwidth}
		\includegraphics[width=\textwidth]{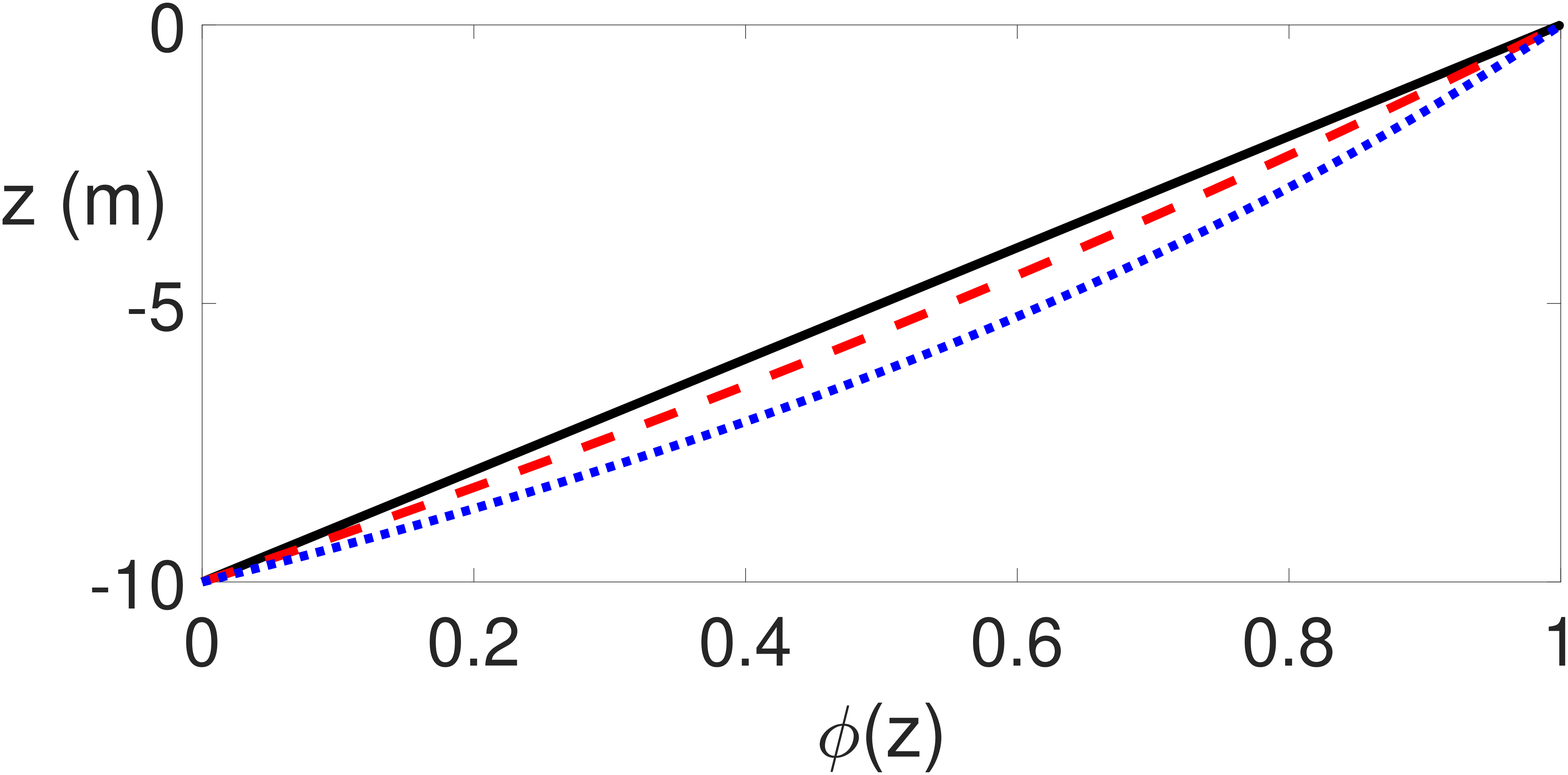}
		\caption{$\theta = \pi$ (upstream)}
		\label{fig:theta_pi}
	\end{subfigure}
	\begin{subfigure}[b]{0.48\textwidth}
		\includegraphics[width=\textwidth]{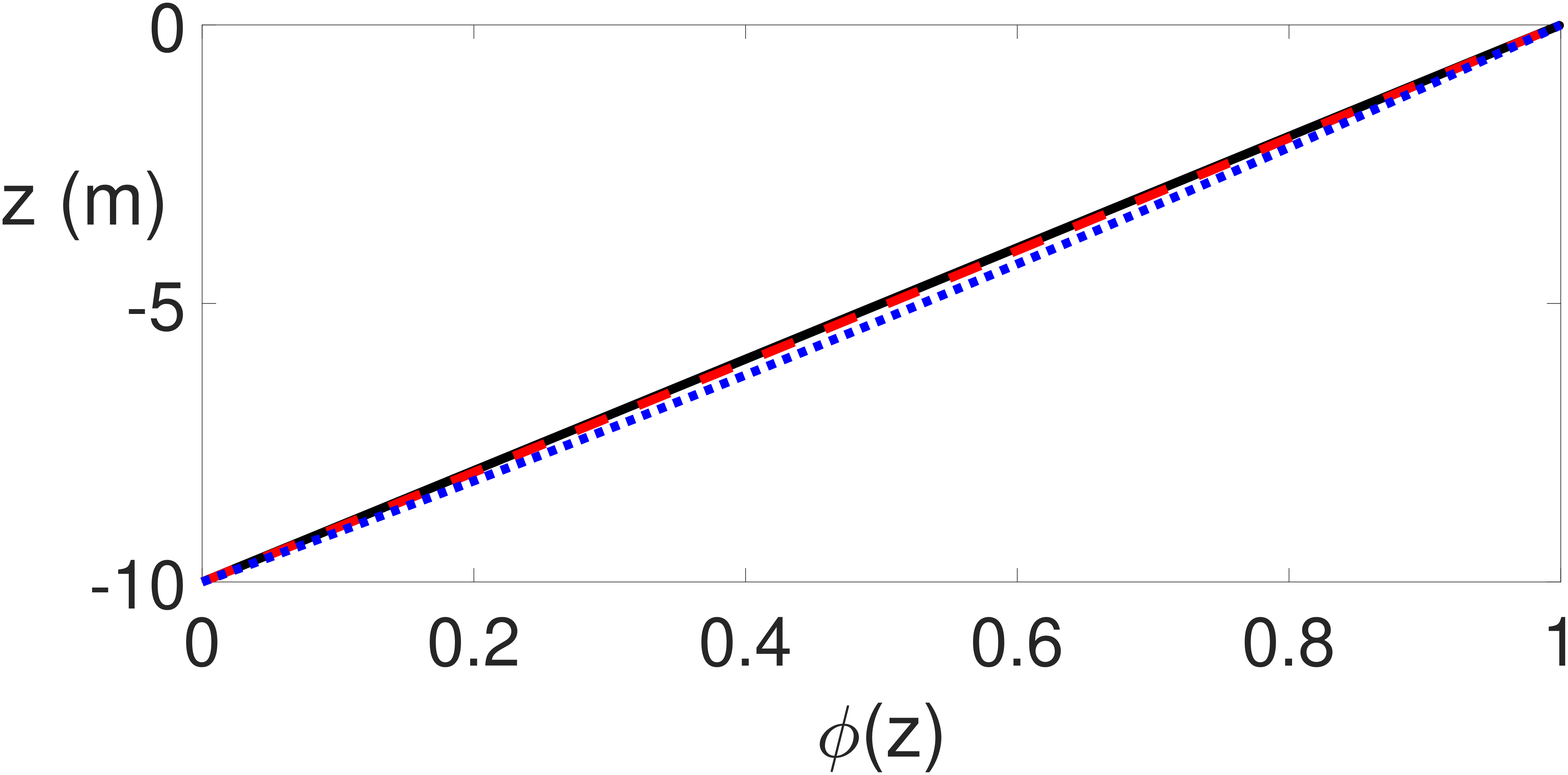}
		\caption{$\displaystyle \theta = \pi/2$ (orthogonal)}
		\label{fig:theta_half_pi}
	\end{subfigure}
	
	\caption{Plots of the modal function \eqref{eq:phi_with_shear} for $\gamma = 0\ m s^{-1}$ (black, solid), $\gamma = 2\ m s^{-1}$ (red, dash) and $\gamma = 5\ m s^{-1}$ (blue, dot). Here, $g = 9.8\ m s^{-2}$ and $h = 10\ m$. 
	}    
	\label{fig:Johnson_modal}
\end{figure}

It is also instructive to examine the two-dimensional structure of the modal function \eqref{eq:phi_with_shear}. Before we can do this, the parameter $\Lambda$ must be determined. We normalize $\phi$ by setting $\phi = 1$ at $z = 0$. This gives 
\begin{equation}
	\Lambda = \rho g,
	\label{eq:L_homo}
\end{equation}  
which is independent of $\theta$. Thus,
\begin{equation}
	\phi =  \frac{g (m^2+m'^2) (z+h)}{\displaystyle s \left [\displaystyle s - \gamma \frac{z+h}{h}  (m \cos \theta - m' \sin \theta)\right ] },
	\label{eq:phi_with_shear_2}
\end{equation}
and the modal function is shown in Figure \ref{fig:Johnson_modal} for $\theta = 0,\displaystyle  \frac{\pi}{2}, \pi$, i.e. in the downstream, orthogonal and upstream directions, respectively.  The current has the effect of a similar magnitude in the downstream and upstream directions, while the effect in the orthogonal direction to the current is expectedly weak. We conclude that  the vertical structure of the wave field is shifted towards the surface in the downstream direction, and towards the ocean bottom in the upstream direction. 

\bigskip


\section{Derivation of modal equations from the formulation for plane waves }

The aim of this section is to arrive at the modal equations (\ref{eq:ME1}) - (\ref{eq:MEC2}) starting from the equations for the plane waves. This allows us to clarify the roles of the general solution of the directional adjustment equation and its envelope, provides a natural way of describing important properties of the ring waves (e.g. their group speed) and allows us to outline an analytical approach to constructing more general hybrid wavefronts consisting of an arc of a ring wave and two tangent plane waves.

Since the terms of interest are linear and in the long-wave regime,
it is sufficient to consider only the linear long-wave equations. 
Relative to a background shear flow $u_{0}(z)$ and a background density 
field $\rho_{0}(z)$, the equations are in the same domain $-h < z < 0$, and in
standard notation, 
\begin{eqnarray}
	& & \rho_0 (u_t + u_0 u_x +  u_{0z}w )  +  p_x  = 0 \,,   \label{eq1}  \\
	& & \rho_0 (v_t + u_0 v_x ) +   p_y  = 0 \,, \label{eq2} \\
	& & p_{z}+g\rho = 0 \,, \label{eq3} \\
	& & \rho_t  + u_0 \rho_{x } +  w\rho_{0z}  =  0 \,,  \label{eq4}  \\
	& & u_{x} + v_{y} + w_{z}  = 0 \,,  \label{eq5}  \\
	& & \zeta_t + u_{0} \zeta_x - w =0 \,. \label{eq6} 
\end{eqnarray}
The boundary conditions are
\begin{eqnarray}
	& &  p -g\rho_{0} \zeta = 0  \quad  \hbox{at} \quad z = 0 \,,  \label{bc2} \\
	& & w =0  \quad \hbox{at} \quad z=-h\,. \label{bc1} 
\end{eqnarray}
Since the only inhomogeneity is the $z$-dependence in $u_{0}, \rho_{0}$, 
it is convenient  to look at the linear long-wave theory in Fourier space, for a disturbance proportional to 
$\exp{(ikx +ily - ik\tilde ct)}$, i.e.
$$
(u,v,w,p,\rho, \zeta) =  (\tilde u, \tilde v, \tilde w, \tilde p,\tilde \rho, \tilde \zeta) e^{i (k x + l y - k \tilde c t)} + c.c.
$$

Then equations (\ref{eq1}) - (\ref{eq6})  become, after eliminating $\tilde w, \tilde \rho$,
\begin{eqnarray}
	& & \rho_0 \left [-ik(\tilde c-u_0)(\tilde u + u_{0z }\tilde \zeta)\right ] + ik\tilde p =0 \,,   \label{f1} \\
	& &   \rho_0 \left [-ik(\tilde c-u_0)\tilde v\right ]    +   il \tilde p   =0 \,, \label{f2} \\
	& &  ik\tilde u + il\tilde v  - ik\left [(\tilde c-u_0) \tilde \zeta\right ]_z =0 \,, \label{f4} \\
	& &  \rho_0 N^2 \tilde \zeta + \tilde p_z   = 0 \,. \label{f3} 
\end{eqnarray}
Next we use (\ref{f2}), (\ref{f4}) to eliminate $\tilde u, \tilde v$ and so obtain in place of (\ref{f1}), (\ref{f2}),
\be
\rho_{0}\left [-ik(\tilde c-u_0)^2 \tilde \zeta_z  \right ] + ik \left (1 + \frac{l^2}{k^2 }\right ) \tilde p=0 \,. \label{ff1} \\
\ee
Together with (\ref{f3}) these equations form two equations for $\tilde \zeta, \tilde p$. 
The final step is to  eliminate $\tilde p$ between
(\ref{f3}) and (\ref{ff1}) to obtain
\be\label{zeq} 
\left [\rho_{0}(\tilde c-u_0)^2 \tilde \zeta_z \right ]_z + \rho_{0}N^2 \left (1 + \frac{l^2}{k^2} \right )\tilde \zeta  =  0  \,.
\ee
The boundary conditions (\ref{bc2}), (\ref{bc1}) are similarly reduced to
\begin{eqnarray}
	& &  (\tilde c-u_0)^2 \tilde \zeta_z =  \left (1+ \frac{l^2}{k^2} \right )  g \tilde \zeta \quad  \hbox{at} \quad z = 0  \,,  \label{bc4}  \\
	& & \tilde \zeta =0  \quad \hbox{at} \quad z=-h\,. \label{bc3} 
\end{eqnarray}
Next, we can write $\tilde \zeta = A(k, l)\phi (z)$ and so
\be\label{modal}
\left [\rho_0 (\tilde c-u_0)^2 \phi_z \right ]_z + \rho_0 N^2 \left (1 + \frac{l^2}{k^2}\right )\phi = 0  \,, 
\ee
subject to the boundary conditions
\be\label{modalbc}
(\tilde c-u_0)^2 \phi_z = g \left (1 + \frac{l^2 }{k^2 } \right )\phi  \quad \hbox{at} \quad z =0 \,,  \quad \hbox{and} \quad 
\phi = 0  \quad \hbox{at} \quad z = -h \,.
\ee
The speed $\tilde c$ and the modal function now retain a dependence on $k, l$
which is removed in the one-dimensional case when $l =0$.  
The KP equation follows when $l^2 \ll k^2$ and again this reduces to the usual modal equation
where at leading order $\tilde c$ is a constant. 
In the general case when there is a shear flow $u_{0}(z) \ne  \hbox{const}$
the dispersion relation is not isotropic.

It is useful to note that the integral identity readily obtained from the modal equation in the case of continuous stratification,
\be\label{intdisp}
\mathcal{D} (\omega, {\bf k}) = \int_{-h}^{0}\, \rho_{0}\,
\left [k^2 (\tilde c-u_0)^2 \phi_{z}^2 - N^{2}(k^2 + l^2)\phi^2\right ]\, dz - [\rho_0 g (k^2 + l^2)\phi^2 ]_{z=0} = 0 \,,
\ee
can be regarded as the dispersion relation, recalling that $\omega =k \tilde c$ and
$k (\tilde c - u_0) = \omega - ku_{0}$.

More generally, $k, l$ may be defined as local wavenumbers, depending on 
$x, y$, and then the wavefronts become the curves
\be\label{wf}
S (x, y, t) = const \,, \quad \mbox{where} \quad  k=S_x \,, \quad l = S_y \,,
\quad \omega = - S_t \,.
\ee
They can be determined by solving the equations
\be\label{wf1}
k_t + \omega_x = 0\,, \quad l_t + \omega_y = 0 \,, \quad k_y = l_x \,,
\quad \hbox{where} \quad \omega = \omega (k, l) \,. 
\ee
Here, the third equation is only required at $t=0$ since
the first two equations imply that $(k_y - l_x)_t = 0$.
Next, in order to change to a reference frame moving with a known speed $c$, we need to use a Galilean transformation which in effect
replaces $u_{0} $ with $u_{0} - c$.


Let us now use the above to recover the modal equations for the ring waves described in the previous section. In polar coordinates, the wavefronts are described by 
$$S=S(r, \theta, t) = const,$$ 
where $x = r\cos{\theta}, y= r\sin{\theta}$.
Then, we define 
\be\label{wfr}
\gamma = S_r \,, \quad \sigma = \frac{S_{\theta}}{r} \,, \quad \mbox{where} 
\ee
\be\label{wfr1}
\gamma_t  + \omega_r = 0 \,, \quad r\sigma_t + \omega_{\theta}  = 0 \,,
\quad \gamma_{\theta}= (r\sigma )_r \,. 
\ee
The local wave vector ${\bf k} = (k, l)$ in Cartesian coordinates, becomes the local wave vector in polar coordinates,
${\bf k} =  \hat \gamma \hat{r} + \hat \sigma\hat{\theta }$
where $\hat{r} = (\cos{\theta}, \sin{\theta})$ and $\hat{\theta} = (-\sin{\theta}, \cos{\theta})$
are unit vectors in the radial and polar angle directions, respectively. 
Hence $k = \hat \gamma \cos{\theta} - \hat \sigma \sin{\theta}$ and
$l = \hat \gamma \sin{\theta} + \hat \sigma \cos{\theta}$. 

We can define 
$\hat \gamma= \kappa \cos{\beta}, \hat \sigma = \kappa \sin{\beta}$ so that
$k= \kappa \cos{\alpha}, l = \kappa \sin{\alpha }$, and
$\kappa = |{\bf k}| = (\hat \gamma^2 + \hat \sigma^2 )^{1/2}$ is the wave vector magnitude. Here,  
$\alpha = \theta + \beta$, where $\beta$ is the angle between  the vectors ${\bf k}$ and $\hat{r}$. 
Then, the modal equations ({\ref{modal}), (\ref{modalbc})  become
	\be\label{modal1}
	\left [\rho_0 k^2 (\tilde c - u_0)^2 \phi_z \right ]_z + \rho_0 N^2 \kappa^2  \, \phi = 0  \,, \
	\ee
	\be\label{modalbc1}
	k^2 (\tilde c - u_0)^2 \phi_z = g\,\kappa^2  \,\phi  \quad \hbox{at} \quad z =0 \,,  \quad \hbox{and} \quad
	\phi = 0  \quad \hbox{at} \quad z = -h \,. 
	\ee
	Here 
	$ \kappa , k$ can be expressed in terms of
	$\hat \gamma, \hat \sigma ; \theta $. The dispersion relation can be expressed in the form,
	\be\label{disp}
	\omega (\hat \gamma, \hat \sigma ; \theta) = 
	k \tilde c(k, \kappa^2),
	\ee
	or in the integral form, in the case of continuous stratification:
	\be\label{intdisp1}
	\begin{split}
		& \mathcal{D} (\omega, \gamma, \sigma ; \theta) = \int_{-h}^{0}\, \rho_{0}\,
		\left [k^2 (\tilde c - u_0)^2 \phi_{z}^2 - N^{2} \kappa^2 \phi^2\right ]\, dz  -
		[\rho_0 g \kappa^2 \phi^2 ]_{z=0} = 0 \,, \\
		&  k (\tilde c - u_0) = \omega - ku_{0} = \omega -  (\hat \gamma \cos{\theta} - \hat \sigma \sin{\theta})u_{0} \,,\  \kappa^2 = \hat \gamma^2 + \hat \sigma^2.
	\end{split}
	\ee
	
	For instance, if we choose to write, as in the previous section,
	\be\label{ringS}
	S = m r - s t \,, 
	\ee
	where $m = m(\theta)$ and $s$ is a constant speed in the absence of the shear flow, then
	$\hat \gamma= m$, $\hat \sigma = m'$ and $\omega = s$.
	The wavefronts of the ring waves are given by $m r - s t = \hbox{const}$ , with
	\be\label{rw}
	\kappa^2 = m^2 + m'^2 \,, \quad k = m \cos{\theta } - m' \sin{\theta }  \,, \quad  l = m \sin \theta + m' \cos \theta.
	\ee
	The modal equations (\ref{modal1}), (\ref{modalbc1}) take the form
	\be\label{modal2}
	(\rho_0 \hat F^2 \phi_z )_z + \rho_0 N^2 (m^2 + m'^2)\, \phi = 0  \,, 
	\ee
	\be\label{modalbc2}
	\hat F^2  \phi_z = g\,(m^2 + m'^2)\phi  \quad \hbox{at} \quad z =0 \,, \quad \hbox{and} \quad \phi = 0  \quad \hbox{at} \quad z = -h \,, 
	\ee
	where
	\be\label{F}
	\hat F = - k (\tilde c - u_0)  = -s + u_{0}(m \cos{\theta } - m' \sin{\theta } ) \,,
	\ee
	which in the reference frame moving with the speed $c$ becomes
	\be\label{F1}
	\hat F = - k (\tilde c - u_0)  = -s + (u_{0} - c) (m \cos{\theta } - m' \sin{\theta } ) \,.
	\ee
	These equations are equivalent to the modal equations (\ref{eq:ME1}) - (\ref{eq:MEC2}) from the previous section. 
	
	The   dispersion relation (\ref{disp}) with $\kappa, k $ given by (\ref{rw}) takes the form
	\be\label{rw1}
	s = (m \cos{\theta } - m' \sin{\theta })\  \tilde c \left [m \cos{\theta } - m' \sin{\theta }, m^2 + m'^2 \right ],
	\ee
	and may be written, in the case of continuous stratification, in the integral form (\ref{intdisp1}) as
	\be\label{intdisp2}
	\mathcal{D} (s, m, m'; \theta )= \int_{-h}^{0}\, \rho_{0}\,
	[\hat F^2 \phi_{z}^2 - N^{2}(m^2 + m'^2)\phi^2]\, dz  -
	[\rho_0 g (m^2 + m'^2)\phi^2 ]_{z=0} = 0 \,. \\
	\ee
	
	In general,  (\ref{rw1}) forms a rather complicated ordinary differential equation for $m (\theta)$ since $\hat F = \hat F(z; \theta)$.
	We will analyse its solutions for the cases of a two-layered fluid with the linear current and a power-law upper-layer current in the next sections. 
	
	Here, we consider another simple example. 
	Suppose that $N = \hbox{const}$.   Then, for $u_{0} (z) = 0$, and in the Boussinesq and rigid-lid approximations, the solution of the modal equations is given by 
	\be
	\phi = \Lambda \sin \frac{n \pi z}{h}, \quad \mbox{where} \quad n = 1,2,3, \dots,
	\ee
	where
	\be
	m^2 + m'^2 = \left ( \frac{n \pi s}{Nh} \right )^2.
	\label{mN}
	\ee
	The parameter $\Lambda$ can be used to normalise the modal function to be equal to one at some level of interest.
	Here, since there is no shear flow, $m=1$, and then (\ref{mN}) implies that
	\be
	s = \frac{Nh}{\pi n}, \quad \mbox{where} \quad n = 1,2,3, \dots,
	\ee
	describing the speeds of the concentric ring waves. It is instructive also to consider the general solution of (\ref{mN}), which is given by
	\be
	m = a \cos \theta + b(a) \sin \theta, \quad \mbox{where} \quad a^2 + b^2 =  \left ( \frac{n \pi s}{Nh} \right )^2,
	\ee
	as these solutions describe plane waves propagating at an arbitrary angle:
	\be
	S = m r - s t = [a \cos \theta + b(a) \sin \theta] r - s t = ax + b y - st,
	\ee
	where 
	\be
	s^2 =  \frac{Nh}{\pi n} \sqrt{a^2 + b^2}, \quad a = k, b = l, \quad \mbox{and} \quad {\bf k} = (k, l).
	\ee
	Looking for a singular solution, we re-parametrise the general solution as
	\be
	m = \frac{n \pi s}{Nh} \cos (\Theta - \theta) \quad (a = \frac{n \pi s}{Nh} \cos \Theta, b = \frac{n \pi s}{Nh} \sin \Theta),
	\ee
	and then find the envelope of this general solution, by requiring that
	$\displaystyle
	\frac{dm}{d \Theta} = 0,
	$
	which immediately yields $m = 1$.
	The case when $u_0(z) = U_0 = \hbox{const}$ can be reduced to the previous case by a Galilean transformation, and therefore again describes concentric ring waves in a reference frame moving with the speed $c=U_0$.

	Next, it is useful to obtain the group speed and the wave action conservation law for the ring waves. This can be done working with the local wave vector introduced above. Indeed, the wave action conservation law is expressed by
	\be\label{action}
	\mathcal{A}_t + \nabla \cdot ({\bf c}_g \mathcal{A} ) = 0 \,, 
	\ee
	where the group velocity ${\bf c}_g = (\omega_k , \omega_l )$ and 
	$\mathcal{A}$ is the wave action density (e.g. \cite{W}), given in the long-wave limit by
	\be\label{actiond}
	\mathcal{A} = \int_{-h}^{0} \, \rho_{0} (\tilde c - u_0) |A|^2 \phi_{z}^2 \, dz\, .
	\ee		
	Since $\omega =\omega (k,l)=  k\tilde c = k\tilde c(k, k^2 + l^2)$, we have
	\be\label{group}
	{\bf \tilde c}_g  = (\tilde c + k \tilde c_{\tilde k} + 2 k^2 \tilde c_{\tilde \xi},  \ 2 k l \tilde c_{\tilde \xi}), \quad \mbox{where} \quad \tilde \xi = \kappa^2 = k^2 + l^2.
	\ee
	In the case of continuous stratification we can use the integral form (\ref{intdisp}) to obtain 
	\be\label{group1}
	\mathcal{D}_{\omega} {\bf \tilde c}_g + \nabla_{\bf k} \cdot \mathcal{D} = 0 \,.
	\ee
	Finally, for the ring waves,  we should replace $k$ and $l$ with  the local wavenumbers given by the formulae (\ref{rw}).

	We will finish this section by outlining the construction of more general {\it hybrid solutions} which are formed by a part of an outward propagating ring wave and two tangent plane waves shown in Figure \ref{fig:hybrid}.  Similarly looking wavefronts are often present in satellite images of internal waves, see, for example, Figure 13 in \cite{A}.  They were considered for surface waves in the absence of a current by \cite{OS}.  Other related hybrid solutions have been discussed by \cite{CK,KKMS,OSt, RHB} (see also the references therein). 
	
	Here, we will consider only a simple case of surface waves on the current $\displaystyle u_0 = \gamma \frac{z+h}{h}$, but similar solutions can be constructed for all examples of surface and internal waves discussed in our paper. Here, we are concerned only with the kinematics of such hybrid solutions, addressing the issue of finding the analytical description of the wavefront shown in Figure \ref{fig:hybrid}.
	
	For this current, the directional adjustment equation has the form (\ref{eq:gammaza}). Let $m = m_0 + \tilde m$, where $m_0$ is the singular solution of (\ref{eq:gammaza}). Let $\tilde m (\theta_0) = - \alpha$, with $\alpha > 0$.  Substituting this into the equation (\ref{eq:gammaza}) we obtain
	\begin{eqnarray}
		&&\tilde m'^2 = \sigma^2 - (\sigma + \tilde m)^2, \quad \mbox{where} \quad \sigma = \sqrt{1 + \left ( \frac{\gamma}{2 s} \right )^2};\\
		&&\tilde m(\theta_0) = - \alpha.
	\end{eqnarray}
	This problem has two solutions
	\be
	\tilde m = - 2 \sigma \sin^2 \left (\frac{\theta - \theta_0}{2} \pm \arcsin \sqrt{\frac{\alpha}{2 \sigma}} \right ),
	\ee
	yielding the explicit formulae for two particular members of the general solution:
	\begin{eqnarray}
		&&m = \tilde a \cos \theta + \tilde b \sin \theta, \quad \mbox{where}  \label{Eq1} \\
		&&\tilde a = - \frac{\gamma}{2s} + \sigma \cos \left (2 \arcsin \sqrt{\frac{\alpha}{2 \sigma}} \pm \theta_0\right ), \\
		&& \tilde b = \pm  \sigma \sin \left (2 \arcsin \sqrt{\frac{\alpha}{2 \sigma}} \pm \theta_0\right ).
		\label{Eq2}
	\end{eqnarray}

	The physical nature of the two new solutions is clear if we recall that
	$
	A = A(m r - st),
	$
	and therefore, using (\ref{Eq1}) for $m(\theta)$, we obtain
	\begin{equation}
		A = A(k x + l y - s t), \quad \mbox{where} \quad {\bf{k}} = (k, l) = (\tilde a, \tilde b).
		\label{Eq3}
	\end{equation}
	The two plane waves are a part of the general solution of the equation (\ref{eq:gammaza}). They are tangent to the ring wave and intersect at the point $C$ as shown in Figure~\ref{fig:hybrid}. If the wavefront of the ring wave is given by $\displaystyle r = \frac{st}{m(\theta)}$, where $m(\theta)$ is the singular solution of  (\ref{eq:gammaza}),  then the distance from the origin to the intersection point of the two tangent lines is given by 
	\be
	|OC| = \frac{st}{m(\theta_0)-\alpha},
	\ee
	which completely defines the tangent lines in terms of convenient parameters $\theta_0$ and $|OC|$, which can extracted from observational data.  Such a hybrid wavefront may propagate as a whole, with matched speeds and slopes at the junctions. Stability of outward propagating localised hybrid waves in the absence of a current was shown by \cite{OS}.

	\begin{figure}
	\centering
	\begin{subfigure}[b]{0.49\textwidth}
		\includegraphics[width=\textwidth]{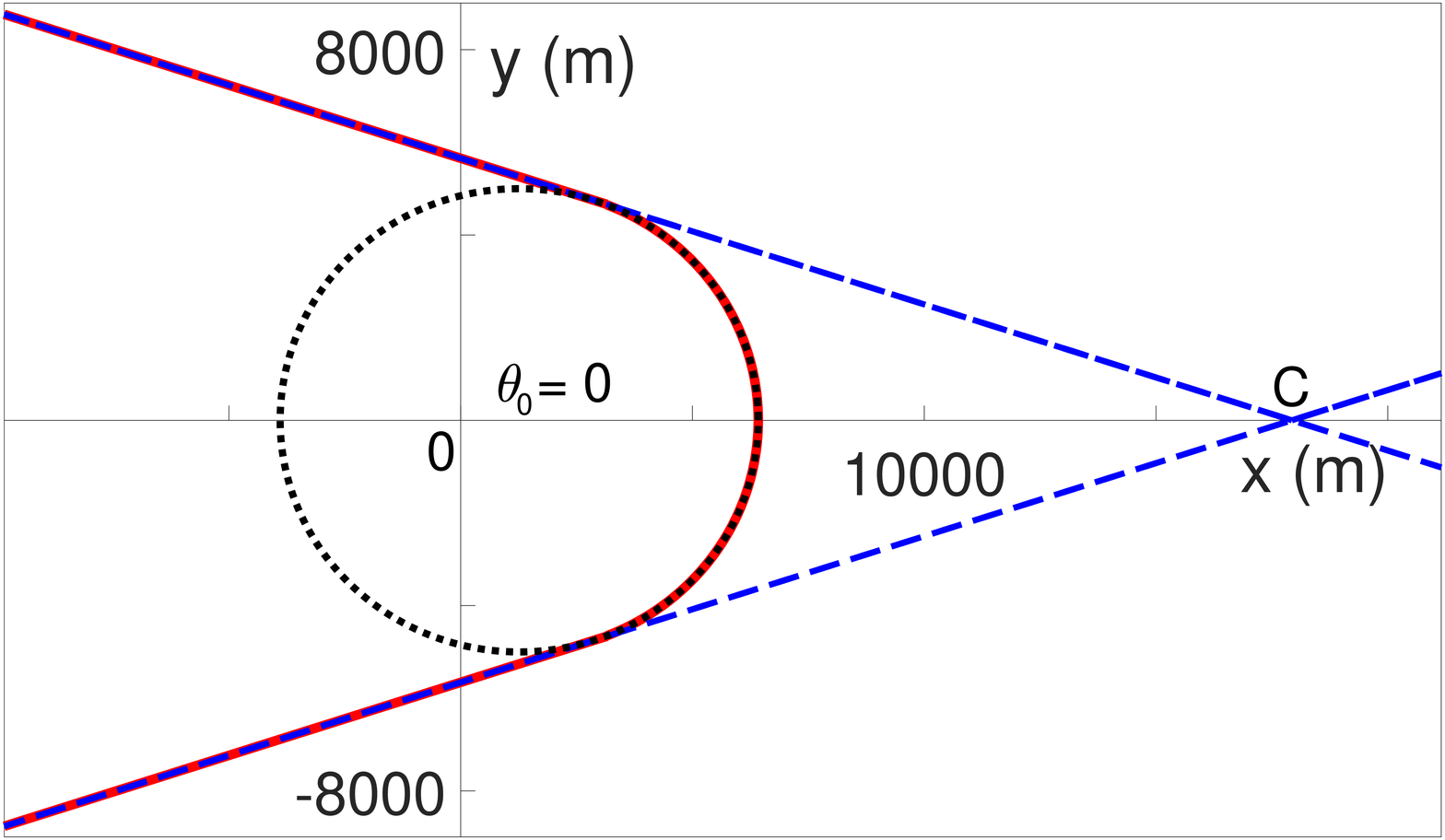}
		\caption{Symmetric wavefront}
		\label{fig:hybrid_sym}
	\end{subfigure}
	~ 
	\begin{subfigure}[b]{0.49\textwidth}
		\includegraphics[width= \textwidth]{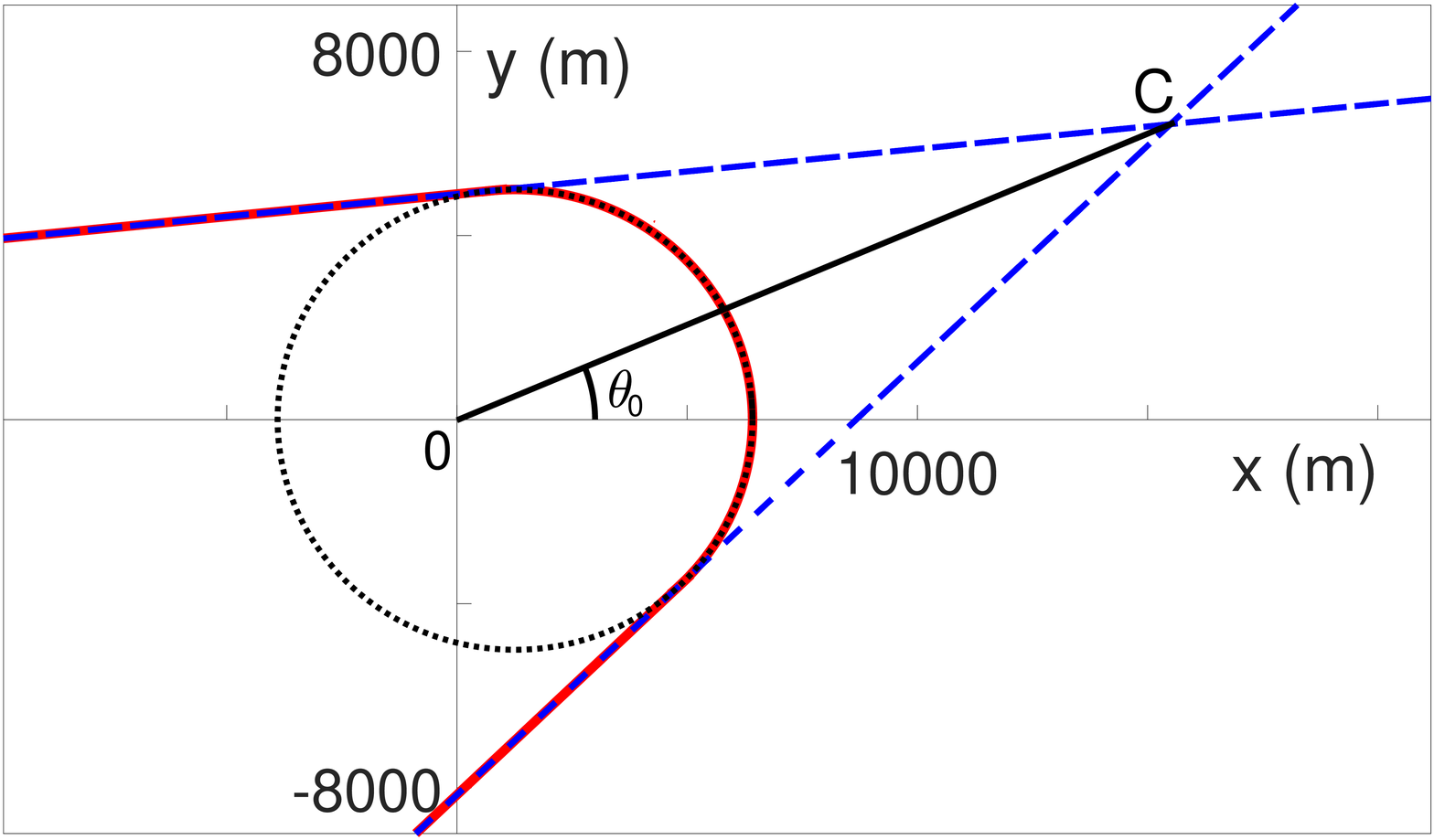}
		\caption{Asymmetric wavefront}
		\label{fig:hybrid_asym}
	\end{subfigure}
	\caption{Wavefronts of hybrid solutions for (a) $\theta_0 = 0$ (symmetric) and (b) $\theta_0 \ne 0$ (asymmetric). Here, $\alpha = 0.5$ and $\gamma = 5\ ms^{-1}$. The blue (dash) lines represent the tangent lines, the black (dot) curve is the ring wave and the red (solid) curve is the wavefront of a hybrid solution. Here, $g = 9.8\ m s^{-2}$,  $h = 10\ m$ and $r m(\theta) = 5000\ m$. }   
	\label{fig:hybrid}
\end{figure}

	\section{Two-layer fluid with a linear shear current}
	
	\subsection{Problem formulation and modal equations}

	We now consider a two-layered fluid with the  free surface and the shear flow given by $\displaystyle u_0(z) = \gamma \frac{z+h}{h}$ for some positive constant $\gamma$, as shown in Figure \ref{fig:2_layer_gamma_z}. This is a generalisation of the example considered in the previous section.

	\begin{figure}
		\begin{center}
			\includegraphics[width=0.5 \textwidth]{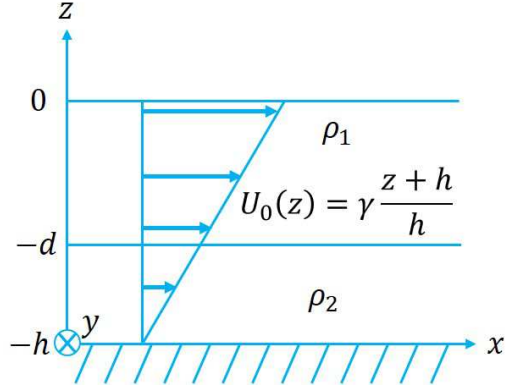}
			\caption{Two-layer model with a linear shear current.}
			\label{fig:2_layer_gamma_z}
		\end{center}
	\end{figure}

	The modal equations are given by
	\begin{eqnarray}
		&&\Bigg( \frac{\rho_0 \hat F^2}{m^2+m'^2} \phi_{z} \Bigg)_z - g \rho_{0z} \phi = 0,
		\label{ME} \\
		&&\frac{\hat  F^2}{m^2+m'^2}\phi _z - g \phi = 0 \hspace{0.4cm} \text{at} \quad  z=0,
		\label{BC1} \\
		&& \phi = 0 \hspace{0.4cm} \text{at} \quad z=-h,
		\label{BC2}
	\end{eqnarray}
	where $\rho_0 = \rho_2 H(z+h) + (\rho_1 - \rho_2) H(z+d)$ and $H(z)$ is the Heaviside function.
	We choose $c$ to be equal to the speed of the shear flow at the bottom which gives $c=u_0(-h)=0$, thus 
	\begin{equation}
		\hat  F= \hat  F(z; \theta)= -s+\gamma \frac{z+h}{h} (m\cos \theta - m' \sin \theta).
		\label{eq:F2}
	\end{equation}
	When the value of $\theta$ is not specified, we shall write $\hat  F(z; \theta)$ as simply $\hat  F(z)$ where $-h \le z \le 0$, for brevity.
	On solving \eqref{ME} - \eqref{BC2} we find that
	\begin{equation}
		\phi_1=\frac{\Lambda_1}{\rho_1 g}\Bigg[1 + \frac{g (m^2+m'^2)z}{\hat  F(0) \hat  F(z)}\Bigg], \quad -d < z< 0,
		\label{eq:phi1_2layer}
	\end{equation}
	and
	\begin{equation}
		\phi_2 = \frac{\Lambda_2 (m^2+m'^2) (z+h)}{\rho_2 \hat  F(-h) \hat  F(z)}, \quad -h < z < -d, 
		\label{eq:phi2_2layer}
	\end{equation}
	where $\Lambda_{1,2}$ are parameters dependent on $\theta$. Requiring the continuity of $\phi$ at the interface between the two layers ($z = -d$), we obtain \begin{align}
		\Lambda_2 =  \frac{ \rho_2 \hat  F[-h] [\hat  F(0) F(-d) - g d (m^2 + m'^2)]}{\rho_1 g \hat  F[0] (h-d) (m^2 + m'^2)} \Lambda_1.
		\label{eq:M}
	\end{align}
	Thus, the solution to the modal equations in the two layers is given, in dimensional form, by
	\begin{eqnarray}
		&& \phi_1 = \Lambda \Bigg[1+\frac{g (m^2+m'^2)z}{\hat  F(0)\hat  F(z)}\Bigg], \quad -d < z< 0, \label{eq:pp1} \\ 
		&& \phi_2 = \Lambda \Bigg[\frac{\hat  F(0)\hat  F(-d) - g d(m^2+m'^2)}{\hat  F(0)\hat  F(z)}\Bigg]\frac{z+h}{h-d}, \quad -h < z < -d, \label{eq:pp2}
	\end{eqnarray}
	where $\Lambda = \Lambda_1 / \rho_1 g$.
	
	Integrating \eqref{ME} across the interface from $z = -d - \epsilon$ to $z=-d+\epsilon$ and considering the limit $\epsilon \to 0$, we obtain the jump condition
	\begin{equation}
		[\rho_0 \phi_z]  \frac{\hat  F^2(-d) }{m^2+m'^2} = g [\rho_0]  \phi(-d),
		\label{eq:JC1.1}
	\end{equation}
	which provides the directional adjustment equation  for the ring waves, i.e. an equation defining both the speed in the absence of the current, and the speed modifying function $m(\theta)$ for the ring wave at all angles to the direction of the current, when the current is present:
	\begin{align}
		&(\rho_2 - \rho_1) g^2 d (h-d) (m^2 + m'^2)^2  - \rho_2 g \hat  F(-d) [\hat  F(-h) d + \hat  F(0) (h-d) ](m^2 + m'^2) \nonumber \\ 
		&+\rho_2 \hat  F(-h) \hat  F(0) \hat F^2(-d)=0.
		\label{eq:DRgammaz}
	\end{align} 
	
	To find the wave speed $s$ in the absence of a shear flow we set $\gamma = 0$ and $m=1$. The dispersion relation takes the form of a bi-quadratic equation in $s$,
	\begin{equation}
		\rho_2 s^4 -\rho_2 g h s^2+(\rho_2-\rho_1) g^2 d (h-d)=0.
		\label{eq:s4}
	\end{equation}
	Thus the wave speed in the absence of a shear flow is given by
	
	\begin{equation}
		s^2  =\frac{\rho_2 g h  \pm \sqrt{\Delta_1}}{2\rho_2},
		\label{eq:s1}
	\end{equation}
	where 
	\begin{align}
		\Delta_1 &= (\rho_2 g h)^2-4\rho_2(\rho_2-\rho_1)g^2d(h-d) \nonumber \\ 
		&\geq (\rho_2 g h)^2-4\rho_2(\rho_2-\rho_1) g^2 \frac{h^2}{4} =\rho_1\rho_2 g^2 h^2 > 0.
	\end{align}	
	The upper sign corresponds to the surface mode and the lower sign to the slower internal mode. For example, if $\rho_1=1000\ kg\ m^{-3}$, $\rho_2=1020\ kg\ m^{-3}$, $h = 10\ m$  and $d=5\ m$, we obtain $s_{sur} \approx 9.88\ m s^{-1}$ and $s_{int} \approx 0.69\ m s^{-1}$.  In the estimate, we used the maximum of the function $d (h-d)$ on the interval $0 \le d \le h$.		
	
	With the shear flow present, equation \eqref{eq:DRgammaz} constitutes a nonlinear first-order differential equation for the function $m(\theta)$. We have 
	\begin{equation}
		m^2+m'^2=\frac{\rho_2 g \hat  F(-d)[d \hat  F(-h)] + (h-d) \hat  F(0)]\pm \sqrt{\Delta_2}}{2(\rho_2-\rho_1) g^2 d (h-d)},
		\label{eq:DRgammaz2}
	\end{equation}
	where
	\begin{equation}
		\Delta_2 = \rho_2 g \hat  F(-d) [d \hat  F(-h) + (h-d) \hat  F(0)]^2 - 4(\rho_2-\rho_1) g^2 d (h-d) \rho_2 \hat  F(-h) \hat  F(0) \hat  F^2(-d).
		\label{eq:Delta2}
	\end{equation}
	We can show the positivity of $\Delta_2$ in the absence of a shear flow. Indeed, when $\gamma =0$,
	\begin{align}
		\Delta_2 &= s^4 g^2 [\rho_2^2 h^2 - 4\rho_2(\rho_2-\rho_1)d(h-d)] \geq s^4g^2[\rho_2^2 h^2-4\rho_2(\rho_2-\rho_1) \frac{h^2}{4}] \nonumber \\
		&=s^4 g^2 \rho_1\rho_2 h^2 > 0,
	\end{align}
	by the same argument presented above in proving that $\Delta_1 > 0$. By continuity, this inequality will hold in the case of a sufficiently weak shear flow that we consider here. 
	
	We recall that the generalised Burns condition \citep{J90} for surface waves in a homogeneous fluid with this linear shear flow is given by the equation \eqref{eq:gammaza},
	as discussed in Section 2. We note that this equation can be recovered from equation \eqref{eq:DRgammaz} in the limit $d \rightarrow 0$. 
	
	\subsection{Singular solution for the interfacial ring waves: rigid-lid approximation}
	
	We now impose the rigid-lid approximation at the surface to eliminate surface waves: 
	\begin{equation}
		\phi= 0 \hspace{1cm}\text{at} \hspace{1cm} z=0.
		\label{eqRL2}
	\end{equation}

	The modal functions in the top and bottom layers now are, respectively, 
	\begin{align}
		\phi_1 &= \frac{\Lambda (m^2+m'^2) z}{\hat  F(0) \hat  F(z)}, \quad -d < z < 0, 
		\label{eqn:modal1_rl} \\
		\vspace{0.2cm}
		\phi_2 &= \frac{\Lambda (-d) (m^2+m'^2)(z+h)}{\hat  F(0) \hat  F(z) (h-d)}, \quad -h < z < -d, 
		\label{eqn:modal2_rl} 
	\end{align}
	where, as before, $\Lambda$ is a parameter depending on $\theta$.
	The jump condition at the interface again provides the directional adjustment equation:
	\begin{align}
		m^2+m'^2 &= \frac{\hat  F(-d) [\rho_1 \hat  F(0) (h-d) + \rho_2 \hat  F(-h) d]}{(\rho_2 - \rho_1) g d (h-d)}.
		\label{RL}
	\end{align}
	Assuming that $\displaystyle \frac{\rho_2-\rho_1}{\rho_2} \ll 1$, one can show that, to leading order, the right-hand side of this equation is given by 
	\be 
	\frac{\rho_2 h \hat  F^2(-d)}{(\rho_2 - \rho_1) g d (h-d)} > 0,
	\ee
	 which by continuity will continue to hold for a sufficiently small $\gamma$.

	Setting $\gamma=0$ and $m=1$, the speed of the waves in the absence of a shear flow is found as
	\begin{equation}
		s^2=\frac{(\rho_2-\rho_1) g d (h-d)}{\rho_1(h-d)+\rho_2 d} \geq 0,
		\label{eq:s_my_model}
	\end{equation}
	and the adjustment equation \eqref{RL} can be written in the form
	\begin{equation}
		m^2+m'^2= 1 + \frac{[\rho_1 (2 h - d) + \rho_2 d] (-s) + \rho_1  (h-d) \gamma M}{(\rho_2-\rho_1) g d h} \gamma M,
		\label{dsrel2}
	\end{equation}
	where $M=m\cos \theta -m' \sin \theta$. 
	The general solution 
	has the form \eqref{eq:GS}, where
	\begin{align}
		a^2+b^2 &= 1 + \alpha^2 a^2 -  \beta^2 a, 
	\end{align}
	and
	\be
	\alpha^2 = \frac{\gamma ^2 \rho_1  (h-d)}{(\rho_2-\rho_1) g d h}, \qquad \beta^2 = \frac{\gamma s [\rho_1 (2 h - d) + \rho_2 d]}{(\rho_2-\rho_1) g d h}.
	\ee
	The right-hand side is equal to $1$ for $\gamma = 0$, and by continuity it will remain positive for sufficiently weak currents considered here. Solving for $b$, we find
	\be
	b = \pm \sqrt{1 - (1-\alpha^2) a^2 - \beta^2 a},
	\ee
	where 
	\be
	a \in [a_1, a_2], \quad a_{1,2} = -\frac{\beta^2 \pm \sqrt{\beta^4 + 4 (1-\alpha^2)}}{2 (1-\alpha^2)}.
	\ee
	Here, we assume that the locus of parameters $a$ and $b$ is an ellipse, i.e. $1 - \alpha^2 > 0$ implying 
	\be
	\gamma^2 < \frac{(\rho_2 - \rho_1) gdh}{\rho_1  (h-d)},
	\ee
	i.e. we consider an {\it elliptic regime}, when a part of the ring can propagate upstream in the reference frame moving with the speed of the current at the bottom (\cite{K}). 
	
	The general solution can be found in the form
	\begin{equation}
		m(\theta)=a\cos \theta \pm \sqrt{1 - (1-\alpha^2) a^2-\beta^2 a}\sin \theta,
		\label{gs1}
	\end{equation}
	and reparametrised as
	\begin{equation}
		m(\theta)= \frac{-\beta^2 + \sqrt{4 (1-\alpha^2) + \beta^4} \cos \phi}{2 (1-\alpha^2)} \cos \theta + \frac{\sqrt{ 4 (1-\alpha^2) + \beta^4}}{2 \sqrt{1-\alpha^2}} \sin \phi \sin \theta.
		\label{gs2}
	\end{equation}
	Then, the singular solution is found by requiring $\displaystyle \frac{dm}{d \phi} = 0$, which yields
	\begin{equation}
		\tan \phi = \sqrt{1-\alpha^2} \tan \theta.
	\end{equation}
	Finally, the singular solution corresponding to the outward propagating ring wave takes the form
	\begin{equation}
		m = \frac{1}{2 (1-\alpha^2)}  \left [ - \beta^2 \cos \theta +  \sqrt{[4 (1-\alpha^2) + \beta^4] [\cos^2 \theta + (1-\alpha^2) \sin^2 \theta]} \right ].
		\label{ss1}
	\end{equation}

	\subsection{Singular solutions for surface and interfacial ring waves: free surface}
	
	If we don't make the rigid-lid approximation, the required singular solution can not be found in the form $m = m(\theta)$, but it can be found  in parametric form $m = m(a), \theta = \theta(a)$.
	On substitution of $m(\theta)=a\cos\theta +b(a)\sin\theta$ into \eqref{eq:DRgammaz2} we obtain
	\begin{align}
		a^2+b^2 &= \frac{\rho_2 g (-s+\frac{\gamma}{h} (h-d) a)[d (-s) + (h-d) (-s + \gamma  a)] \pm \sqrt{\Delta_3}}{2(\rho_2-\rho_1) g^2 d (h-d)},
		\label{DR_FS}
	\end{align}
	where
	\begin{eqnarray}
		\Delta_3 &=& (\rho_2 g (-s+\frac{\gamma}{h} (h-d) a)[d (-s) + (h-d) (-s + \gamma  a)])^2  \nonumber \\
		&-& 4 (\rho_2 - \rho_1) g^2 d (h-d) \rho_2 (-s) (-s + \gamma  a) (-s + \frac{\gamma}{h} (h-d) a)^2.
		\label{delta_3}
	\end{eqnarray}
	We can show that when $\gamma = 0 $, $\Delta_3 \ge \rho_1 \rho_2 s^4 g^2 h^2 > 0$. Thus, by continuity, the solutions will exist for a sufficiently weak shear flow.
	
	
	
	The singular solution $m=m(\theta)$ takes the form: 
	\begin{empheq}[left=\empheqlbrace]{align}
		&m(\theta)=a\cos\theta +b(a)\sin\theta, \label{para1}\\
		&b'(a)=-1/\tan\theta, \label{para2}\\
		&a^2+b^2(a)=\frac{\rho_2 g (-s+\frac{\gamma}{h} (h-d) a)[d (-s) + (h-d) (-s + \gamma  a)] \pm \sqrt{\Delta_3}}{2(\rho_2-\rho_1) g^2 d (h-d)},\label{para3}
	\end{empheq}
	where the upper sign corresponds to the interfacial mode, and the lower sign - to the surface mode.
	
	Let us denote
	\begin{equation}
		a^2+b^2(a)=  \frac{\rho_2 g (-s+\frac{\gamma}{h} (h-d) a)[d (-s) + (h-d) (-s + \gamma  a)] \pm \sqrt{\Delta_3}}{2(\rho_2-\rho_1) g^2 d (h-d)} = Q.
		\label{abQ}
	\end{equation}
	The solution of the inequality $b^2=Q-a^2 \geq 0$ determines the domain of $a \in [a_{min},a_{max}]$. We are interested in an outward propagating ring wave, thus we require $m(\theta) > 0$. In order to keep the positivity of $m(\theta)$ everywhere, $a$ must take both positive and negative values, therefore the interval $[a_{min},a_{max}]$ should be chosen such that it contains the point $a=0$. Following \cite{KZ16a} (for details see Appendix B), we find that
	\begin{equation}
		m(a)=-\frac{aQ_a-2Q}{\sqrt{(Q_a-2a)^2+4b^2}},
		\label{eq:k(a)_final}
	\end{equation}
	and
	\begin{equation}
		\text{sign}(b)=
		\begin{cases}
			1 \hspace{0.54cm} \text{if}\hspace{0.13cm} \theta \in (0,\pi),\\
			-1 \hspace{0.23cm}\text{if}\hspace{0.13cm} \theta \in (\pi,2\pi).
		\end{cases}
		\label{cases}
	\end{equation}
	Therefore, if $\theta \in (0,\pi)$, then
	\begin{align}
		b&=\sqrt{Q-a^2},\\
		\tan\theta &=-\frac{2\sqrt{Q-a^2}}{Q_a-2a},
	\end{align}
	and we let
	\begin{equation}
		\theta = 
		\begin{cases}
			\displaystyle \arctan \Bigg( -\frac{2\sqrt{Q-a^2}}{Q_a-2a} \Bigg)  \hspace{1.2cm} \text{if}\hspace{0.13cm} Q_a-2a<0,\\
			\displaystyle \arctan \Bigg( -\frac{2\sqrt{Q-a^2}}{Q_a-2a} \Bigg) + \pi \hspace{0.41cm}\text{if}\hspace{0.13cm} Q_a-2a>0.
		\end{cases}
	\end{equation}
	Likewise, if $\theta \in (\pi,2\pi)$, then, the solution is obtained using the symmetry of the problem, and is explicitly given by
	\begin{align}
		b&=-\sqrt{Q-a^2},\\
		\tan\theta &=\frac{2\sqrt{Q-a^2}}{Q_a-2a},
	\end{align}
	where we let
	\begin{equation}
		\theta = 
		\begin{cases}
		\displaystyle	\arctan \Bigg( \frac{2\sqrt{Q-a^2}}{Q_a-2a} \Bigg) +\pi \hspace{1cm} \text{if}\hspace{0.13cm} Q_a-2a>0,\\
		\displaystyle	\arctan \Bigg( \frac{2\sqrt{Q-a^2}}{Q_a-2a} \Bigg) + 2\pi \hspace{0.75cm}\text{if}\hspace{0.13cm} Q_a-2a<0.
		\end{cases}
	\end{equation}

	

	\begin{figure}
		\centering
		\begin{subfigure}[b]{0.49 \textwidth}
			\includegraphics[width=\textwidth]{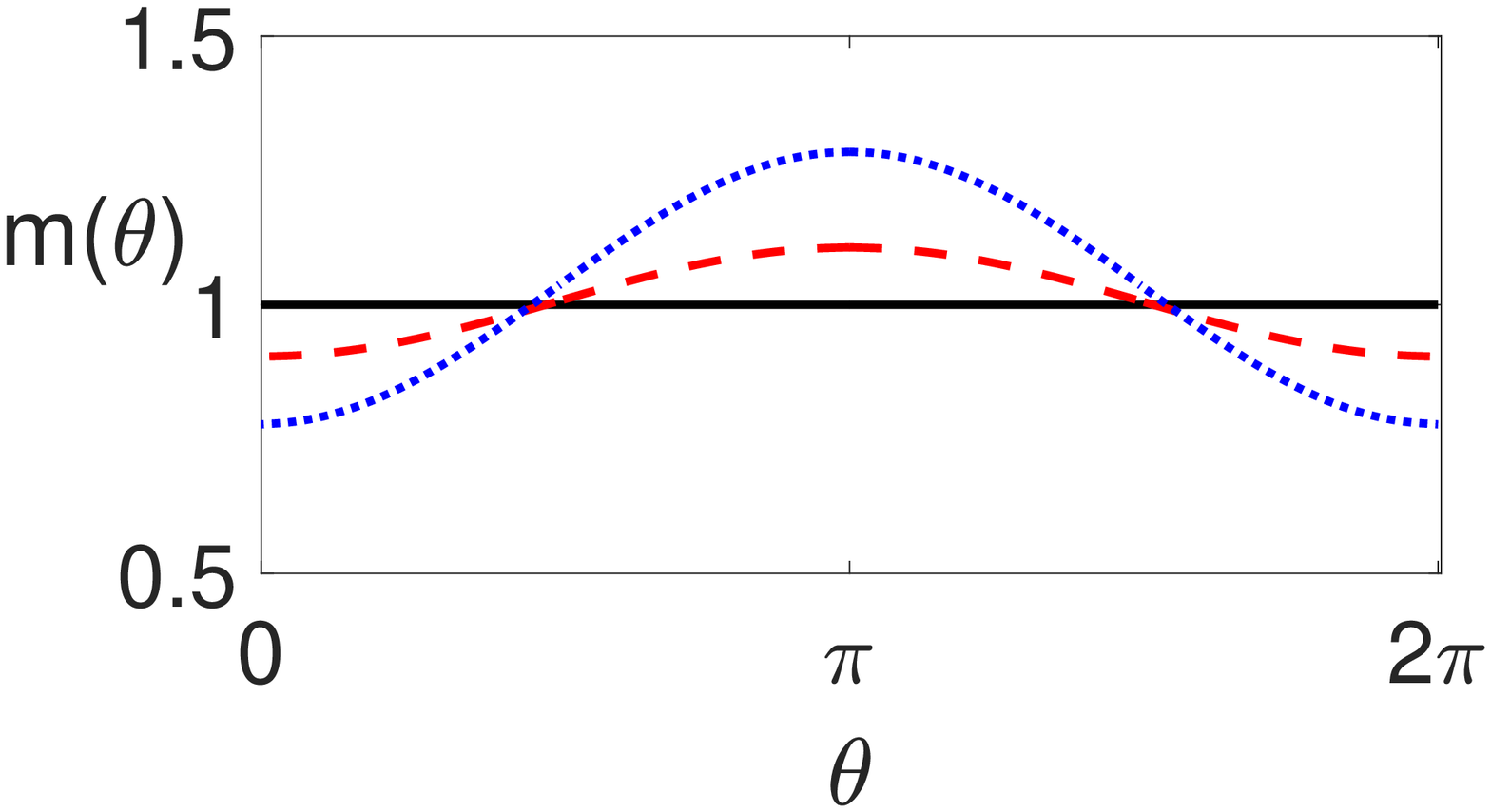}
			\caption{Surface mode: $\gamma = 0\ ms^{-1}$ (black, solid), $\gamma = 2\ ms^{-1}$ (red, dash) and $\gamma = 5\ ms^{-1}$  (blue,\hspace{0.7cm} dot).}
			\label{fig:t_a_surface}
		\end{subfigure}
		~ 
		\begin{subfigure}[b]{0.49 \textwidth}
			\includegraphics[width=\textwidth]{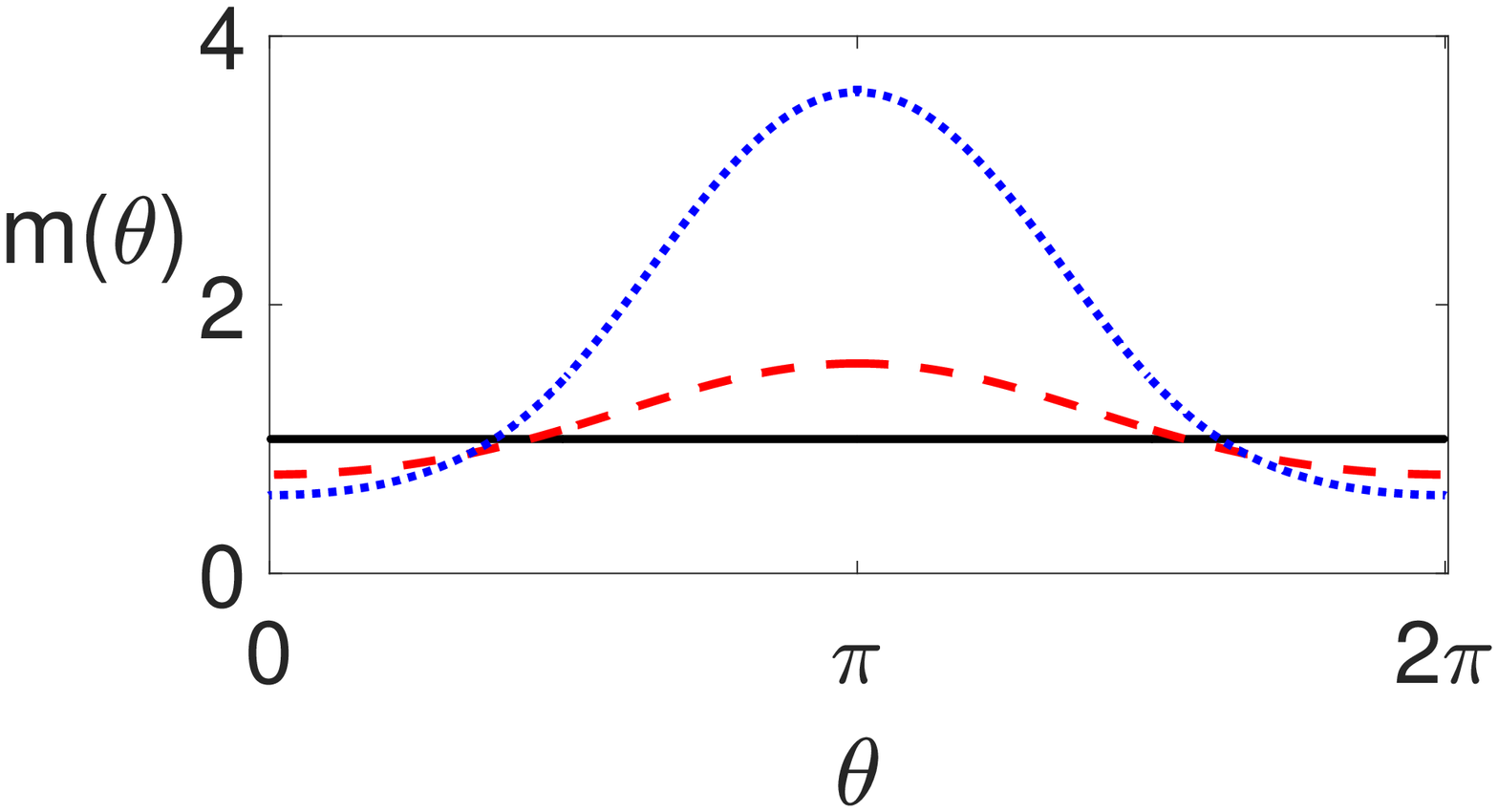}
			\caption{Interfacial mode: $\gamma = 0\ ms^{-1}$ (black, solid), $\gamma = 0.5\ ms^{-1}$ (red, dash) and $\gamma = 1\ ms^{-1}$ (blue, dot).}
			\label{fig:t_a_interface}
		\end{subfigure}
		\caption{Plots of the function $m(\theta)$ for the surface (a) and the interfacial (b) modes.}   
		\label{fig:m_t}
	\end{figure}
	
	The function $m(\theta)$ is shown in Figure \ref{fig:m_t} for both the surface and interfacial mode for a range of different shear flow strengths before it is used to plot the wavefronts of the surface and interfacial waves described by $rm(\theta)=5000\ m$ in Figures \ref{fig:2_layer_gamma_z_surface} and \ref{fig:2_layer_gamma_z_interfacial}. We keep $\rho_1=1000\ kg\ m^{-3}$, $\rho_2=1020\ kg\ m^{-3}$ and $d=5\ m$ as before.

	\begin{figure}
		\centering
		\begin{subfigure}[b]{0.49\textwidth}
			\includegraphics[width=\textwidth]{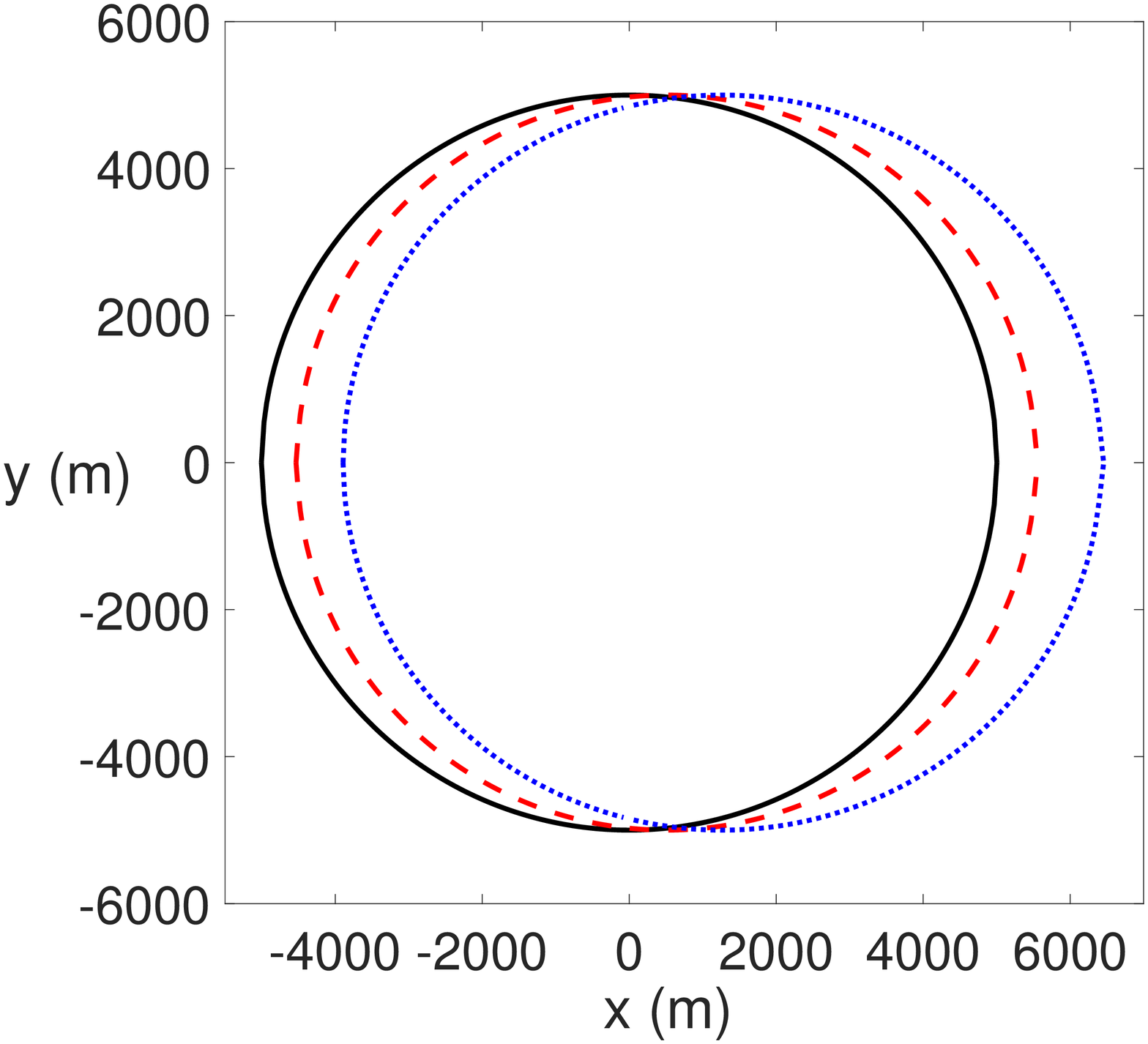}
			\caption{Surface mode: $\gamma = 0\ m s^{-1}$ (black, solid), $\displaystyle{\gamma = 2\ m s^{-1}}$ (red, dash) and $\gamma = 5\ m s^{-1}$ \hspace{1cm} (blue, dot).}
			\label{fig:2_layer_gamma_z_surface}
		\end{subfigure}
		~ 
		\begin{subfigure}[b]{0.49\textwidth}
			\includegraphics[width= \textwidth]{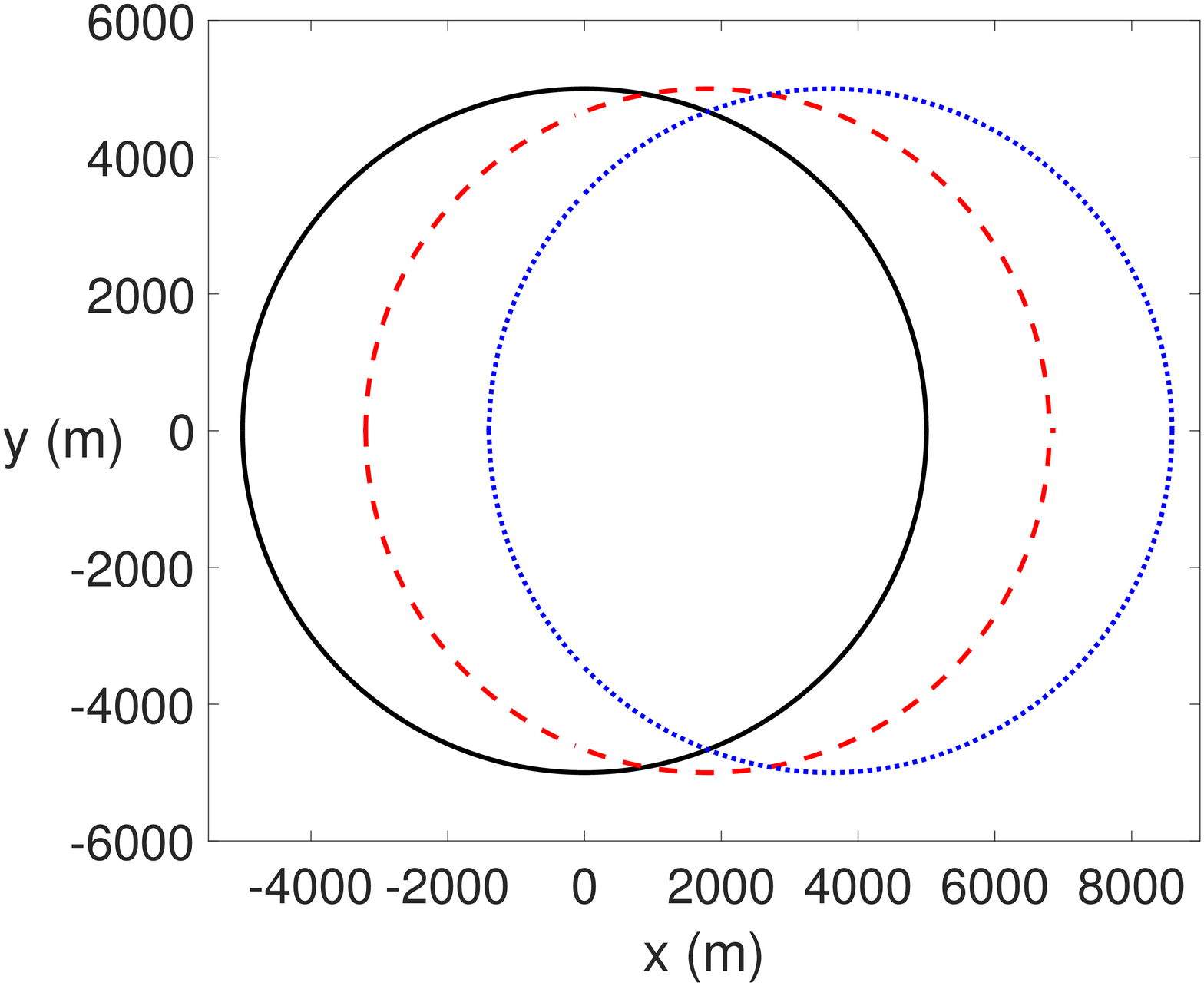}
			\caption{Interfacial mode:  $\gamma = 0\ ms^{-1}$ (black, solid), $\gamma = 0.5\ m s^{-1}$ (red, dash) and $\gamma = 1\ m s^{-1}$ (blue, dot).}
			\label{fig:2_layer_gamma_z_interfacial}
		\end{subfigure}
		\caption{Plots of the wavefronts for the surface (a) and the interfacial (b) modes.}   
		\label{fig:wf}
	\end{figure}
	
		For these values, the approximate solution \eqref{ss1} for the internal waves that arose from applying the rigid-lid approximation is compared to the exact solution given by \eqref{para1} - \eqref{para3} in Figure \ref{fig:comp} for $\gamma=0.5\ ms^{-1}$ and $\gamma = 1\ ms^{-1}$.  The solutions are very close, although the agreement is slightly worse for larger values of $\gamma$.

	\begin{figure}
		\begin{center}
			\includegraphics[width=0.6 \textwidth]{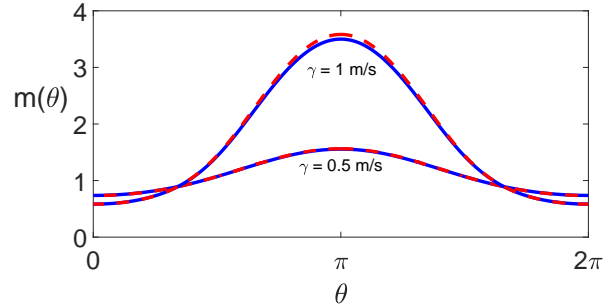}
		\end{center}
		\caption{Plots of the function $m(\theta)$ for the interfacial mode: rigid-lid approximation (blue, solid) and exact (red, dash).}
		\label{fig:comp}
	\end{figure}
	
			It is clear from Figure \ref{fig:2_layer_gamma_z_surface} that the surface wavefronts here share the same qualitative characteristics as the surface wavefronts presented Section 2, i.e. the wavefronts become elongated in the direction of the shear flow as the strength of the shear flow increases. 
	It is also clear from Figure \ref{fig:2_layer_gamma_z_interfacial} that the interfacial wavefronts do not become elongated. To quantify the weak deformation of the wavefront in that case we shall use the global and local measure introduced in Section 2.

 The relative distance between points on the wavefronts in the upstream and downstream direction given previously by \eqref{D1} are plotted, for both modes, in Figure \ref{fig:2_layer_distance}  as a function of $\gamma$. The speed of the wavefronts in the downstream and upstream directions is shown in Figure \ref{fig:2_layer_speed}, while the relative curvature is shown in Figure \ref{fig:2_layer_curvature}.    
	
	\begin{figure}
		\centering
		\begin{subfigure}[b]{0.49\textwidth}
			\includegraphics[width=\textwidth]{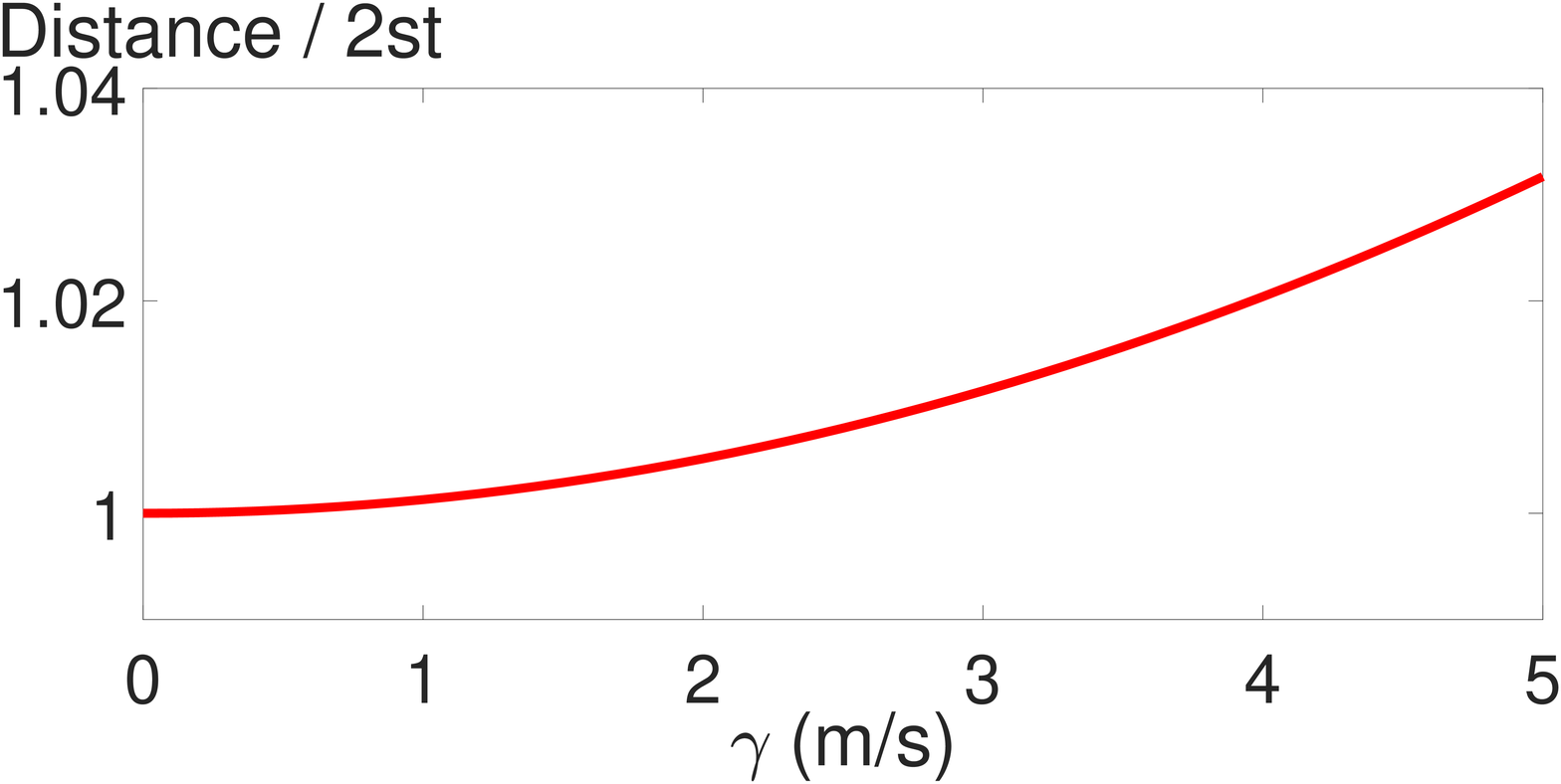}
			\caption{Surface mode.}
			\label{fig:2_layer_surface_distance}
		\end{subfigure}
		~ 
		\begin{subfigure}[b]{0.49\textwidth}
			\includegraphics[width= \textwidth]{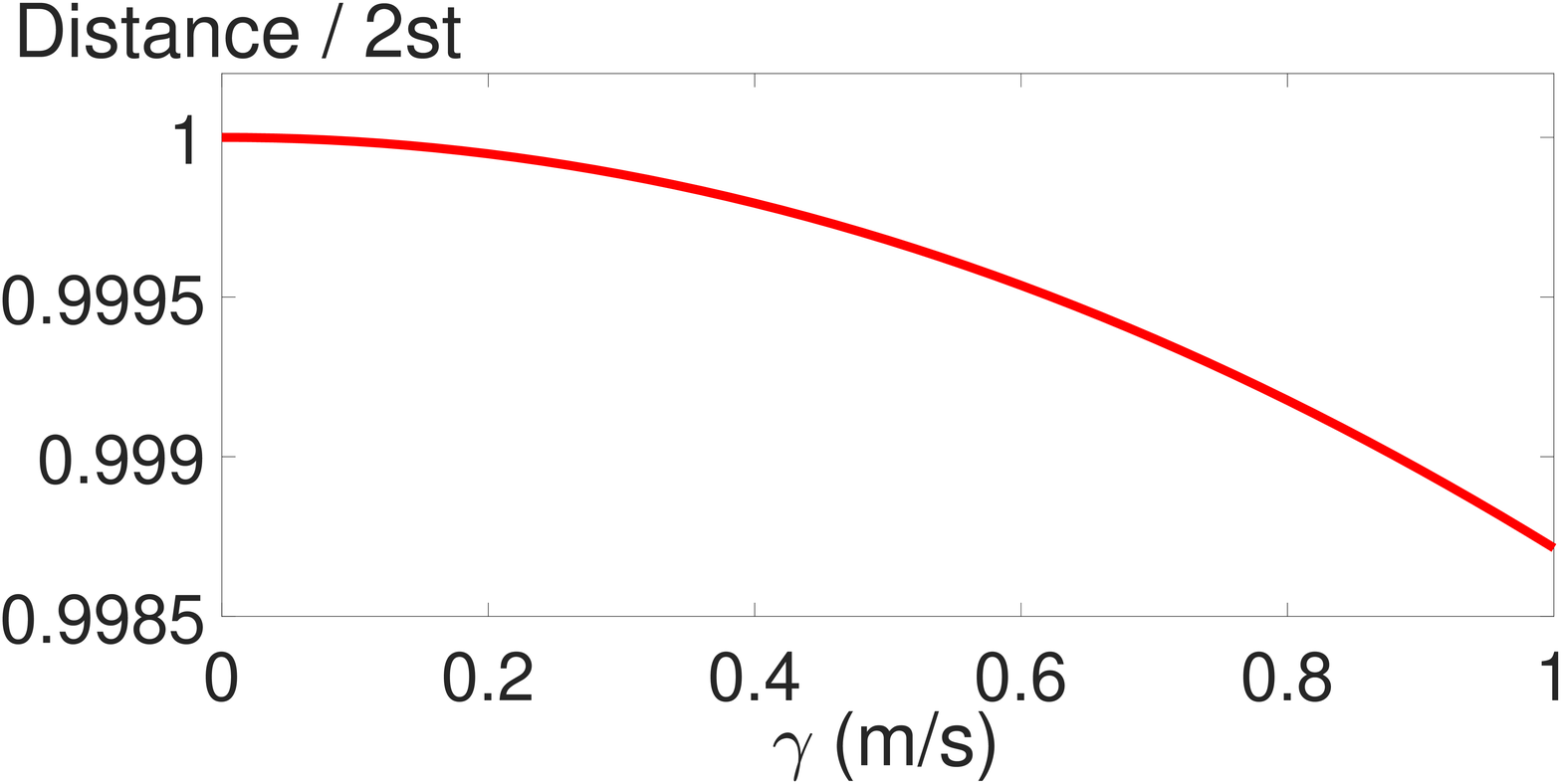}
			\caption{Interfacial mode.}
			\label{fig:2_layer_interfacial_distance}
		\end{subfigure}
		\caption{Relative distance between the points on the wavefronts of (a) surface ring waves and (b) interfacial ring waves in downstream and upstream directions as a function of $\gamma$. 
		}   
		\label{fig:2_layer_distance}
	\end{figure}
	
	\begin{figure}
		\centering
		\begin{subfigure}[b]{0.49\textwidth}
			\includegraphics[width=\textwidth]{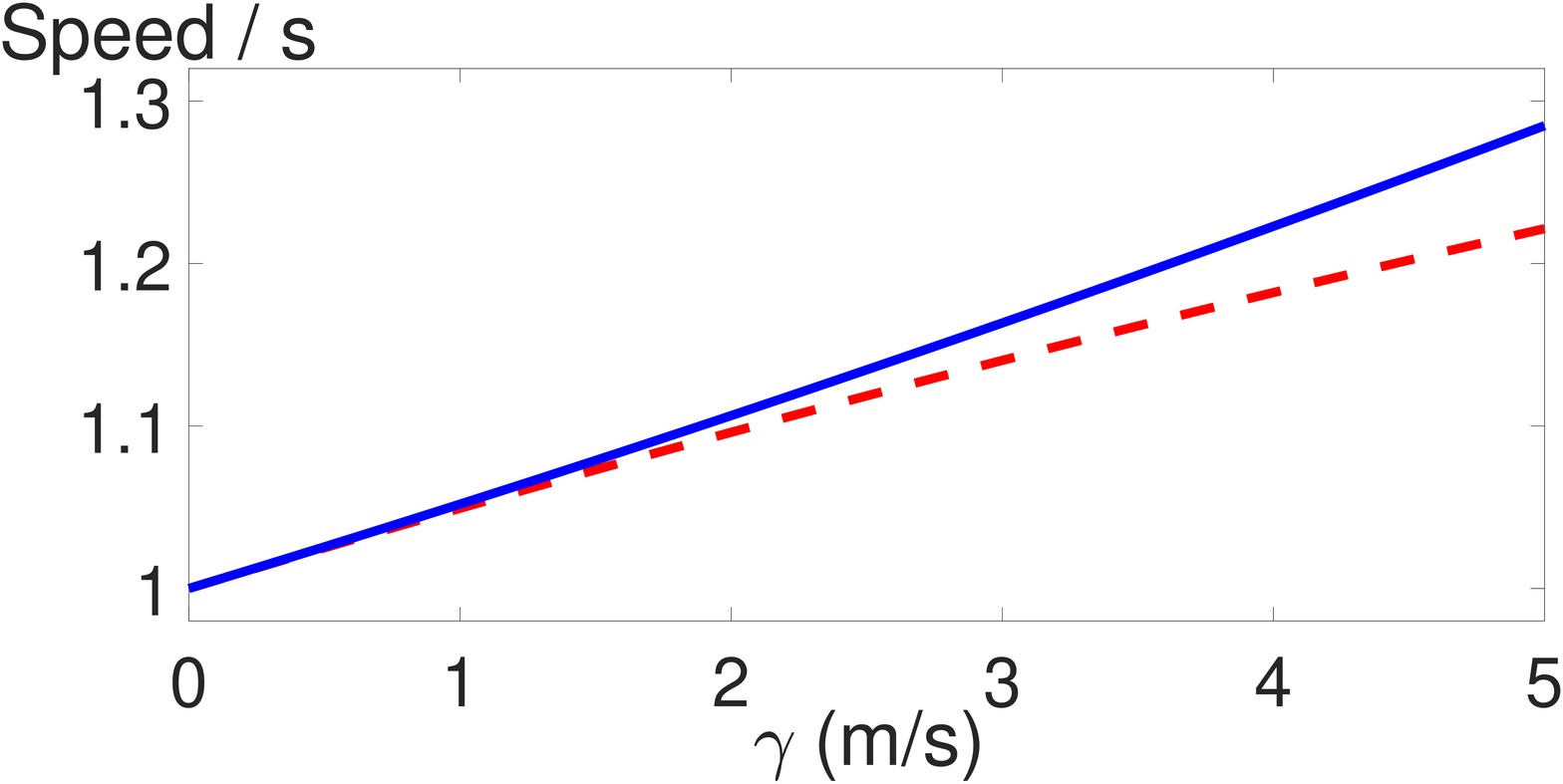}
			\caption{Surface mode.}
			\label{fig:2_layer_surface_speed}
		\end{subfigure}
		~ 
		\begin{subfigure}[b]{0.49\textwidth}
			\includegraphics[width= \textwidth]{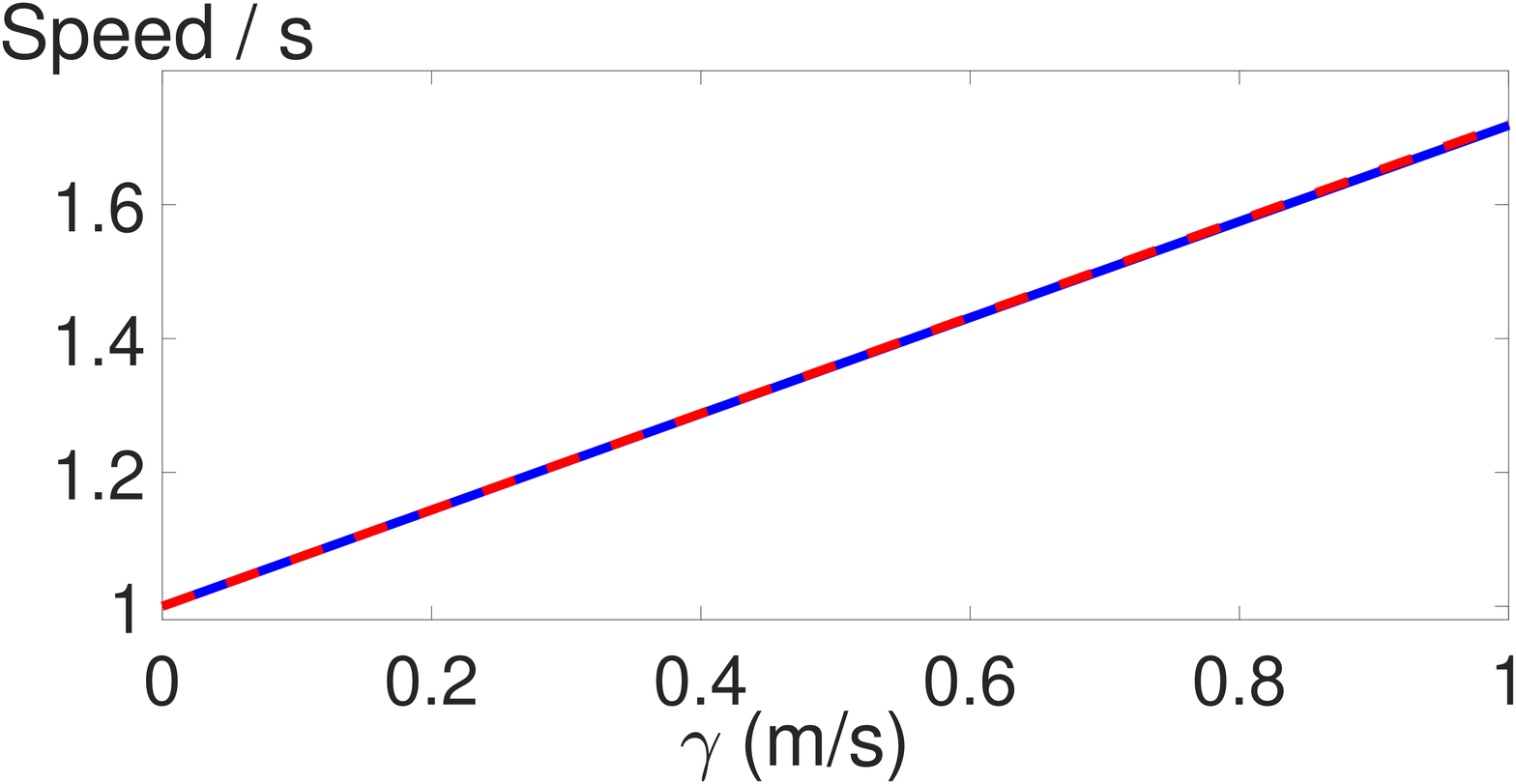}
			\caption{Interfacial mode.}
			\label{fig:2_layer_interfacial_speed}
		\end{subfigure}
		\caption{Relative speed of the wavefronts of (a) surface ring waves and (b) interfacial ring waves as a function of $\gamma$. The blue (solid) curve is for $\theta = 0$ and the red (dash) curve for $\theta = \pi$. 
		}   
		\label{fig:2_layer_speed}
	\end{figure}
	
	\begin{figure}
		\centering
		\begin{subfigure}[b]{0.49\textwidth}
			\includegraphics[width=\textwidth]{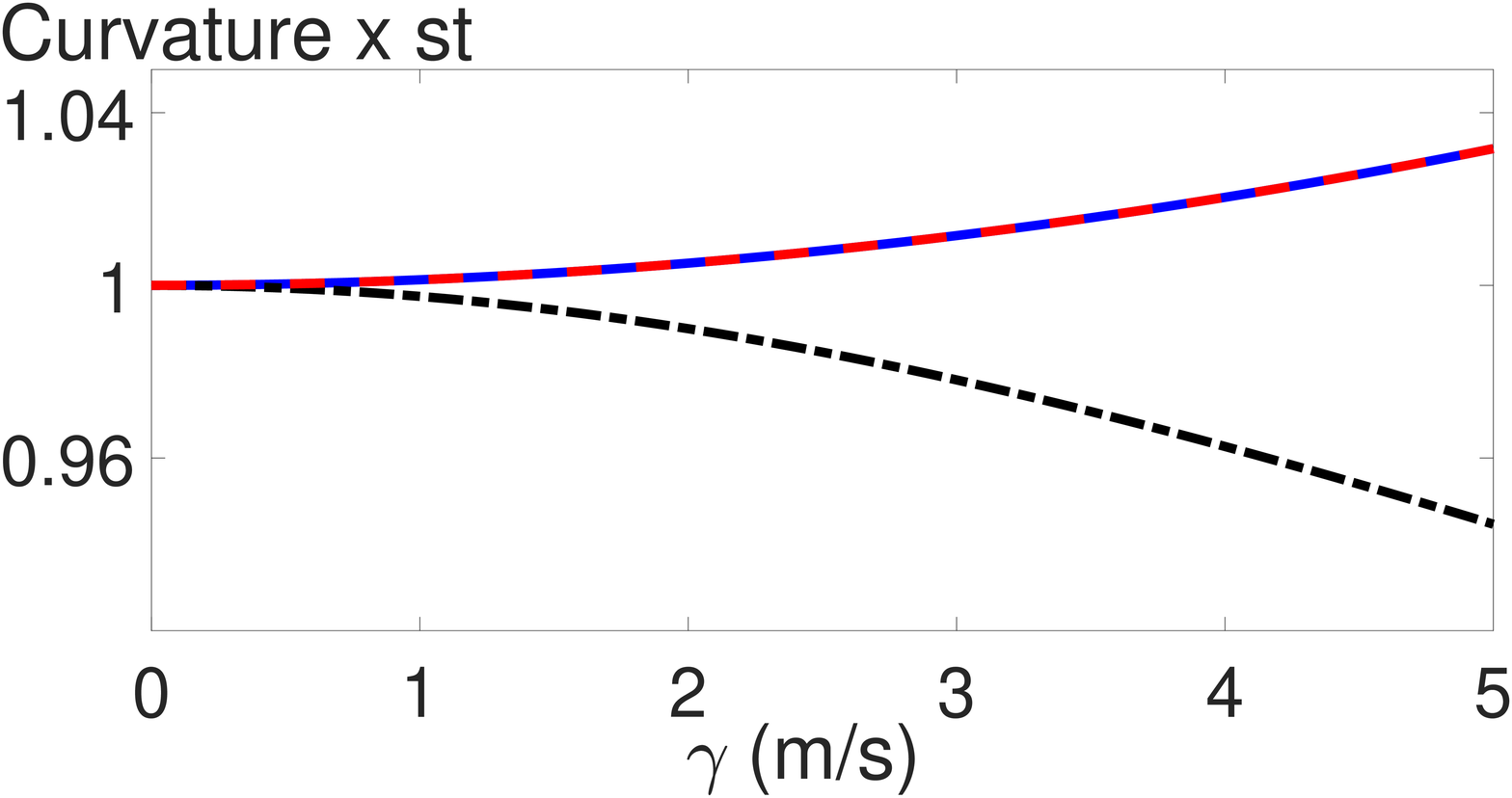}
			\caption{Surface mode.}
			\label{fig:2_layer_surface_curvature}
		\end{subfigure}
		~ 
		\begin{subfigure}[b]{0.49\textwidth}
			\includegraphics[width= \textwidth]{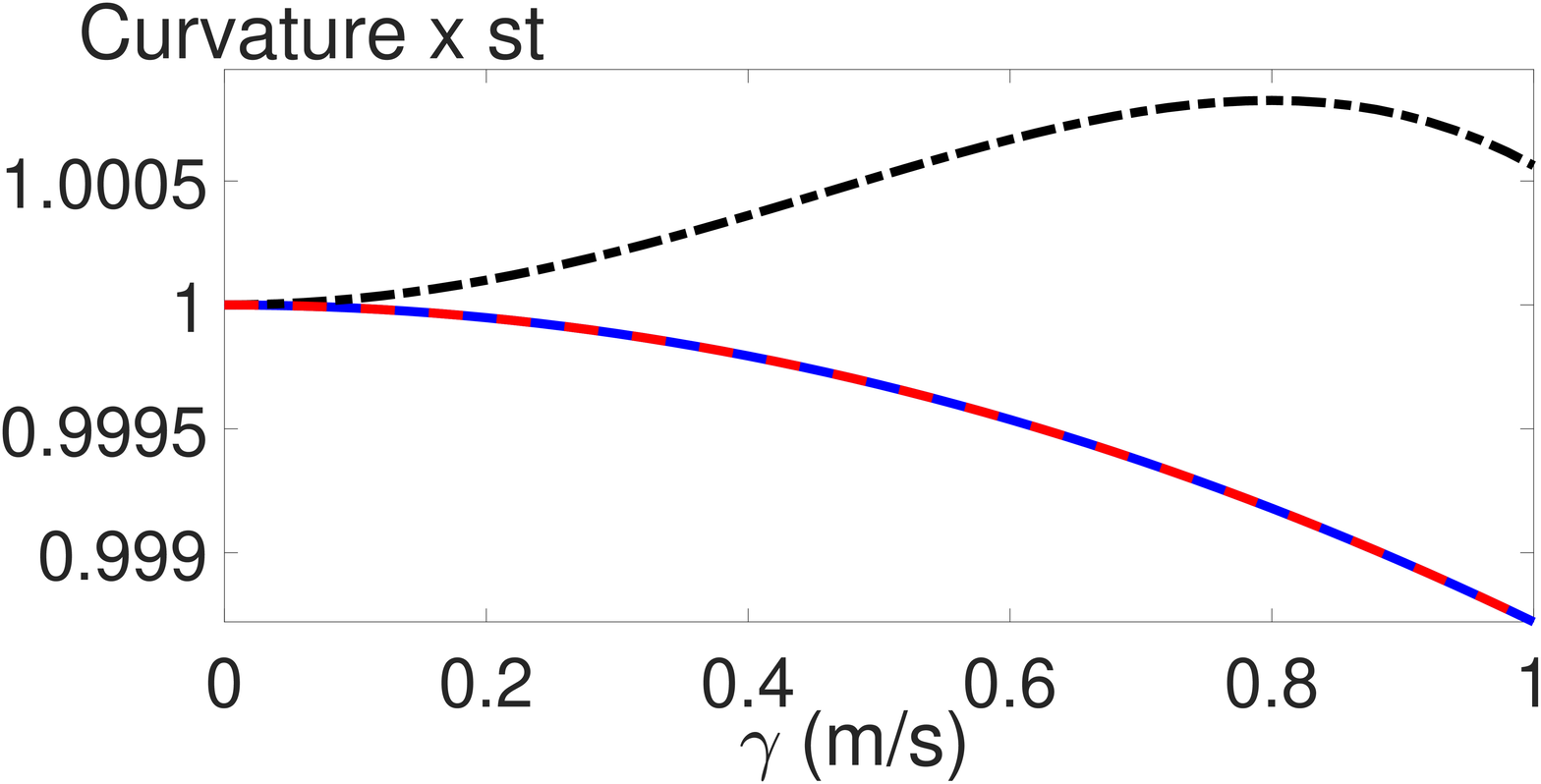}
			\caption{Interfacial mode.}
			\label{fig:2_layer_interfacial_curvature}
		\end{subfigure}
		\caption{Relative curvature of the wavefronts of (a) surface ring waves and (b) interfacial ring waves in different directions as a function of $\gamma$. The blue (solid) curve is for $\theta = 0$, the black (dash-dot) curve for $\theta = \frac{\pi}{2}$ and the red (dash) curve for $\theta = \pi$. 
		}   
		\label{fig:2_layer_curvature}
	\end{figure}
	

	The relative distance between the points on the wavefronts  in the downstream and upstream directions increases for the surface ring waves which indicates elongation of the wavefront in the direction of the flow, and gently decreases for the interfacial waves, thus indicating a slight squeezing of the wavefront in the direction of the flow. Note that the Figure \ref{fig:2_layer_curvature} for the curvature of the wavefronts  is more instructive than the Figure \ref{fig:2_layer_speed} for the speeds in the downstream and upstream directions, indicating again elongation of the surface ring wave and a small squeezing of the wavefront of the interfacial wave. Unlike the strong squeezing reported in the case of a piecewise-constant current, the effect is very weak in that case, and the interfacial wavefront is mainly convected by the flow. It is known that the plane interfacial waves on a constant vorticity current are stable (e.g. \cite{CEGP,BV}), and there is no long-wave instability unlike the case of the piecewise-constant current discussed earlier. We think that this explains the difference in the behaviour of the wavefronts of interfacial ring waves in these two examples. in Section 5 we shall shed more light on the deformation of wavefronts of interfacial waves by considering a family of power-law upper-layer currents approaching the piecewise-constant current as a limiting case.

	To plot the modal functions, we need to choose the parameter $\Lambda$ by normalizing $\phi$. We do this by requiring $\phi(0;\theta) = 1$ for the case of the surface mode, and $\phi(-d;\theta) = 1$ for the interfacial mode. From the modal functions with the free surface \eqref{eq:pp1}, \eqref{eq:pp2} and rigid-lid \eqref{eqn:modal1_rl}, \eqref{eqn:modal2_rl} one can see that applying the normalization for the interfacial mode at $z = -d$ allows us to find $\Lambda$ as a function of $\theta$, i.e. the normalization can be performed simultaneously for all directions.
	
	The exact solutions for the surface mode in the two-layer case is shown in Figure \ref{fig:modal_fs_sm} along with the counterpart of this solution in the case of a homogeneous fluid shown previously in Figure \ref{fig:Johnson_modal}. 
	\begin{figure}
		\centering
		\begin{subfigure}[b]{1\textwidth}
			\includegraphics[width=0.49\textwidth]{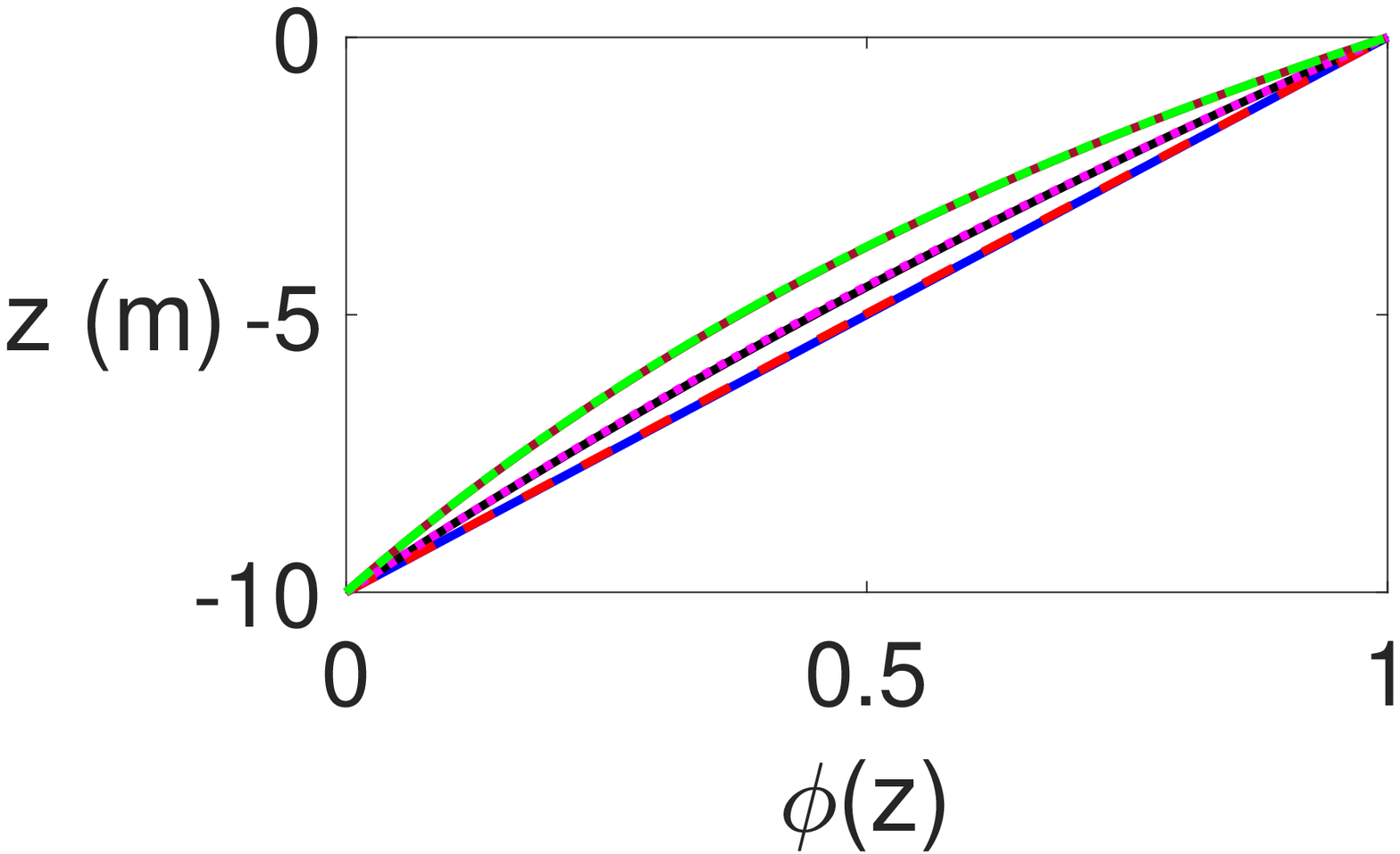}
						\includegraphics[width=0.49\textwidth]{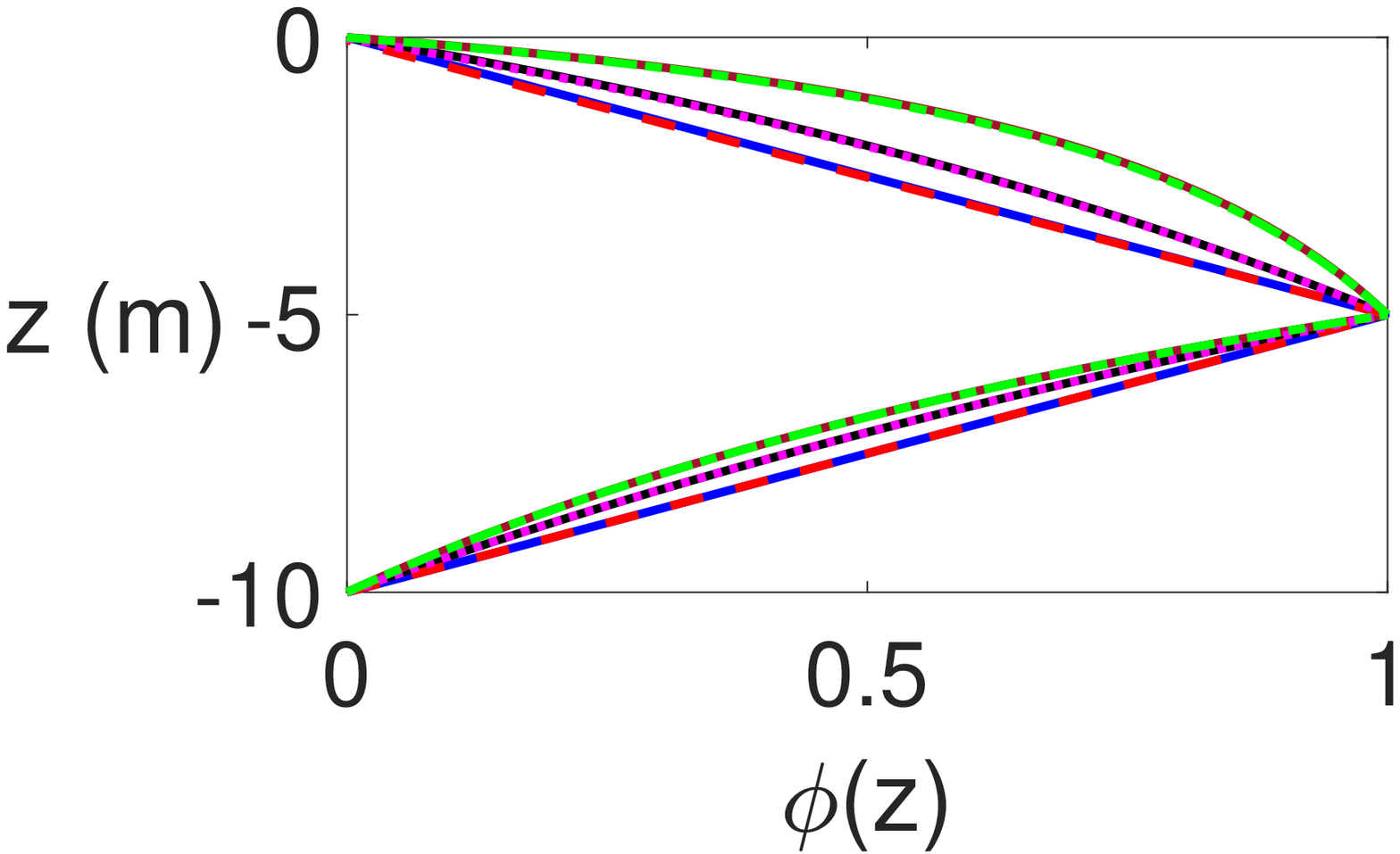}
			\caption{$\theta = 0$ }
			\label{fig:modal_fs_sm_0}
		\end{subfigure}
		
		\begin{subfigure}[b]{1\textwidth}
			\includegraphics[width=0.49\textwidth]{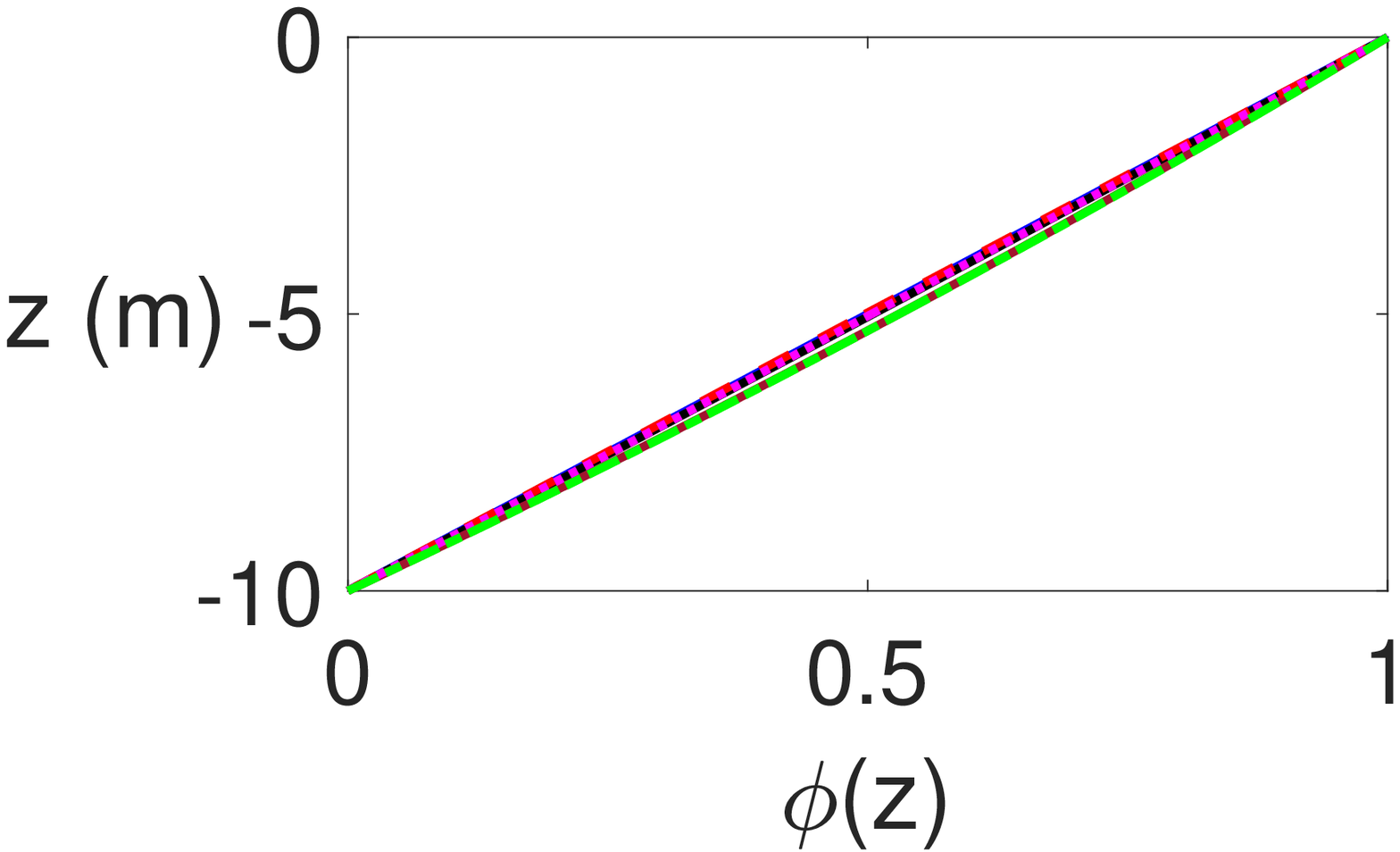}
					\includegraphics[width=0.49\textwidth]{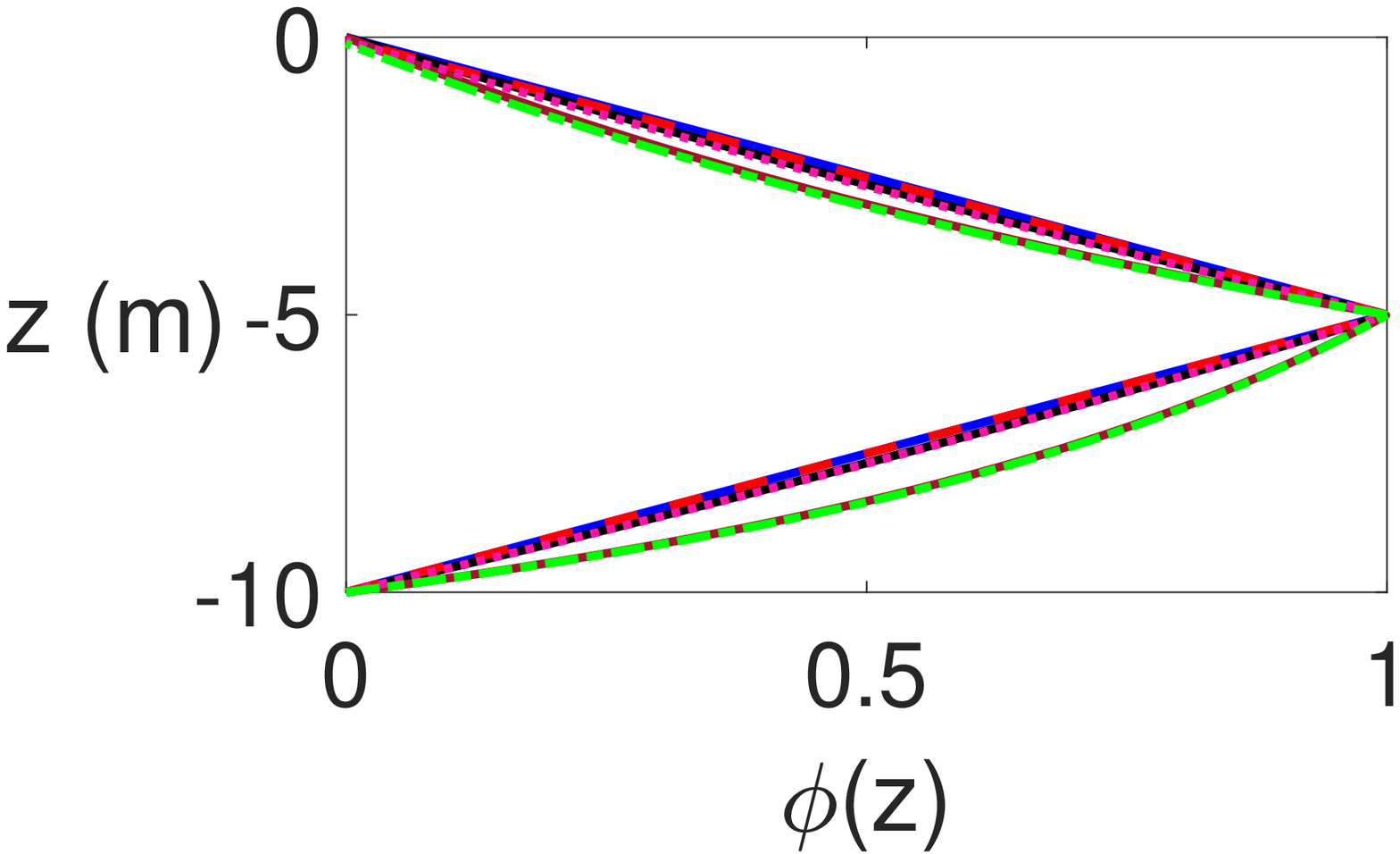}
			\caption{$\theta = \pi/2$}
			\label{fig:modal_fs_sm_half_pi}
		\end{subfigure}
		
		\begin{subfigure}[b]{\textwidth}
			\includegraphics[width=0.49\textwidth]{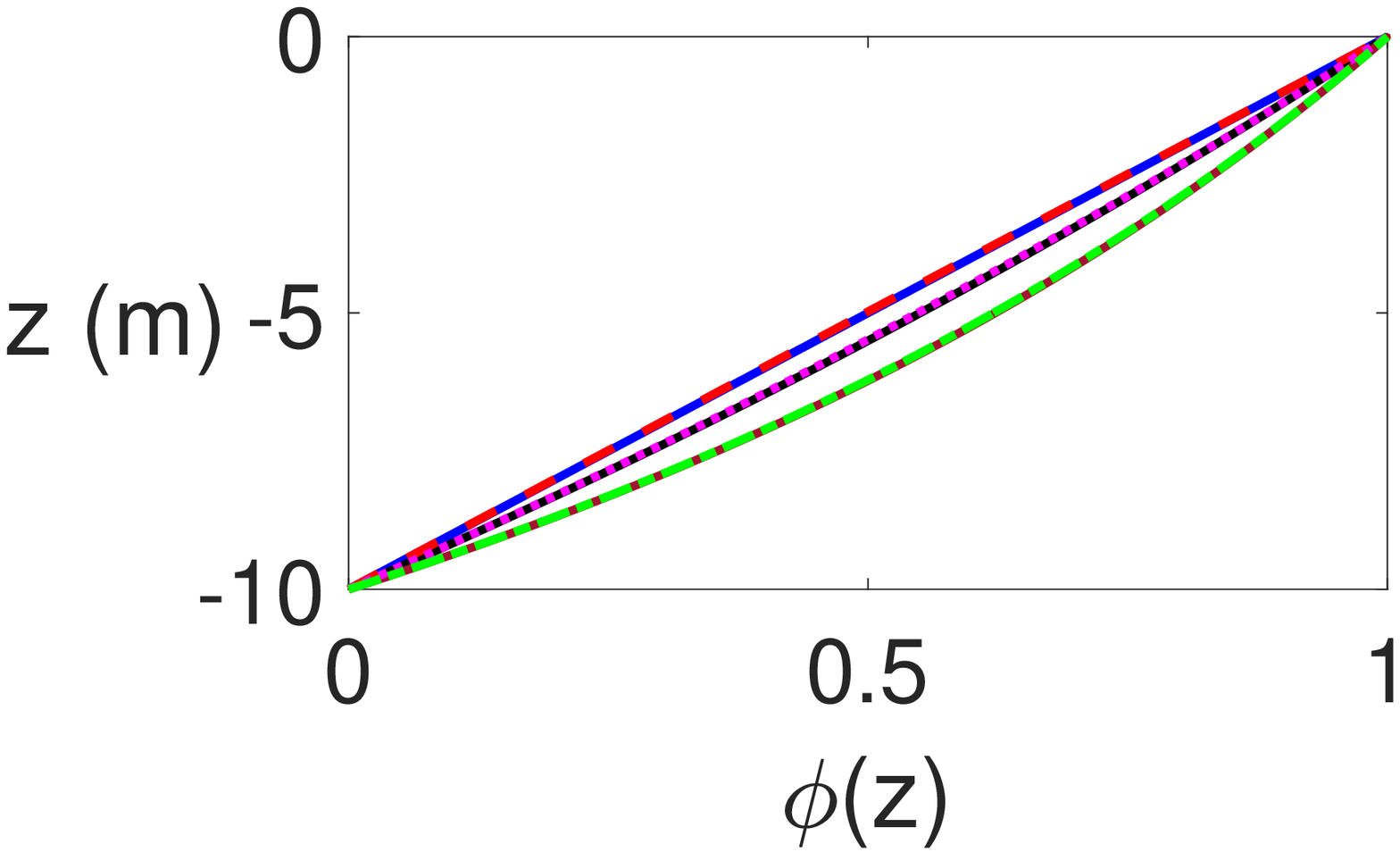}
						\includegraphics[width=0.49\textwidth]{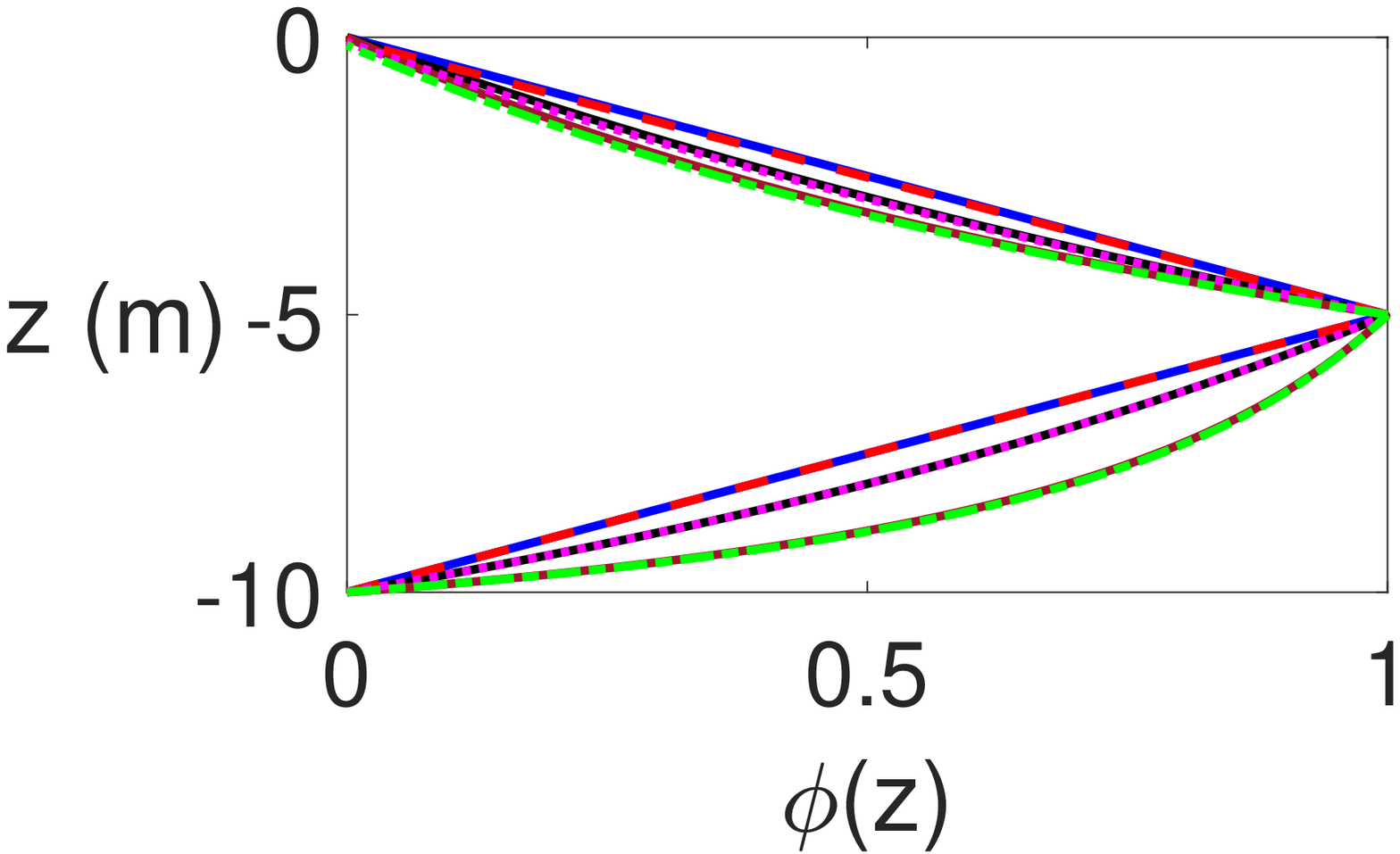}
			\caption{$\theta = \pi$}
			\label{fig:modal_fs_sm_pi}
		\end{subfigure}
		\caption{Plots of the modal functions of the surface mode for a homogenous fluid (HF) and a two layer fluid with a free surface (2L) (left column) with shear flow strengths of $\gamma = 0\ m s^{-1}$ (HF blue, solid) (2L red, dash), $\gamma = 2\ m s^{-1}$ (HF black, solid) (2L pink, dot) and $\gamma = 5\ m s^{-1}$ (HF brown, solid) (2L green, dash dot), and plots of the modal function of the interfacial mode with the free surface (FS) and in the rigid-lid (RL) approximation (right column) with shear flow strengths of $\gamma = 0\ ms^{-1}$ (RL blue, solid) (FS red, dash), $\gamma = 0.5\ ms^{-1}$ (RL black, solid) (FS pink, dot) and $\gamma = 1\ ms^{-1}$ (RL brown, solid) (FS green, dash dot), for different values of $\theta$.} 
		\label{fig:modal_fs_sm}
	\end{figure}

	Also in Figure \ref{fig:modal_fs_sm} are the modal functions for the interfacial mode obtained with the rigid-lid approximation and also with the free surface condition. 
	There is virtually no difference between the two solutions, which should be expected as we have shown in Figure \ref{fig:comp} that $m(\theta)$ for the rigid-lid and exact solution are in good agreement. 

	For both modes, the effect of the parallel shear flow is expectedly at its least in the orthogonal direction. The vertical structure of the wave field is shifted towards the ocean surface in the upstream direction, and towards the ocean bottom in the downstream direction. It gradually changes between these two extremes as $\theta$ changes from $0$ (downstream) to $\pi$ (upstream). For the interfacial mode, the variation of $\phi$ in the upstream direction is larger in the bottom layer, but  in the downstream direction it is larger in the upper layer.  Overall, there are significant differences in the behaviour of all modal functions in the downstream, orthogonal and upstream directions, and the wave field becomes strongly three-dimensional with the increasing strength of the shear flow.


	
	\section{Two-layer fluid with a power-law upper-layer current}
	
	In this section we consider a family of upper-layer currents 

\begin{eqnarray}
	u_0 (z) = \left \{ 
	\begin{array}{c}
		\displaystyle \gamma \left (\frac{z+d}{d} \right )^{\alpha}, \quad \mbox{if} \quad -d < z < 0, \\
		0, \quad \mbox{if} \quad  -h < z < -d,
	\end{array}
	\right .
			\label{eqn:U-L_C}
\end{eqnarray}
	see Figure \ref{fig:power}. The current tends towards a piecewise constant current in the limit $\alpha \to 0$.
	
	Here, we will consider surface waves in the approximation of a homogeneous fluid with that current, and interfacial waves in the rigid-lid approximation, extending the analysis developed in \cite{K}.  The emphasis is on the global and local characterisation of the deformation of the wavefronts introduced in Section 2. We note that the problem can be also handled analytically with the free surface condition, but the formulae become more cumbersome and are not shown here.

	\begin{figure}
	\centering
	\begin{subfigure}[b]{0.49\textwidth}
		\includegraphics[width=\textwidth]{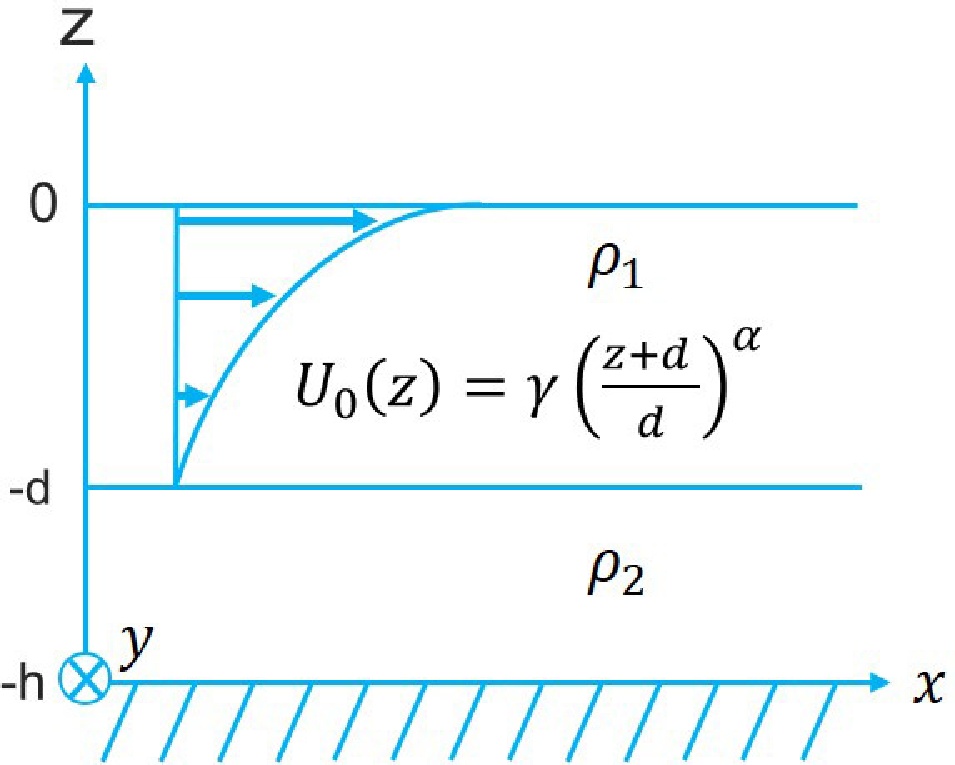}
		\caption{Two-layer fluid with an upper-layer current ($\alpha > 1$).\vspace{0.35cm}}
		\label{fig:two-layer}
	\end{subfigure}
	~ 
	\begin{subfigure}[b]{0.49\textwidth}
		\includegraphics[width= \textwidth]{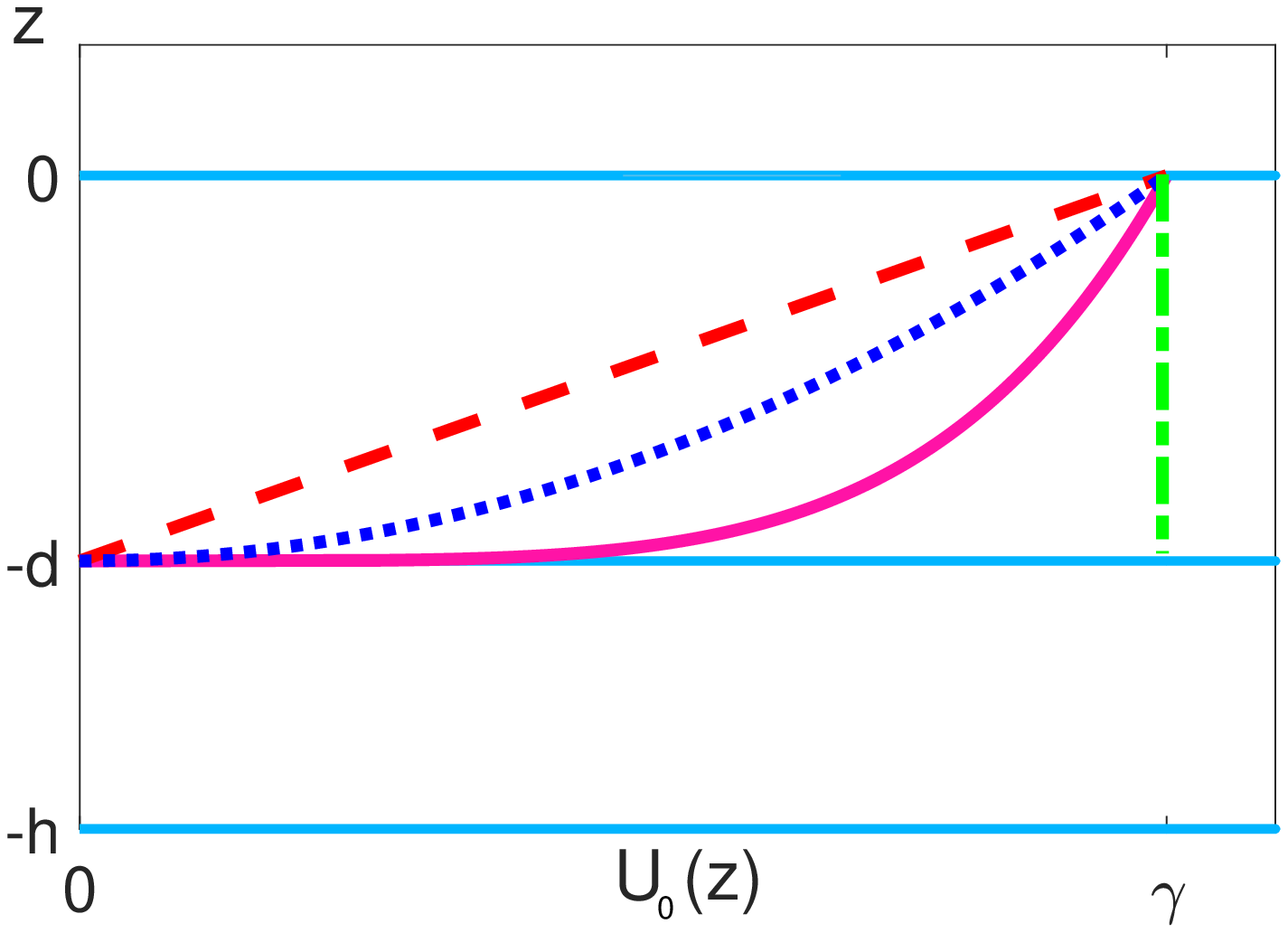}
		\caption{Upper-layer current with $\alpha = 1$ (red, dash), $\alpha = 1/2$ (blue, dot) and $\alpha = 1/5$ (pink, solid) and piecewise-constant current (green, dash-dot).}
		\label{fig:UL_current}
	\end{subfigure}
	\caption{Two-layer model with a power-law upper-layer current  given by \eqref{eqn:U-L_C}.}   
	\label{fig:power}
\end{figure} 
	
	\subsection{Surface waves in a homogeneous fluid}
	
	Solving the modal equations (\ref{eq:ME1}) - (\ref{eq:MEC2}) with $N^2 = 0$ and requiring the continuity of the modal function at $z=-d$, we obtain, in respective layers,
	\begin{eqnarray}
		&& \phi_1(z) = \frac{\Lambda}{g} \left [1 - g (m^2 + m'^2) \int_{z}^0 \frac{dz}{\hat  F_1^2}\right ], \quad -d < z < 0, \\
		&& \phi_2(z) = \frac{\Lambda \left [\displaystyle 1 - g(m^2 + m'^2) \int_{-d}^0 \frac{dz}{\hat  F_1^2}\right ]}{g (h-d)} (z+h), \quad -h < z < -d ,
		\label{phi} 
	\end{eqnarray}
	where $\displaystyle \hat F_1 = -s + \gamma \left (\frac{z+d}{d} \right )^{\alpha} (m \cos \theta - m' \sin \theta)$, and $\Lambda$ is a parameter which can be used to normalise the modal function to be equal to 1 at the surface.
	Requiring the continuity of the derivative with respect to $z$, $\phi_{1z} = \phi_{2z}$ at $z = -d$, we obtain an adjustment equation for $m(\theta)$:
	\begin{equation}
		\left (1 + \frac{gh}{h-d} \int_{-d}^0 \frac{dz}{\hat  F_1^2}  \right ) (m^2 + m'^2) = \frac{h}{h-d}.
		\label{Eqm}
	\end{equation}
	Here, the integral can be expressed in terms of the hypergeometric function ${}_2F_1$ (e.g., {\it Wolfram Mathematica 12.1.1.0}):
	\be
	\int_{-d}^0 \frac{dz}{\hat F_1^2} = \frac{d}{s^2}\  {}_2F_1 \left (2, \frac{1}{\alpha}, 1 + \frac{1}{\alpha}, \frac{\gamma}{s} (m \cos \theta - m' \sin \theta) \right ),
	\ee
	yielding the equation
	\begin{equation}
		m^2 + m'^2 = \frac{h}{\displaystyle h-d + \frac{dgh}{s^2}  {}_2F_1 \left (2, \frac{1}{\alpha}, 1 + \frac{1}{\alpha}, \frac{\gamma}{s} (m \cos \theta - m' \sin \theta) \right )}.
		\label{Eqm1}
	\end{equation}
	First, when $\gamma = 0$, we have $m = 1$ and $\displaystyle {}_2F_1\left (2, \frac{1}{\alpha}, 1 + \frac{1}{\alpha}, 0\right ) = 1$, yielding 
	$
	s^2 = gh.
	$
	Next, the general solution of (\ref{Eqm1}) can be found in the form
	$
	m = a \cos \theta + b(a) \sin \theta,
	$
	where
	\be
	a^2 + b^2 = \frac{h}{\displaystyle h-d+d\  {}_2F_1\left (2, \frac{1}{\alpha}, 1 + \frac{1}{\alpha}, \frac{\gamma}{s} a\right )},
	\ee
	and the singular solution can be found in parametric form by requiring $\displaystyle \frac{dm}{da} = 0$, which yields
	\begin{eqnarray}
		&&b(a) = \sqrt{\frac{h}{h-d + s^2 I(a)} - a^2} , \\
		&&\theta(a) =  \left \{ 
		\begin{array}{c}
			\arctan \frac{\displaystyle 2b [h-d+s^2 I(a)]^2}{\displaystyle 2a [h-d+s^2 I(a)]^2 + s^2 h I'(a)}, \quad \mbox{if} \quad a \in[a_0, a_{max}]\  \left (\displaystyle \theta \in \left [0, \frac{\pi}{2}\right ]\right ), \\
			\arctan \frac{\displaystyle 2b [h-d+s^2 I(a)]^2}{\displaystyle 2a [h-d+s^2 I(a)]^2 + s^2 h I'(a)} + \pi, \quad \mbox{if} \quad a \in[a_{\min}, a_0]\ \left (\displaystyle \theta \in \left [\frac{\pi}{2}, \pi \right ] \right ), 
		\end{array}
		\right .
	\end{eqnarray}
	where
	\begin{eqnarray}
		&&I(a) = \frac{d}{s^2}\  {}_2F_1 \left (2, \frac{1}{\alpha}, 1 + \frac{1}{\alpha}, \frac{\gamma a}{s} \right ),  \label{I}\\
		&&I'(a) = \frac{d}{\alpha a s^2} \left [ \frac{1}{\displaystyle \left (1 - \frac{\gamma a}{s} \right )^2} - {}_2F_1 \left (2, \frac{1}{\alpha}, 1 + \frac{1}{\alpha}, \frac{\gamma a}{s}\right ) \right ], \label{I'}
	\end{eqnarray}
	and $[a_{\min}, a_{\max}]$ is the interval where $b(a)$ is real-valued, i.e. 
	\be
	\frac{h}{h-d + s^2 I(a)} - a^2 \ge 0.
	\ee
	The interval must contain zero in order to have $m(\theta) > 0$ for all $\theta$. It is sufficient to define the solution for $\theta \in [0, \pi]$ because of the symmetry of the problem (see Section 4). The value $a_0$ corresponds to $\displaystyle \theta = \frac{\pi}{2}$, it is found from the equation
	\be
	2a [h-d+s^2 I(a)]^2 + s^2 h I'(a)] = 0.
	\ee
	
	We note that for many values of $\alpha$ the hypergeometric function featured in the solution reduces to elementary functions, e.g. 
	\begin{eqnarray}
		&& {}_2F_1(2,1,2, z) = \frac{1}{1-z} \quad (\alpha = 1), \\
		&& {}_2F_1(2,1/2,3/2, z) = \frac{1}{2(1-z)} + \frac{{\rm arctanh}\ \sqrt{z}}{2 \sqrt{z}} \quad (\alpha = 2), \\
		&& {}_2F_1(2,2,3, z) = \frac{2[-z - \log(1-z) + z \log (1-z)]}{(-1+z)\ z^2} \quad (\alpha = 1/2),
	\end{eqnarray}
	(e.g., {\it Wolfram Mathematica 12.1.1.0}).

\subsection{Internal waves in the rigid-lid approximation}

Solving the modal equations (\ref{ME}) - (\ref{BC2}) with $\rho_0 = \rho_2 H(z+h) + (\rho_1 - \rho_2) H(z+d)$  and the free surface condition replaced with the rigid-lid approximation (\ref{eqRL2}), we obtain, requiring the continuity of the modal function at $z=-d$,
\begin{eqnarray}
	&& \phi_1(z) = - \tilde \Lambda (m^2 + m'^2) \int_{z}^0 \frac{dz}{\hat F_1^2}, \quad -d < z < 0, \label{phi1}\\
	&& \phi_2(z) = - \frac{\tilde \Lambda (z+h)}{h-d} (m^2 + m'^2) \int_{-d}^0 \frac{dz}{\hat F_1^2},  \quad -h < z < -d ,
	\label{phi2} 
\end{eqnarray}
where $\displaystyle \hat F_1 = -s + \gamma \left (\frac{z+d}{d} \right )^{\alpha} (m \cos \theta - m' \sin \theta)$, and $\tilde \Lambda$ is a parameter which is used to normalise the modal function to be equal to 1 at the interface.

The jump condition at $z = -d$ gives an adjustment equation for $m(\theta)$:
\begin{equation}
	(\rho_2 - \rho_1) g (h-d) (m^2 + m'^2) \int_{-d}^0 \frac{dz}{\hat F_1^2}  = \rho_1 (h-d) + \rho_2 s^2 \int_{-d}^0 \frac{dz}{\hat F_1^2},
	\label{Eqmm}
\end{equation}
yielding the equation
\begin{equation}
	m^2 + m'^2 = \frac{\rho_1 (h-d) + \rho_2 s^2 I[M(\theta)]}{(\rho_2 - \rho_1) g (h-d) I[M(\theta)]}, 
	\label{Eqmm1}
\end{equation}
where $M(\theta) = m \cos \theta - m' \sin \theta$ and $\displaystyle I[M(\theta)] = \frac{d}{s^2}\  {}_2F_1\left (2, \frac{1}{\alpha}, 1 + \frac{1}{\alpha}, \frac{\gamma}{s} M(\theta)\right )$.

When $\gamma = 0$, we have $m = 1$ and $\displaystyle {}_2F_1\left (2, \frac{1}{\alpha}, 1 + \frac{1}{\alpha}, 0\right ) = 1$, recovering the formula (\ref{eq:s_my_model}) for the speed $s$ of concentric waves in the absence of any current.

Next, the general solution of (\ref{Eqmm1}) is found in the form
$
m = a \cos \theta + b(a) \sin \theta,
$
where
\be
a^2 + b^2 = \frac{\rho_1 (h-d) + \rho_2 s^2 I(a)}{(\rho_2 - \rho_1) g (h-d) I(a)},
\ee
and the singular solution is found in parametric form by requiring $\displaystyle \frac{dm}{da} = 0$, which yields, for $\theta \in [0, \pi]$,
\begin{eqnarray}
	&&b(a) = \sqrt{\frac{\rho_1 (h-d) + \rho_2 s^2 I(a)}{(\rho_2 - \rho_1) g (h-d) I(a)} - a^2} , \\
	&&\theta(a) =  \left \{ 
	\begin{array}{c}
		\arctan \frac{\displaystyle 2b (\rho_2 - \rho_1) g [I(a)]^2}{\displaystyle 2a (\rho_2 - \rho_1) g  [I(a)]^2 + \rho_1  I'(a)}, \quad \mbox{if} \quad a \in[a_0, a_{max}]\  \left (\displaystyle \theta \in \left [0, \frac{\pi}{2}\right ]\right ), \\[3ex]
		\arctan \frac{\displaystyle 2b (\rho_2 - \rho_1) g [I(a)]^2}{\displaystyle 2a (\rho_2 - \rho_1) g  [I(a)]^2 + \rho_1  I'(a)} + \pi, \quad \mbox{if} \quad a \in[a_{\min}, a_0]\ \left (\displaystyle \theta \in \left [\frac{\pi}{2}, \pi \right ]\right ), 
	\end{array}
	\right .
\end{eqnarray}
where $I(a)$ and $I'(a)$ are given by (\ref{I}) and (\ref{I'}), respectively.
It is sufficient to define the solution for $\theta \in [0, \pi]$ because of the symmetry of the problem.
The interval $[a_{\min}, a_{\max}]$ is the interval where $b$ is real-valued, i.e. 
\be
\frac{\rho_1 (h-d) + \rho_2 s^2 I(a)}{(\rho_2 - \rho_1) g (h-d) I(a)} - a^2 \ge 0.
\ee
The interval must contain zero in order to have $m > 0$ for all $\theta$. The value $a_0$ corresponds to $\theta = \pi/2$ and is found from the condition
\be
2a (\rho_2 - \rho_1) g  [I(a)]^2 + \rho_1  I'(a) = 0.
\ee

When $\alpha = 1$ the singular solution can be rewritten in the form
\be
m = \sqrt{1 + \left ( \frac{\rho_1 \gamma s}{2 (\rho_2 - \rho_1) g d} \right )^2} - \frac{\rho_1 \gamma s}{2 (\rho_2 - \rho_1) g d} \cos \theta.
\ee
}

\begin{figure}
	\centering
	\begin{subfigure}[b]{0.49\textwidth}
		\includegraphics[width=\textwidth]{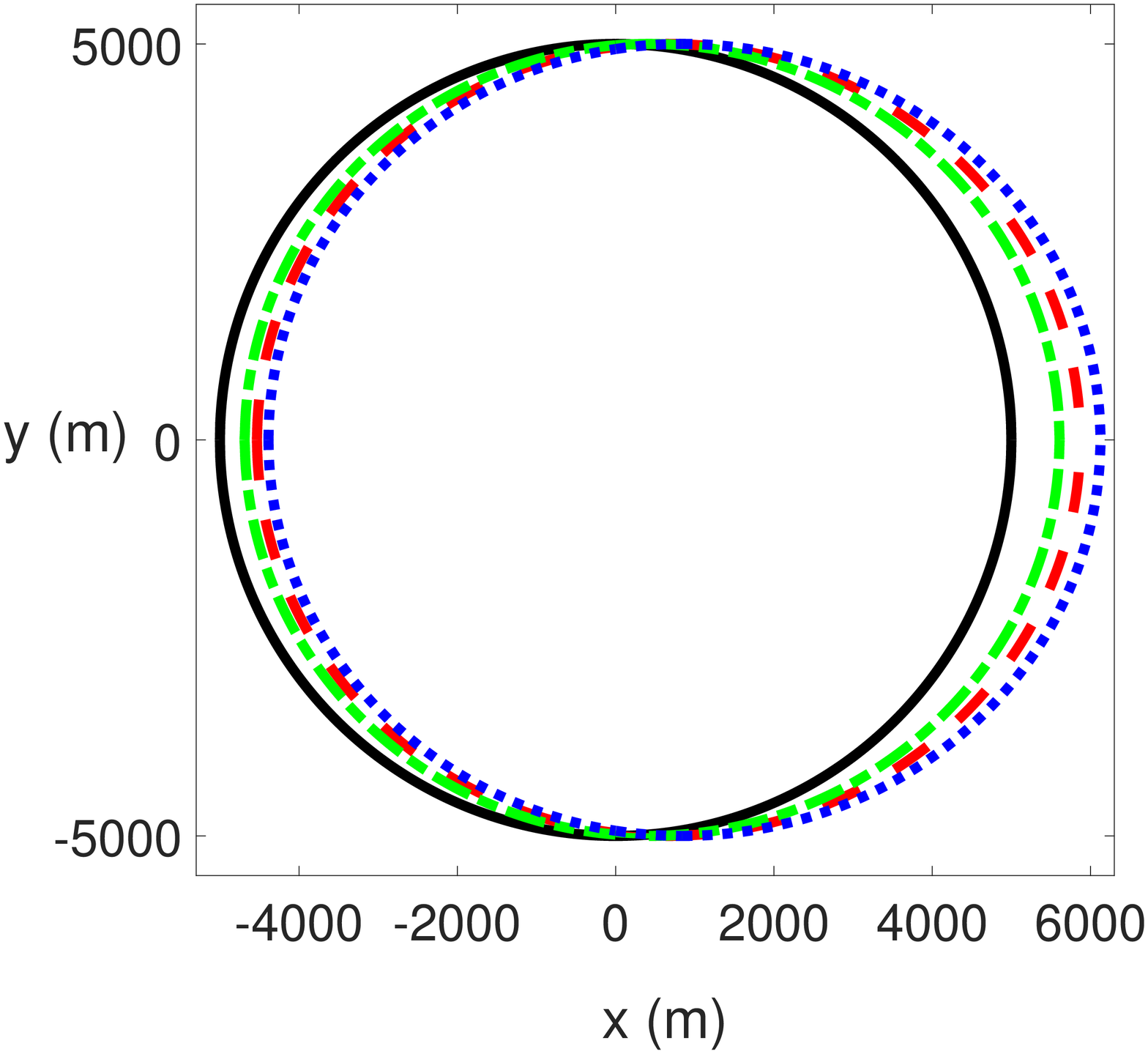}
	\caption{Surface mode: $\gamma = 0\ ms^{-1}$ (black, solid), and $\gamma = 5\ ms^{-1}$ with $\alpha = 0.5$ (blue, dot), $\alpha = 1$ (red, dash) and $\alpha = 2$ (green, dash-dot).}
	\label{fig:HG_surface_wavefronts}
\end{subfigure}
~ 
\begin{subfigure}[b]{0.49\textwidth}
	\includegraphics[width= \textwidth]{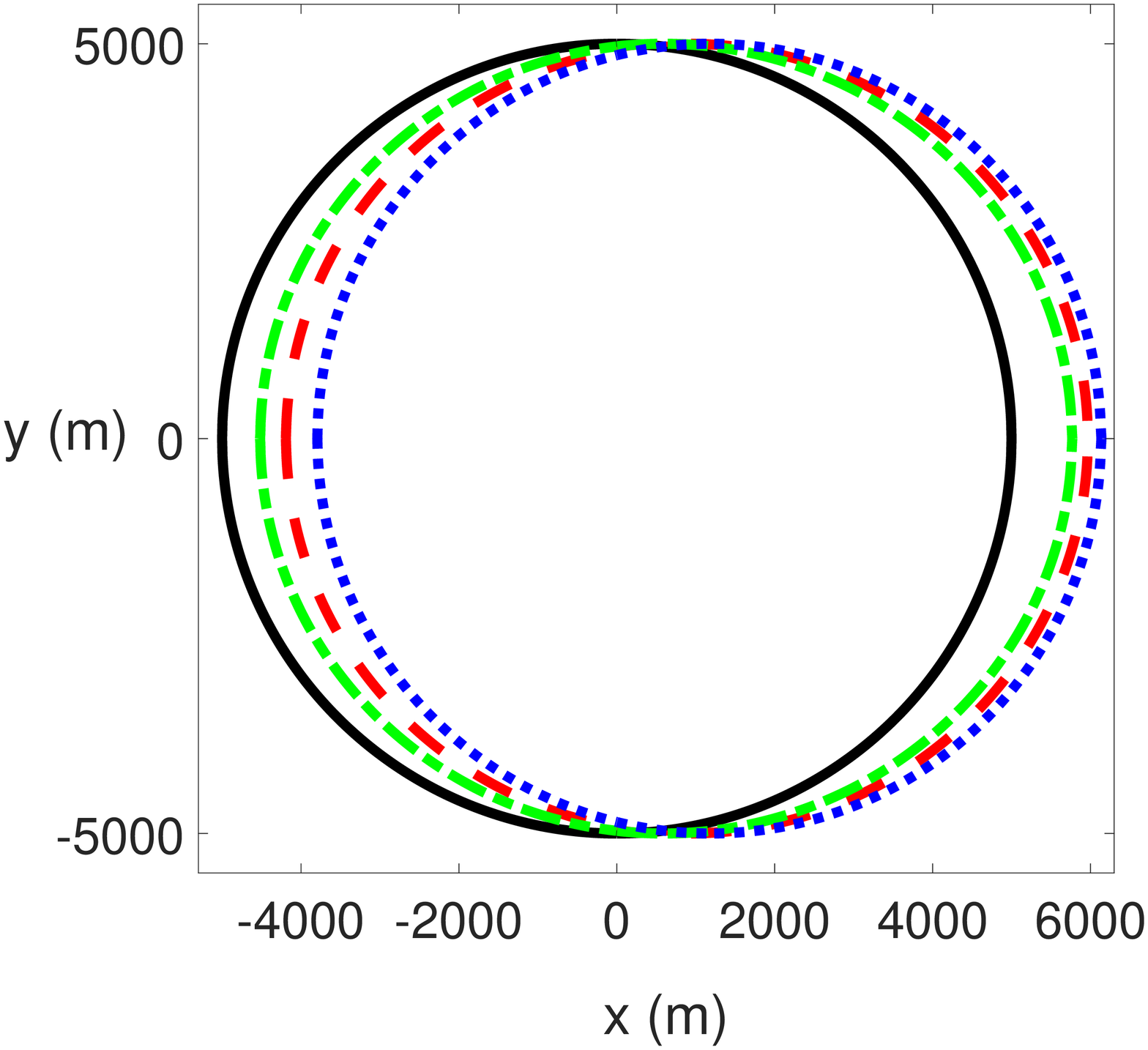}
	\caption{Interfacial mode: $\gamma = 0\ ms^{-1}$ (black, solid), and $\gamma = 0.5\ ms^{-1}$ with $\alpha = 0.5$ (blue, dot), $\alpha = 1$ (red, dash) and $\alpha = 2$ (green, dash-dot).}
	\label{fig:HG_interface_wavefronts}
\end{subfigure}
\caption{Plots of the wavefronts for the surface mode in a homogeneous fluid (a) and interfacial mode of a two-layer fluid with rigid lid approximation (b). The current is given by \eqref{eqn:U-L_C}. }   
\label{fig:HG_wavefronts}
\end{figure}

Plots of the wavefronts described by $rm(\theta) = 5000\ m$ of surface ring waves  in a homogeneous fluid and the interfacial ring waves in a two-layer fluid in the rigid-lid approximation  are shown in Figure \ref{fig:HG_wavefronts} for a set values of $\gamma$ and different values of $\alpha$. We keep $\rho_1=1000\ kg\ m^{-3}$, $\rho_2=1020\ kg\ m^{-3}$ and $d=5\ m$ as before.

The wavefronts of surface ring waves show the same qualitative features as in our previous two examples and become elongated in the direction of the flow. It is not immediately clear from Figure \ref{fig:HG_interface_wavefronts} if the wavefronts at the interface are elongated or squeezed, and we shall use the quantitative measure introduced in Section 2 to clarify that.

\begin{figure}
	\centering
	\begin{subfigure}[b]{0.49\textwidth}
		\includegraphics[width=\textwidth]{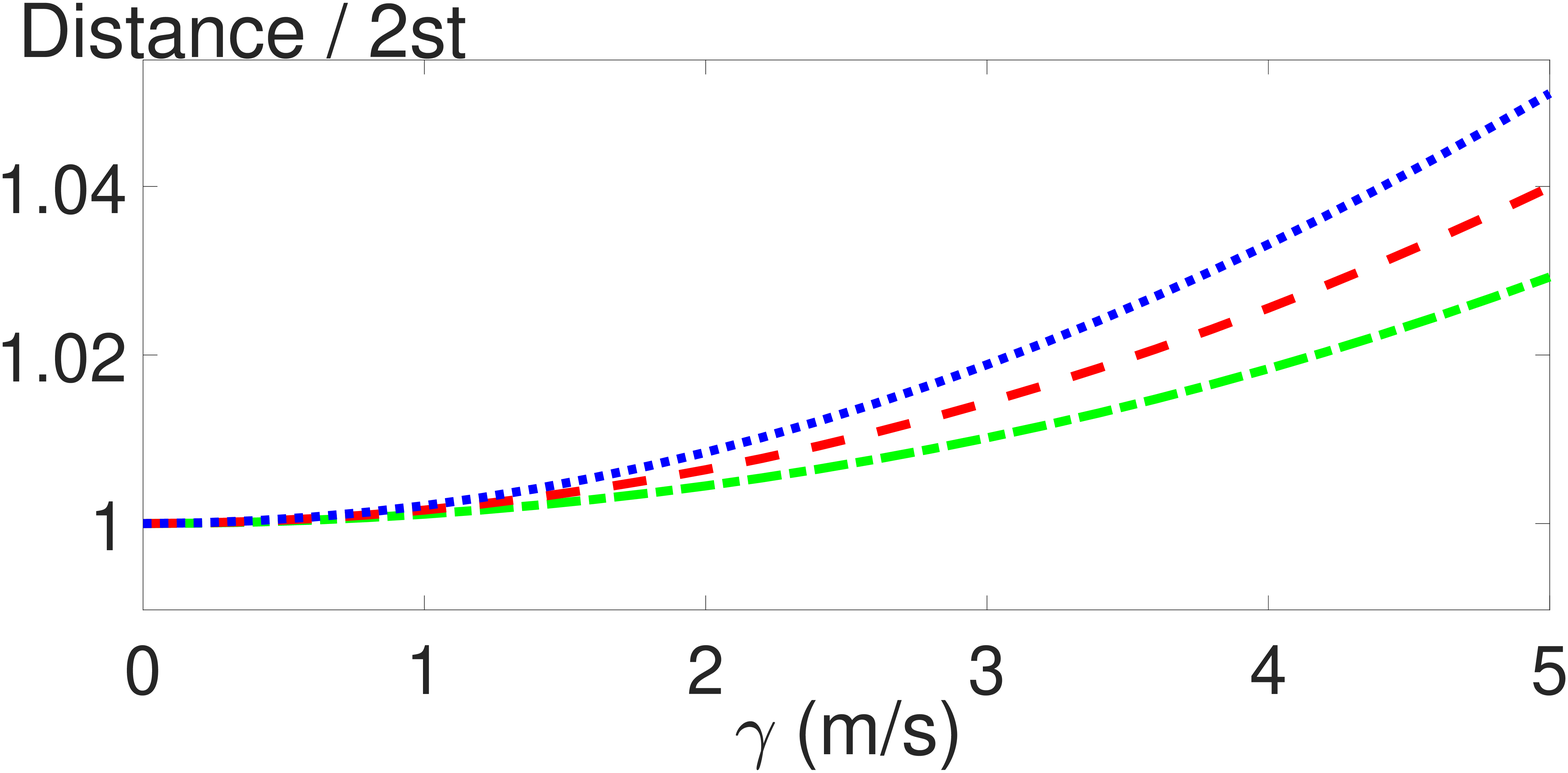}
		\caption{Surface mode.}
		\label{fig:HG_surface_distance}
	\end{subfigure}
	~ 
	\begin{subfigure}[b]{0.49\textwidth}
		\includegraphics[width= \textwidth]{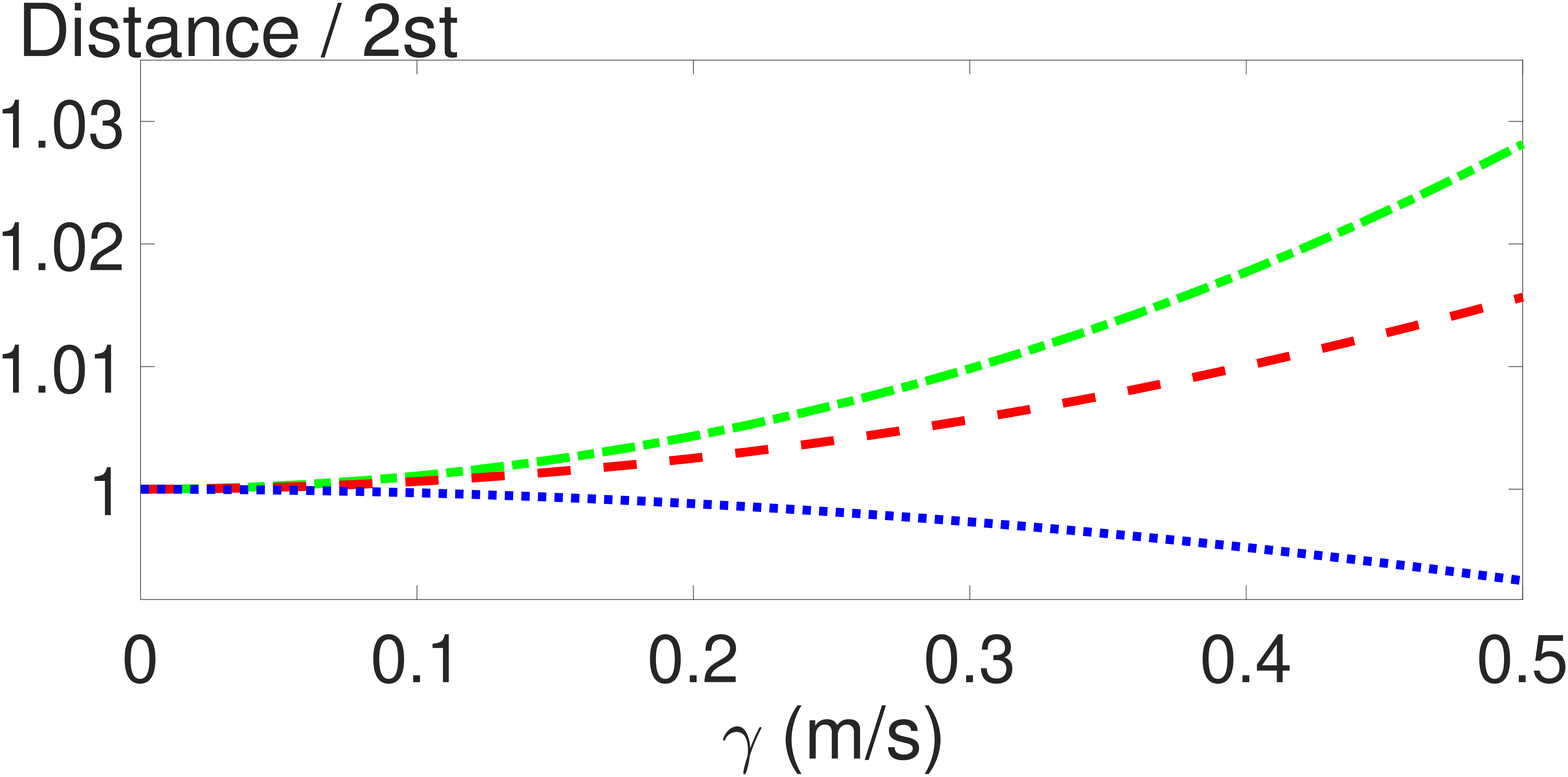}
		\caption{Interfacial mode.}
		\label{fig:HG_interface_distance}
	\end{subfigure}
	\caption{Relative distance between the points on the wavefronts in downstream and upstream directions as a function of $\gamma$ for (a) surface ring waves in a homogeneous fluid and (b) interfacial ring waves in a two-layer fluid with rigid lid approximation with $\alpha = 0.5$ (blue, dot), $\alpha = 1$ (red, dash) and $\alpha = 2$ (green, dash-dot). Here, $g = 9.8\ m s^{-2}$ and $h = 10\ m$.}   
	\label{fig:HG_distance}
\end{figure}

\begin{figure}
	\centering
	\begin{subfigure}[b]{0.49\textwidth}
		\includegraphics[width=\textwidth]{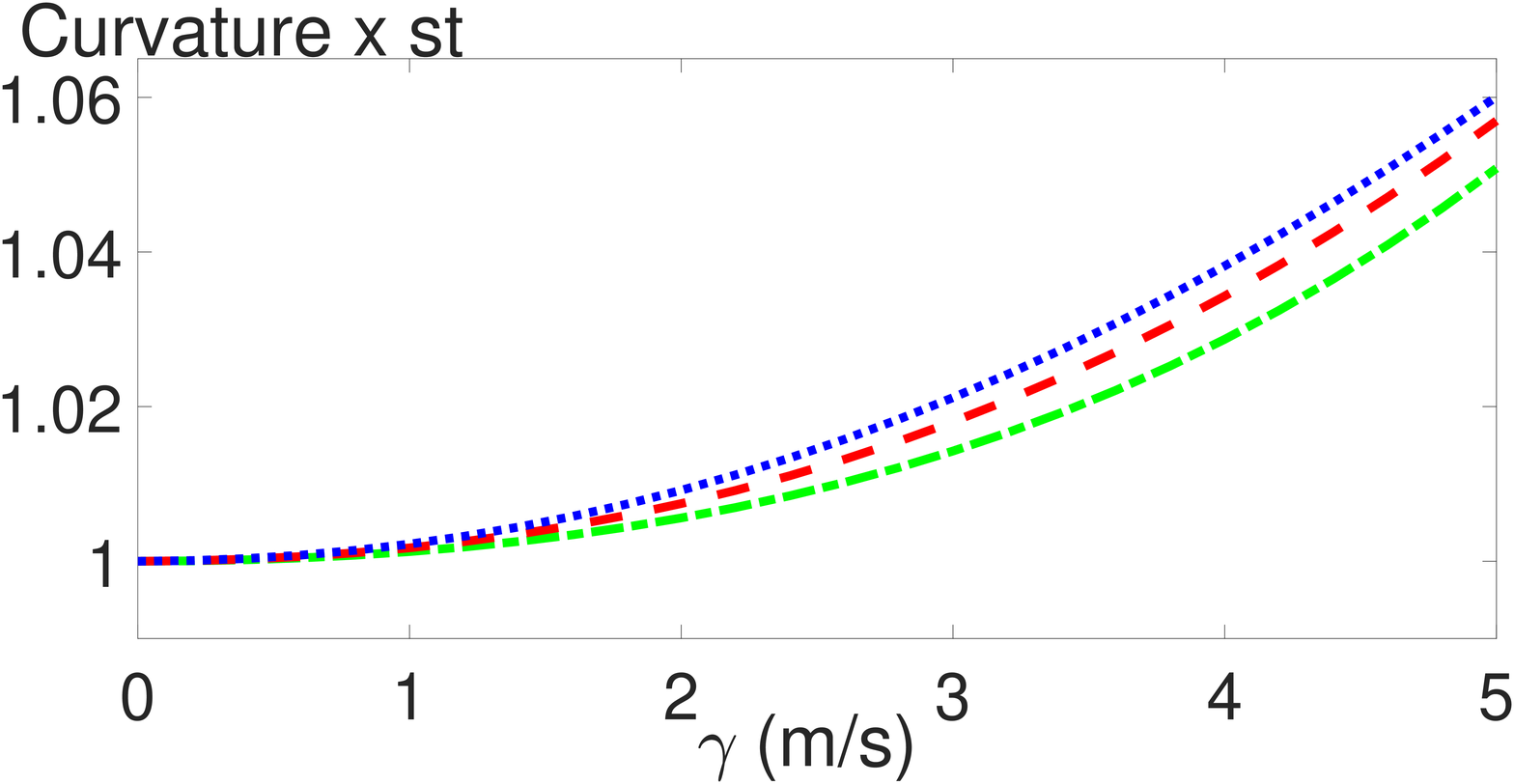}
		\caption{Surface mode.}
		\label{fig:HG_surface_curv_gamma}
	\end{subfigure}
	~ 
	\begin{subfigure}[b]{0.49\textwidth}
		\includegraphics[width= \textwidth]{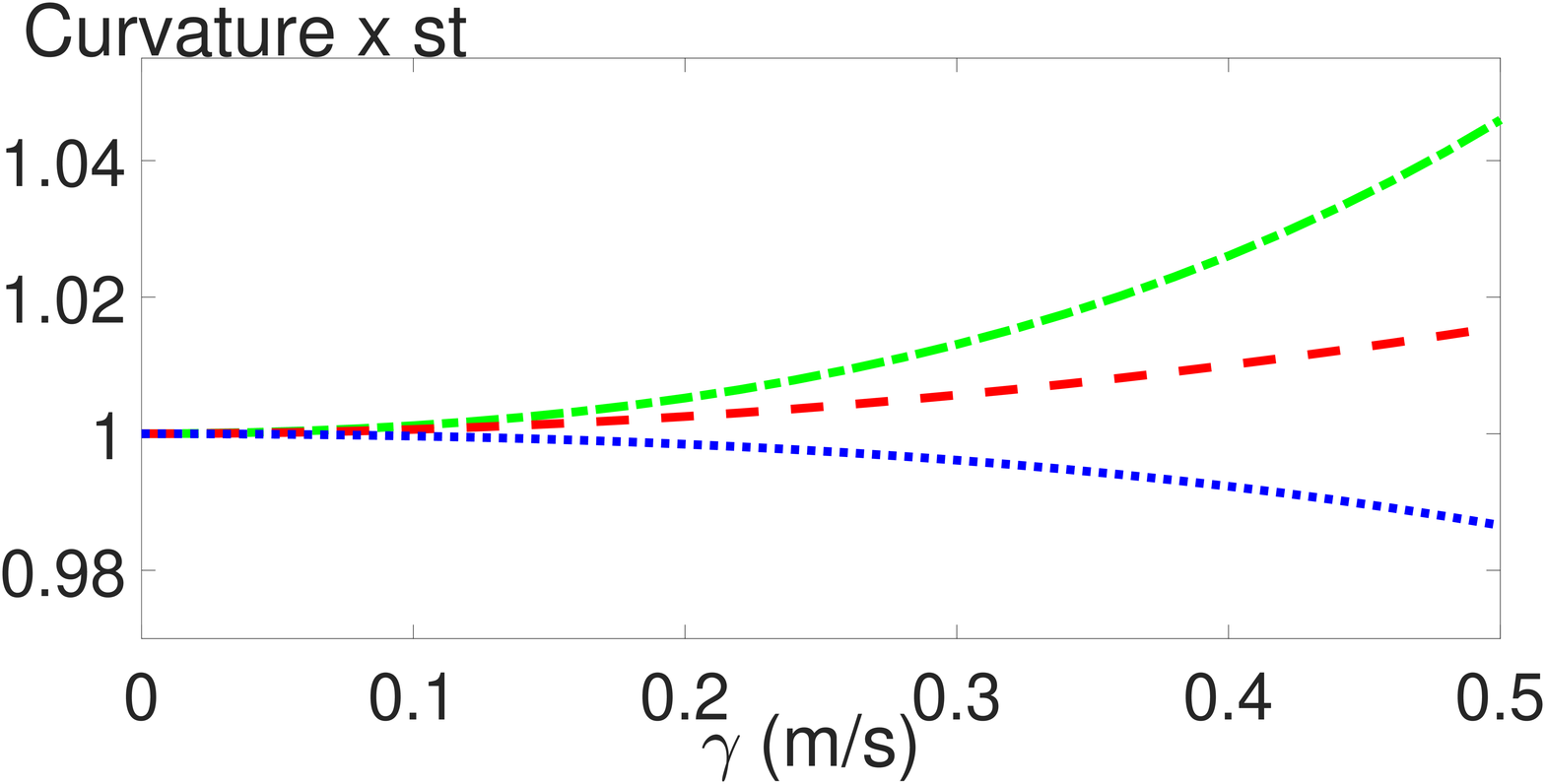}
		\caption{Interfacial mode.}
		\label{fig:HG_interface_curve_gamma}
	\end{subfigure}
	\caption{Relative curvature between the points on the wavefronts in downstream and upstream directions as a function of $\gamma$ for (a) surface ring waves in a homogeneous fluid and (b) interfacial ring waves in a two-layer fluid with rigid lid approximation with $\alpha = 0.5$ (blue, dot), $\alpha = 1$ (red, dash) and $\alpha = 2$ (green, dash-dot). Here, $g = 9.8\ m s^{-2}$ and $h = 10\ m$.}   
	\label{fig:HG_curv_gamma}
\end{figure}

The relative distance and curvature of the wavefronts in the downstream direction (see Section 2 for the definitions) are shown in Figure \ref{fig:HG_distance} and Figure \ref{fig:HG_curv_gamma}, respectively. Whilst the properties of the surface wavefronts show similar features to our previous examples, we note from Figure \ref{fig:HG_interface_distance} and \ref{fig:HG_interface_curve_gamma} that for $\alpha =1$ and $\alpha = 2$ the relative distance and curvature increases with $\gamma$, but both decrease for $\alpha = 0.5$. This indicates that there is some critical value $\alpha_{crit}$ between $0.5$ and $1$ where the wavefronts transition from elongation to squeezing. Thus for $\alpha > \alpha_{crit}$ the wavefronts are elongated and for $\alpha < \alpha_{crit}$ the wavefronts are squeezed.

\begin{figure}
	\centering
	\begin{subfigure}[b]{0.49\textwidth}
		\includegraphics[width=\textwidth]{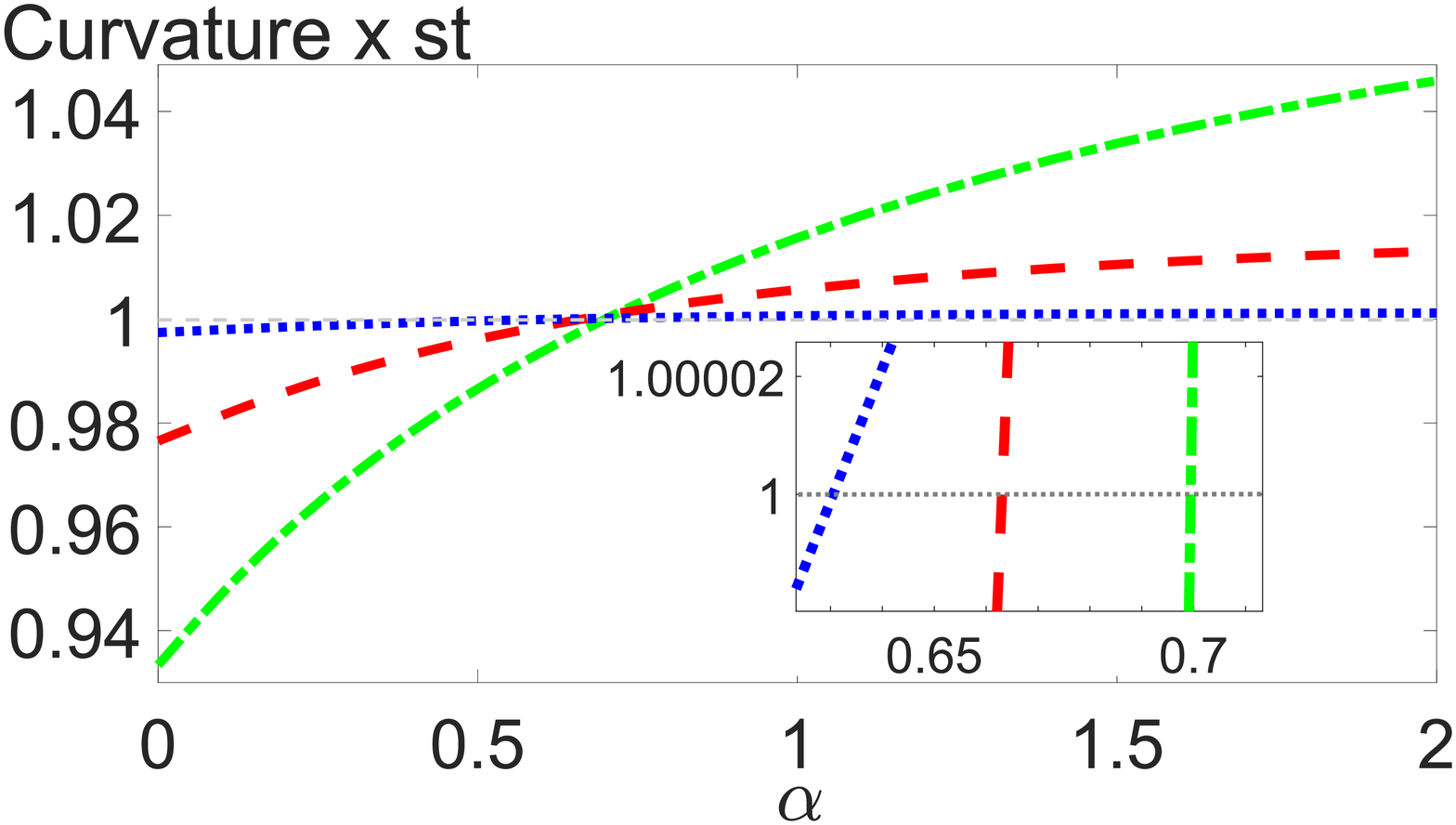}
		\caption{$\gamma = 0.1\ ms^{-1}$ (blue, dot), $\gamma = 0.3\ ms^{-1}$ (red, dash) and $\gamma = 0.5\ ms^{-1}$ (green, dash-dot) with $d = 0.5\ m$.}
		\label{fig:HG_interface_curv_g}
	\end{subfigure}
	~ 
	\begin{subfigure}[b]{0.49\textwidth}
		\includegraphics[width= \textwidth]{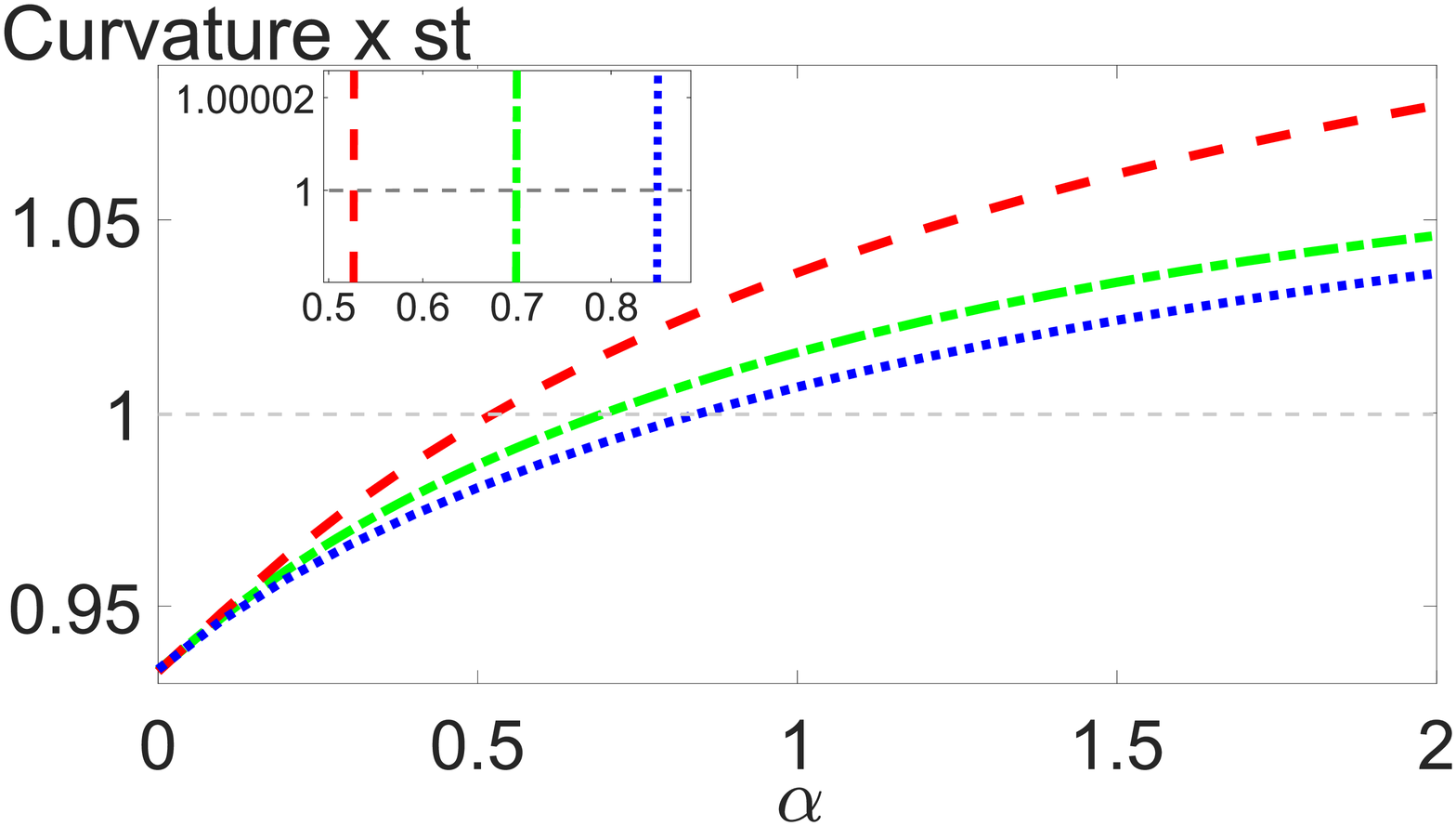}
		\caption{$d = 7\ m$ (blue, dot), $d = 5\ m$ (green, dash-dot) and $d = 3\ m$ (red, dash) \\ with  $\gamma = 0.5\ ms^{-1}$ .}
		\label{fig:HG_interface_curve_d}
	\end{subfigure}

	\begin{subfigure}[b]{0.49\textwidth}
	\includegraphics[width=\textwidth]{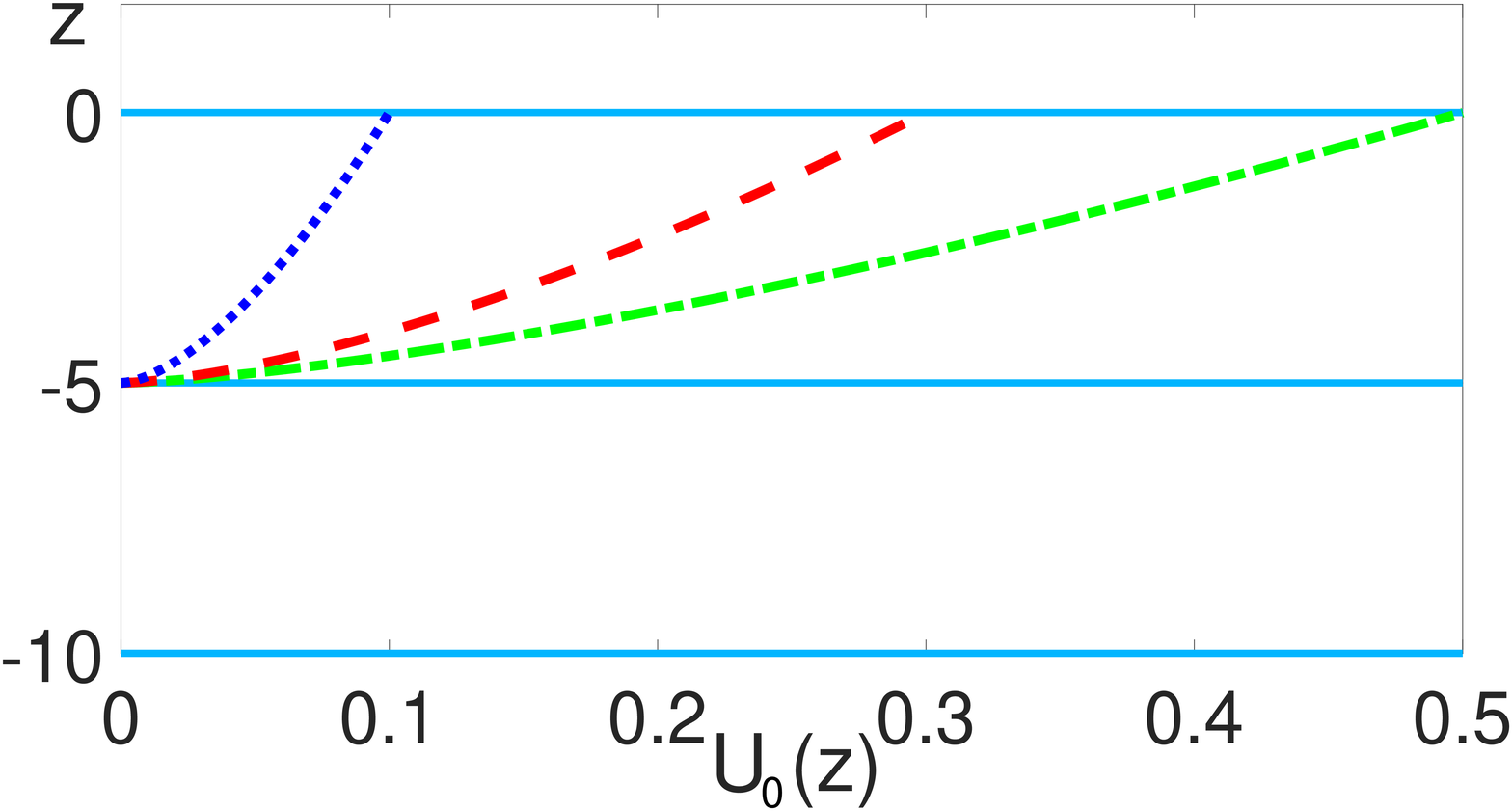}
	\caption{Currents for $d = 0.5\ m$ with $\gamma = 0.1\ ms^{-1}$ (blue, dot), $\gamma = 0.3\ ms^{-1}$ (red, dash) and $\gamma = 0.5\ ms^{-1}$ (green, dash-dot), all with $\alpha = \alpha_{crit}$.}
	\label{fig:current_g}
\end{subfigure}
~ 
\begin{subfigure}[b]{0.49\textwidth}
	\includegraphics[width= \textwidth]{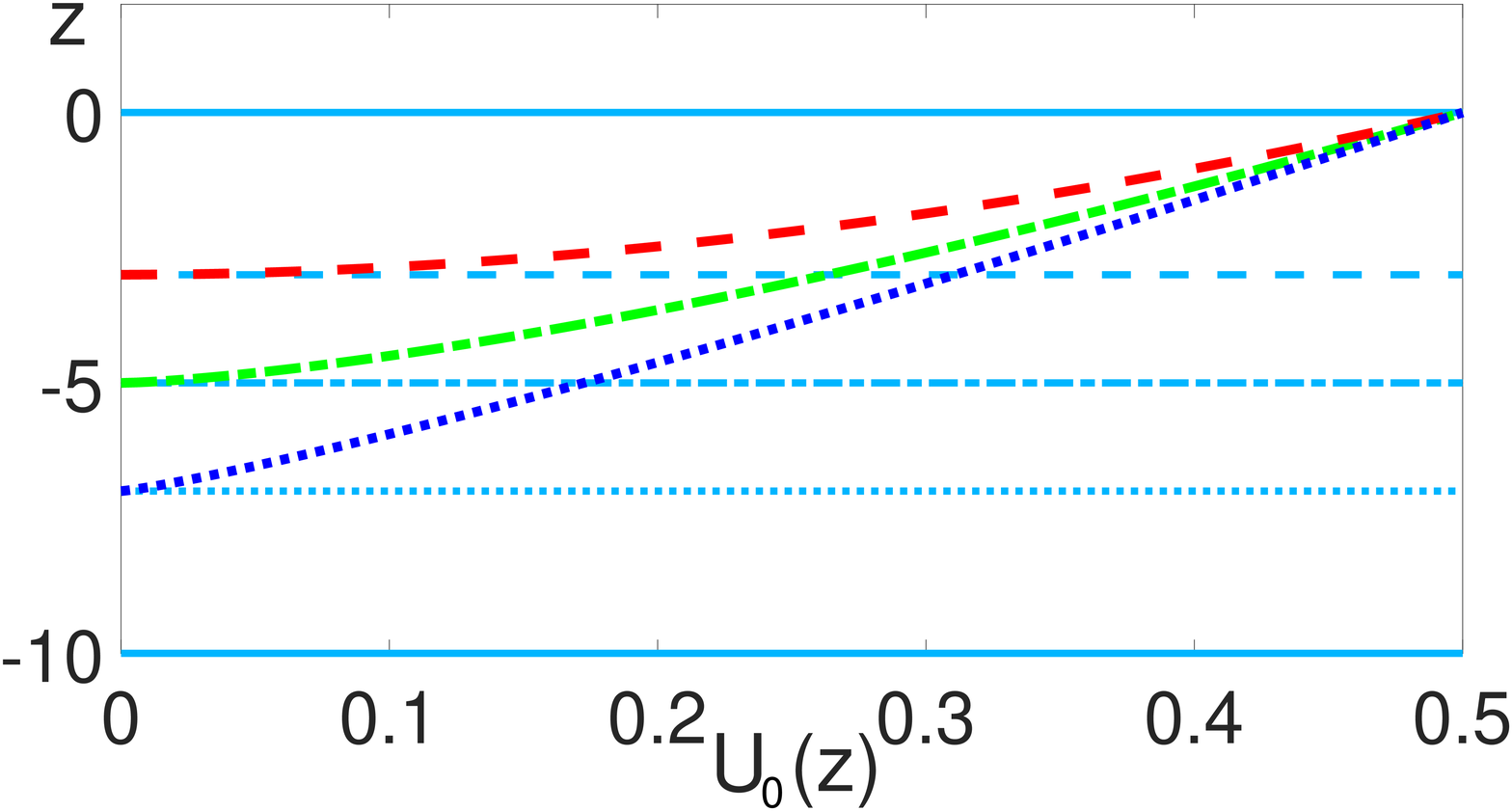}
	\caption{Currents for $\gamma = 0.5\ ms^{-1}$ with $d = 7\ m$ (blue, dot), $d = 5\ m$ (green, dash-dot) and $d = 3\ m$ (red, dash), all with $\alpha = \alpha_{crit}$.}
	\label{fig:current_d}
\end{subfigure}
	\caption{Plots of the relative curvature for interfacial wavefronts as a function of $\alpha$ for (a) different values of $\gamma$ and (b) different values of $d$. The value of $\alpha_{crit}$ is located when the curvature equals 1 (horizontal grey dash). The corresponding currents are shown below in (c) and (d) respectively with $\alpha = \alpha_{crit}$. The light blue (horizontal) lines show the surface, interface and bottom of the fluid in each case.}   
	\label{fig:HG_interface_curve_alpha}
\end{figure}

The value of $\alpha_{crit}$ can be seen in Figure \ref{fig:HG_interface_curve_alpha} where the curvature in the downstream direction is plotted as a function of $\alpha$ for different values of $\gamma$. Squeezing is observed where curvature is less than 1, elongation where curvature is greater than 1, and $\alpha_{crit}$ is located when curvature equals 1. From Figure \ref{fig:HG_interface_curv_g}, the value of $\alpha_{crit}$ is smaller for weaker currents given the same value of $d$, and from Figure \ref{fig:HG_interface_curve_d} we see that the value of $\alpha_{crit}$ is smaller for smaller values of $d$ (i.e. when the interface is closer to the surface). The corresponding currents are shown in Figure \ref{fig:current_g} and Figure \ref{fig:current_d}.

\begin{figure}
	\centering
	\begin{subfigure}[b]{0.49\textwidth}
		\includegraphics[width=\textwidth]{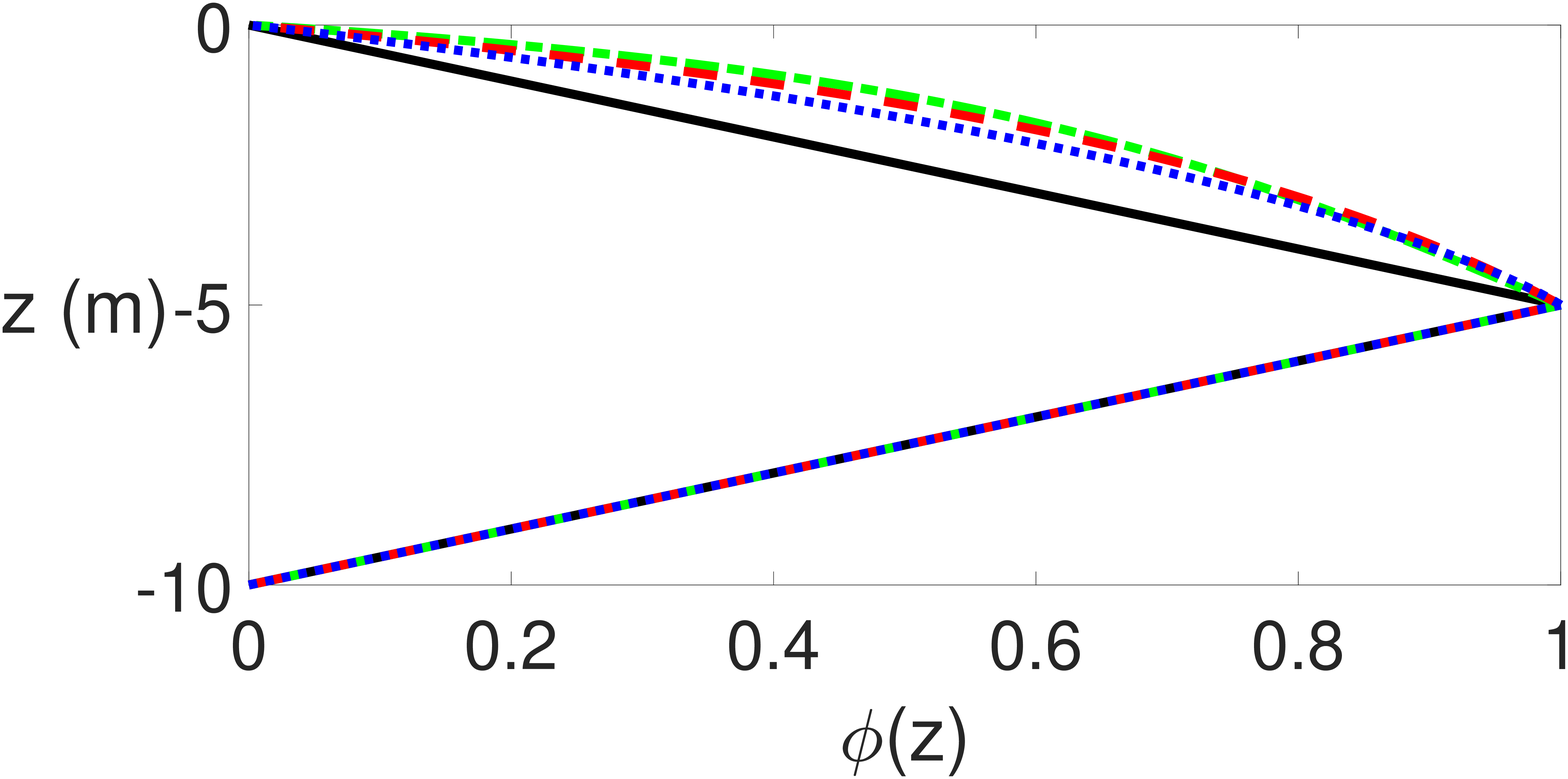}
		\caption{$\theta = 0$ (downstream)}
		\label{fig:HG_0}
	\end{subfigure}
	~ 
	\begin{subfigure}[b]{0.49\textwidth}
		\includegraphics[width=\textwidth]{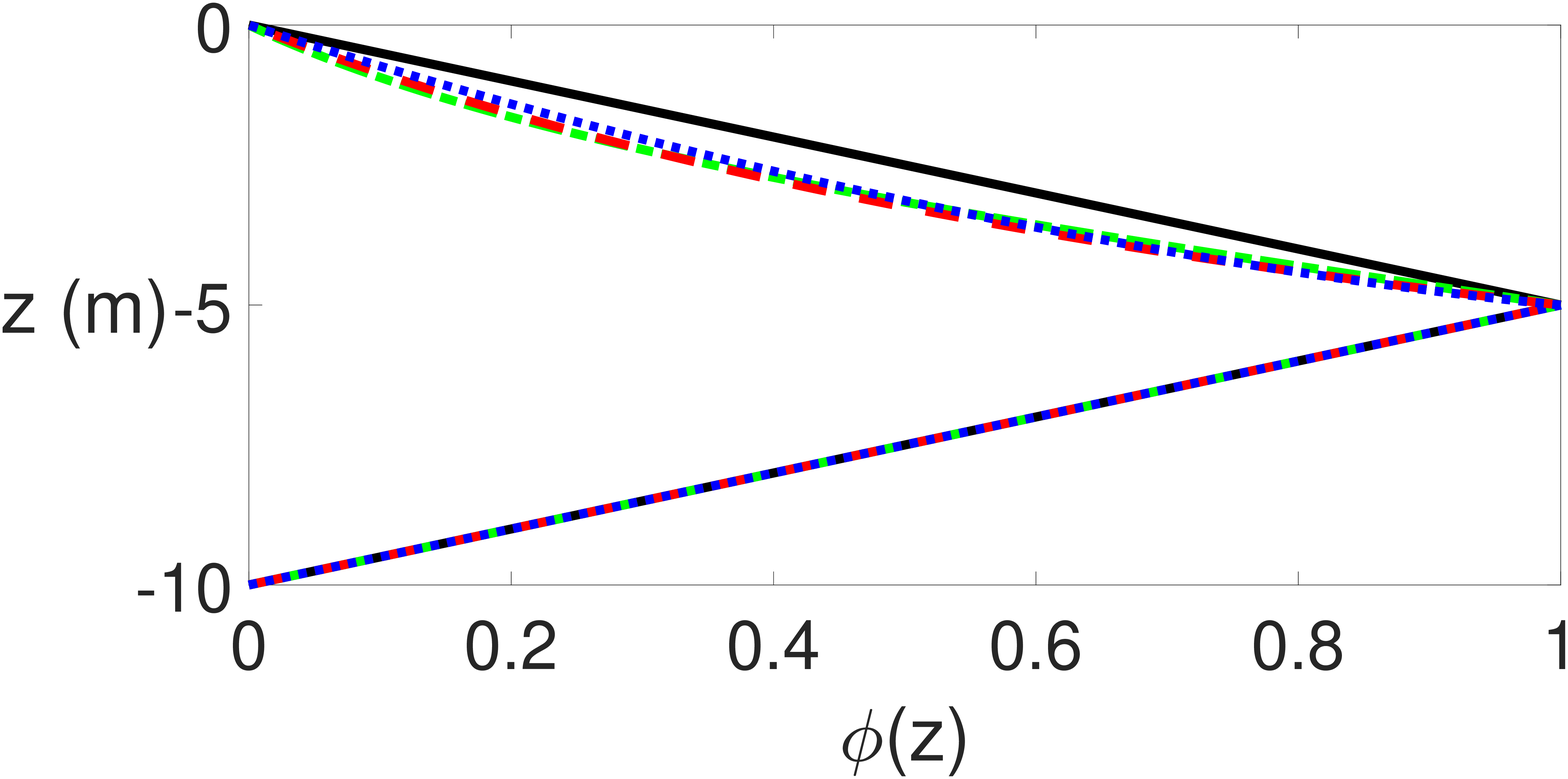}
		\caption{$\theta = \pi$ (upstream)}
		\label{fig:HG_pi}
	\end{subfigure}
	\begin{subfigure}[b]{0.49\textwidth}
		\includegraphics[width=\textwidth]{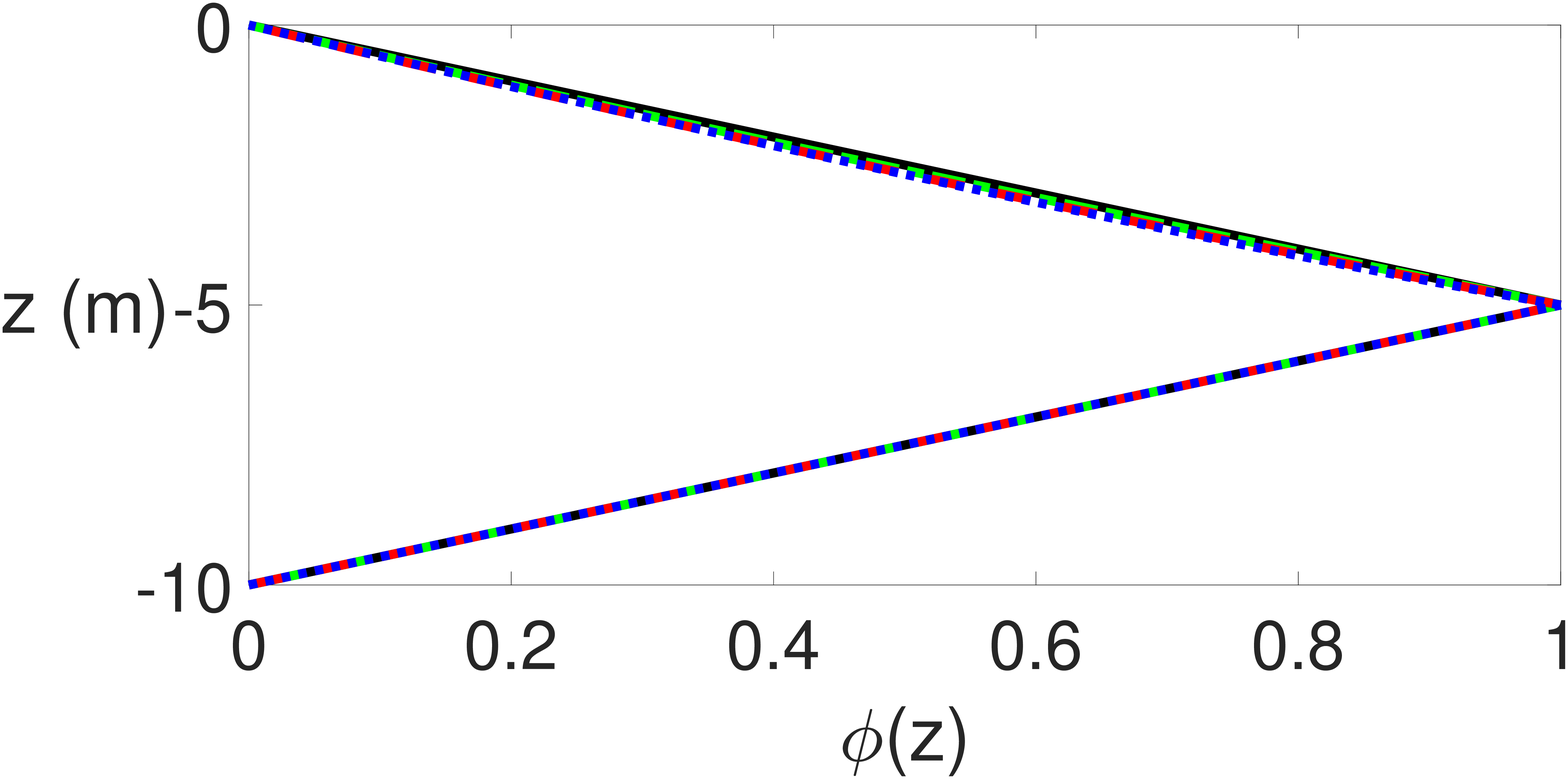}
		\caption{$\theta = \pi/2$ (orthogonal)}
		\label{fig:HG_half_pi}
	\end{subfigure}
	
	\caption{Plots of the modal functions \eqref{phi1} and \eqref{phi2} for $\gamma = 0\ m s^{-1}$ (black, solid) and $\gamma = 0.5\ m s^{-1}$ with $\alpha = 0.5$ (blue, dot), $\alpha = 1$ (red, dash) and $\alpha = 2$ (green, dash-dot). Here, $g = 9.8\ m s^{-2}$ and $h = 10\ m$. 
	}    
	\label{fig:HG_modal_2L}
\end{figure}

The modal function of the two-layer fluid with upper layer current given by equations \eqref{phi1} and \eqref{phi2} is plotted in Figure \ref{fig:HG_modal_2L} in the downstream, upstream and orthogonal directions. Naturally, all functions are almost the same in the lower layer where there is no current. Once again, the effect of the shear flow is least in the orthogonal direction. Similar qualitative features are observed in the downstream and upstream directions as with the linearly increasing current, and the vertical structure is again strongly three-dimensional. 

\begin{figure}
	\centering
	\begin{subfigure}[b]{0.49\textwidth}
		\includegraphics[width=\textwidth]{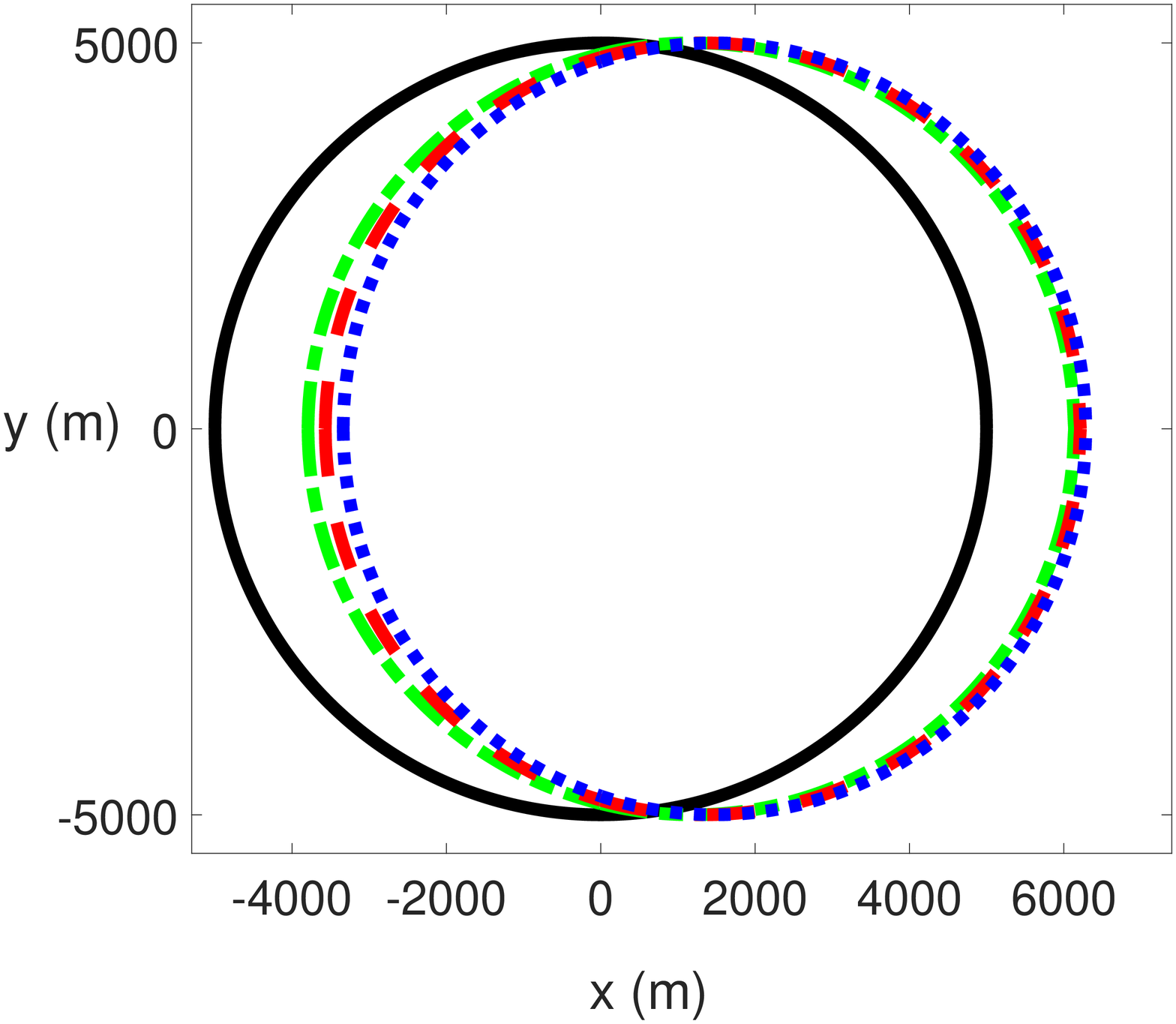}
		\caption{$\rho_2 = 1020\ kg\ m^{-3}$: $\gamma = 0\ m s^{-1}$ (black, solid) and $\gamma = 0.5\ m s^{-1}$ with $\alpha = 1/2$ (green, dash-dot), $\alpha =1/3$ (red, dash) and $\alpha =1/5$ (blue, dot).}
		\label{fig:wavefront_alpha}
	\end{subfigure}
	~ 
	\begin{subfigure}[b]{0.49\textwidth}
		\includegraphics[width= \textwidth]{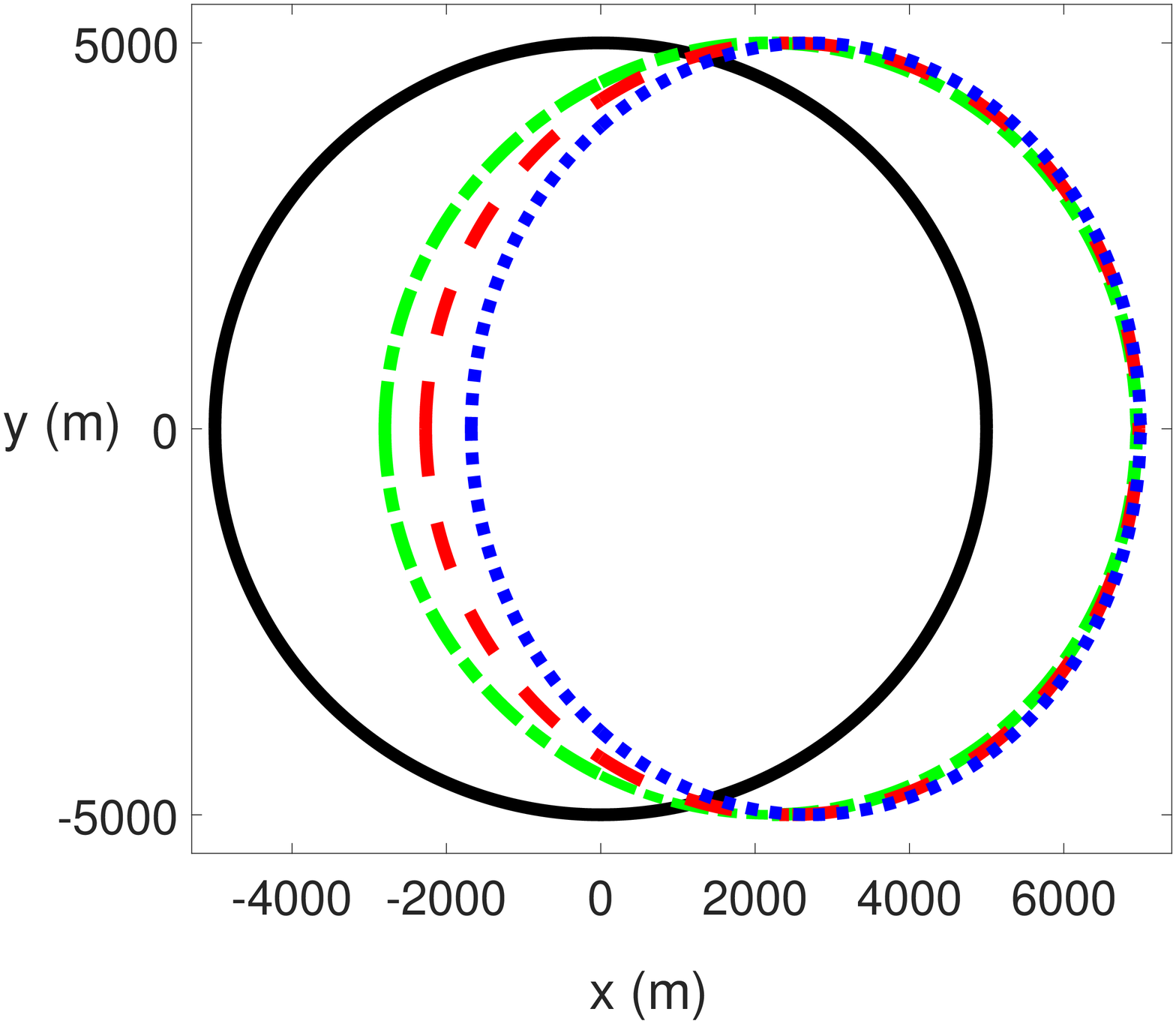}
		\caption{$\rho_2 = 1006\ kg\ m^{-3}$: $\gamma = 0\ m s^{-1}$ (black, solid) and $\gamma = 0.5\ m s^{-1}$ with $\alpha = 1/2$ (green, dash-dot), $\alpha =1/3$ (red, dash) and $\alpha =1/5$ (blue, dot).}
		\label{fig:wavefront_rho}
	\end{subfigure}
	\caption{Plots of the wavefronts for the interfacial mode of a two-layer fluid with rigid lid approximation and different lower densities. The current is given by \eqref{eqn:U-L_C}. }   
	\label{fig:wavefronts_alpha_rho}
\end{figure}

The squeezing of the interfacial wavefronts with decreasing $\alpha$ can be seen in Figure \ref{fig:wavefront_alpha}. The effect is strengthened considerably when the density jump is smaller (i.e. when the interfacial waves are slower) which we achieve by decreasing $\rho_2$ from $1020\ kg\ m^{-3}$ to $1006\ kg\ m^{-3}$. The wavefronts are plotted in Figure \ref{fig:wavefront_rho} for the same values of $\gamma$ and $\alpha$ as in Figure \ref{fig:wavefront_alpha}. In all cases, $ \displaystyle \gamma < \frac{s}{m(0)}$, i.e. there are no critical layers.

\section{Conclusion}

In this study we 
linked the description of the ring waves in cylindrical geometry with the description of the plane waves tangent to the ring wave and propagating at various angles to the shear flow. It was shown, in particular, that geometry of the wavefronts of both sets of waves can be described by one and the same {\it directional adjustment equation} for the {\it speed modifying function} $m(\theta)$. The general solution of this nonlinear first-order differential equation corresponds to the plane waves tangent to the ring wave, while its singular solution (i.e. the envelope of the general solution) describes the ring wave. All considerations of the paper were developed in dimensional form and using the notations suitable for oceanographic applications.

Working with a local wave vector and wave frequency we defined the group speed and wave action conservation law for the ring waves. We also described a convenient analytical procedure for the construction of hybrid wavefronts consisting of a part of a ring wave and two tangent plane waves. This is straightforward in the absence of a current when the ring wave is concentric, but not so when the wavefront of the ring wave is deformed by a shear flow. Such a hybrid wavefront has perfectly matched speeds and slopes at the junctions and may propagate as a whole. Here, we concerned ourselves only with the kinematics of these solutions, in the spirit of the papers by \cite{OS, OSt}. The description of the evolution of nonlinear hybrid waves is an interesting future problem.

The main focus of the paper was on the analysis of the modal equations for several configurations motivated by geophysical applications. The modal equations constitute a new spectral problem, which must be solved in order to describe the wavefronts and vertical structure of these three-dimensional waves, and to calculate the coefficients of the amplitude equation (see Appendix A). 

Two main examples presented in the paper were devoted to the description of surface and interfacial waves in a two-layer fluid with a linear shear current and a power-law upper-layer current.  We obtained solutions of the modal equations $\phi(z; \theta)$ together with the spectral function $\displaystyle \frac{s}{m(\theta)}$, defining the wave speed in different directions to the current.

Assuming weak stratification, we have shown that the modal function for the surface ring wave is captured well in the approximation of a homogeneous fluid, while the rigid-lid approximation works very well for the interfacial modal function. The modal functions have been normalised to be equal to one either on the entire surface (for the surface mode) or at the entire interface (for the interfacial mode), i.e. simultaneously in all directions. 
This normalisation is convenient since the vertical particle displacement at these levels is then described simply by the amplitude function which can be found by solving the appropriate cylindrical Korteweg-de Vries - type equation (see Appendix A).
Our analysis shows that the vertical structure of the ring waves propagating over a parallel shear flow strongly depends not only on the depth, but also on the angle to the current at each depth, with the greatest changes, compared to the case of the waves in the absence of the shear flow, being in the downstream and upstream directions. Moreover, the vertical structure is shifted towards the surface downstream, but towards the ocean bottom upstream. 



The solutions were used to analyse the behaviour of the two-dimensional wavefronts and vertical structure of the wave field for increasing strengths of the shear flow.  We considered only sufficiently weak flows such that there are no critical levels. Both a simple sufficient condition for the absence of the critical levels, and the necessary and sufficient condition  were formulated under the assumption that there are no current reversals, and were satisfied in all examples in the paper. 

We also introduced and used global (distance) and local (curvature) measure for the deformation of the wavefronts, which allowed us to study the transition from the regime of elongating wavefronts to the regime of squeezed wavefronts for interfacial waves in a two-layered fluid over a power-law upper-layer current
\begin{eqnarray}
u_0 (z) = \left \{ 
\begin{array}{c}
	\displaystyle \gamma \left (\frac{z+d}{d} \right )^{\alpha}, \quad \mbox{if} \quad -d < z < 0, \\
	0, \quad \mbox{if} \quad  -h < z < -d,
\end{array}
\right .
\end{eqnarray}
with some positive constants $\gamma$ (surface strength) and $\alpha > 0$. 
 The currents with $\alpha > 1$ could be used to model wind-generated currents, while the currents with $\alpha < 1$ can describe river inflows and exchange flows in straits, with the value of $\alpha$ fitted to observational data. This family of currents tends to a piecewise-constant current as $\alpha \to 0$  (see Figure \ref{fig:UL_current}). 

The solution of the directional adjustment equation, and the modal function, have been presented in terms of the hypergeometric function ${}_2F_1(2, \frac{1}{\alpha}, 1 + \frac{1}{\alpha}, z)$. While all currents were assumed to have one and the same surface strength, they varied in the bulk of the upper layer, and this has a strong effect of the behaviour of the interfacial ring waves \citep{K}. The quantitative measures introduced in the current paper allowed us to study the dependence of the critical value of $\alpha$ on the surface strength of the current, interfacial depth and density jump. Strong squeezing of the interfacial ring waves can be observed for sufficiently small values of $\alpha$, and the effect is stronger for a smaller density jump. This observation invites theoretical studies into the stability of both plane and ring waves on that family of currents since the presence of the strong squeezing could be indicative of the presence of a long-wave instability for stronger currents. Of special interest is the related study of stability of the hybrid solutions consisting of the matched arc of a ring wave and two tangent plane waves. Similarly looking solutions are often present on satellite images of internal waves generated in narrow straits (e.g. Figure 13 in \cite{A}).


\section{Acknowledgements}

We would like to thank Ricardo Barros, Wooyoung Choi, Evgeny Ferapontov, Victor Shrira and Dmitri Tseluiko for useful discussions.

 \section{Appendix A}

In dimensional variables, the amplitude equation has the following form:
\begin{eqnarray}
	&& \mu_1 A_r + \mu_2 A A_\xi + \mu_3 A_{\xi \xi \xi} + \mu_4 \frac{A}{r} + \mu_5 \frac{A_\theta}{r} = 0, \quad \mbox{where} \\
	&& \mu_1 = 2s \int_{-h}^{0} \rho_0 \hat F \phi_z^2 dz, \quad
	\mu_2 = -3m \int_{-h}^{0} \rho_0 \hat F^2 \phi_z^3 dz, \\
	&& \mu_3 = - (m^2 + m'^2)\ m \int_{-h}^0 \rho_0 \hat F^2 \phi^2 dz, \\
	&&\mu_4 = - \int_{-h}^0  \left \{  \frac{\rho_0 \phi_z^2 m (m+m'')}{(m^2+m'^2)^2} \left ( (m^2-3m'^2) \hat F^2 -
{4 m' (m^2 + m'^2) (u_0-c)\sin \theta} \hat F  \right .  \right . \nonumber \\
&&\left . \left .  - \sin^2 \theta (u_0-c)^2(m^2 + m'^2)^2 \right )  \right .  \nonumber \\
  &&\left . +  \frac{2 \rho_0 m}{m^2 + m'^2} \hat F \phi_z \phi_{z\theta} (m' \hat F + (m^2 + m'^2) (u_0-c) \sin \theta ) \right \} ~ \mathrm{d}z,\\
	&& \mu_5 = - \frac{2 m}{m^2 + m'^2} \int_{-h}^0 \rho_0 \hat F \phi_z^2 [m'\hat F + (u_0-c) (m^2 + m'^2) \sin \theta] dz,
\end{eqnarray}
where $\xi = m r - s t$, $\hat F = -s + (u_0-c) (m \cos \theta - m' \sin \theta)$, and $\phi(z; \theta)$ is a solution of the modal equations (\ref{eq:ME1} - \ref{eq:MEC2}).

 \section{Appendix B}

Differentiating equation \eqref{abQ} with respect to $a$ we find
\begin{align}
	&2a+2bb'=Q_a
	\Rightarrow b'=\frac{Q_a-2a}{2b}
	\Rightarrow \tan\theta = -\frac{2b}{Q_a-2a}. \label{eq:curt}
\end{align}
Therefore, $m(a)=m(\theta(a))$ can be written in the form
\begin{align}
	m(a)&=a\cos\theta +b\sin\theta = 
	cos\theta\Bigg(\frac{aQ_a-2Q}{Q_a-2a}\Bigg)\nonumber \\
	&=\text{sign}(\cos\theta)\sqrt{\frac{1}{1+\tan^2\theta}}\Bigg(\frac{aQ_a-2Q}{Q_a-2a}\Bigg)\nonumber \\
	&=\text{sign}(\cos\theta)\text{sign}(Q_a-2a)\Bigg(\frac{aQ_a-2Q}{\sqrt{(Q_a-2a)^2+4b^2}}\Bigg).\label{eq:k(a)}
\end{align}
From equation \eqref{eq:curt}, 
\begin{equation}
	\text{sign}(Q_a-2a)=\frac{\text{sign}(-2b)}{\text{sign}(\tan\theta)},
	\label{eq:sign1}
\end{equation}
thus equation \eqref{eq:k(a)} becomes
\begin{equation}
	m(a) = \text{sign}\Bigg(-2b\frac{\cos\theta}{\tan\theta}\Bigg)\Bigg(\frac{aQ_a-2Q}{\sqrt{(Q_a-2a)^2+4b^2}}\Bigg).
	\label{sign2}
\end{equation}
Since $m(a)>0,$ we obtain
\begin{equation}
	\text{sign}[m(a)]=-\text{sign}\Bigg[b(aQ_a-2Q)\frac{\cos\theta}{\tan\theta}\Bigg]=1.
	\label{sign_k(a)}
\end{equation}
In the case of no shear flow $Q >0, a Q_a - 2 Q < 0$, therefore we can assume that for the small values of $\gamma$ used here these inequalities will remain true. Then, it follows from equation \eqref{sign_k(a)} that
\begin{align}
	-\text{sign}\Bigg[b(aQ_a-2Q)\frac{\cos\theta}{\tan\theta}\Bigg]=\text{sign}\Bigg(b\frac{\cos\theta}{\tan\theta}\Bigg)=1 
	\Rightarrow &\text{sign}(b)=\text{sign}\Bigg(\frac{\cos\theta}{\tan\theta}\Bigg),
\end{align}
giving us (\ref{eq:k(a)_final}) and (\ref{cases}).


\end{document}